\documentclass[aps,twocolumn,nofootinbib,preprintnumbers,superscriptaddress]{revtex4}

\usepackage[dvips]{color}
\usepackage[normalem]{ulem}
\usepackage{amsmath}
\usepackage{enumerate}
\usepackage{amsfonts}
\usepackage{epsfig}
\usepackage{yfonts}

\usepackage{graphicx}
\usepackage{bm}

\def\eg{{\it e.g. }}
\def\ie{{\it i.e.}}

\def\({\left(}
\def\){\right)}
\def\[{\left[}
\def\]{\right]}
\def\<{\langle}
\def\>{\rangle}

\newcommand\half{{\ensuremath{\frac{1}{2}}}}
\newcommand\p{\ensuremath{\partial}}

\newcommand\field[1]{{\ensuremath{\mathbb{{#1}}}}}

\newcommand\vev[1]{{\ensuremath{\left\langle{#1}\right\rangle}}}

\newcommand{\RR}{\field{R}}

\newcommand{\ZZ}{\field{Z}}

\newcommand{\be}{\begin{equation}}
\newcommand{\ee}{\end{equation}}
\newcommand{\bea}{\begin{eqnarray}}
\newcommand{\eea}{\end{eqnarray}}
\newcommand{\bwt}{\begin{widetext}}
\newcommand{\ewt}{\end{widetext}}

\newcommand{\bi}{\begin{itemize}}
\newcommand{\ei}{\end{itemize}}
\newcommand{\ben}{\begin{enumerate}}
\newcommand{\een}{\end{enumerate}}
\newcommand{\bca}{\begin{cases}}
\newcommand{\eca}{\end{cases}}
\newcommand{\bln}{\begin{align}}
\newcommand{\eln}{\end{align}}
\newcommand{\bst}{\begin{split}}
\newcommand{\est}{\end{split}}

\newcommand\al{{\alpha}}
\newcommand\ep{\epsilon}
\newcommand\sig{\sigma}
\newcommand\Sig{\Sigma}
\newcommand\lam{\lambda}

\newcommand\om{\omega}

\newcommand\ga{{\ensuremath{{\gamma}}}}
\newcommand\Ga{{\ensuremath{{\Gamma}}}}

\newcommand\De{{\ensuremath{{\Delta}}}}

\newcommand\ze{{\zeta}}

\newcommand\da{{\dagger}}

\def\th{{\theta}}

\newcommand\ov{\over}
\newcommand\ha{{\half}}

\def\le{\left}
\def\ri{\right}

\newcommand\sD{{\ensuremath{{\mathcal D}}}}

\newcommand\sG{{\ensuremath{{\mathcal G}}}}

\newcommand\sN{{\ensuremath{{\mathcal N}}}}
\newcommand\sO{{\ensuremath{{\mathcal O}}}}
\newcommand\sR{{\ensuremath{{\mathcal R}}}}

\newcommand\ut{{\underline{t}}}
\newcommand\ur{{\underline{r}}}
\newcommand\ui{{\underline{i}}}

\newcommand\bpsi{{\bar \psi}}

\renewcommand{\Im}{\textrm{Im}\,}
\renewcommand{\Re}{\textrm{Re}\,}

\newcommand{\hmq}{\hat \mu_q}
\newcommand{\bone}{{\bf 1}}
\newcommand{\vk}{{\vec k}}

\def\tildem{\tilde m}
\def\yandz{\Phi}

\begin{document}

\title{Emergent quantum criticality, Fermi surfaces, and AdS$_2$}

\preprint{MIT-CTP/4050, NSF-ITP-09xx}

%

\author{Thomas Faulkner}
\affiliation{Center for Theoretical Physics, Massachusetts Institute of Technology,
Cambridge, MA 02139 }
\author{ Hong Liu}
\affiliation{Center for Theoretical Physics,
Massachusetts
Institute of Technology,
Cambridge, MA 02139 }
\author{John McGreevy}
\affiliation{Center for Theoretical Physics, Massachusetts Institute of Technology,
Cambridge, MA 02139 }
\affiliation{KITP, Santa Barbara, CA 93106}
\author{David Vegh}
\affiliation{Center for Theoretical Physics,
Massachusetts
Institute of Technology,
Cambridge, MA 02139 }

\begin{abstract}

Gravity solutions dual to $d$-dimensional field theories at finite charge density have a near-horizon region which is $AdS_2 \times \RR^{d-1}$. The scale invariance of the $AdS_2$ region implies that at low energies the dual field theory exhibits emergent quantum critical behavior controlled by a $(0+1)$-dimensional CFT.
This interpretation sheds light on recently-discovered holographic descriptions of Fermi surfaces, allowing an analytic understanding of their low-energy excitations.
For example, the scaling behavior near the Fermi surfaces is determined by conformal dimensions in the emergent IR CFT. In particular, when the operator is marginal in the IR CFT, the corresponding spectral function is precisely of the
``Marginal Fermi Liquid'' form, postulated to describe the optimally doped cuprates.

\end{abstract}

\today

\maketitle

\tableofcontents

\section{Introduction}

The AdS/CFT correspondence~\cite{AdS/CFT}
has opened new avenues for studying strongly-coupled many-body phenomena by relating certain interacting quantum field theories to classical gravity (or string) systems.
Compared to conventional methods of dealing with many-body problems, this approach
has some remarkable features which make it particularly valuable:
\ben

\item Putting the boundary theory at finite temperature and finite density corresponds to putting a black hole in the bulk geometry.
Questions about complicated many-body phenomena at strong coupling are now mapped to
single- or few-body classical problems in a black hole background.

\item Highly dynamical, strong-coupling phenomena in the dual field theories
can often be understood on the gravity side using simple geometric pictures.

\item At small curvature and low energies a gravity theory reduces to a universal sector:
classical Einstein gravity plus matter fields. Through the duality, this limit typically translates into the strong-coupling and large-$N$ limit
of the boundary theory, where $N$ characterizes the number of species. Thus by working with
Einstein gravity (plus various matter fields) one can extract certain universal properties of a
large number of {\it strongly coupled} quantum field theories.

\een
In this paper following~\cite{Lee:2008xf,Liu:2009dm} we continue the study of non-Fermi liquids using the AdS/CFT correspondence~(see also~\cite{Cubrovic:2009ye}).

Consider a $d$-dimensional conformal field theory (CFT) with a global $U(1)$ symmetry that has an AdS gravity dual. Examples of such theories include the $\sN=4$ super-Yang-Mills (SYM) theory in $d=4$, the $\sN =8$ M2 brane theory in $d=3$,
the $(2,0)$ multiple M5-brane theory in $d=6$, and many others with less supersymmetry. With the help of the AdS/CFT correspondence, many important insights have been obtained into strongly coupled dynamics in these systems, both near the vacuum and at a finite temperature. In particular, as a relative of QCD, thermal $\sN =4$ SYM theory has been used as a valuable guide for understanding the strongly coupled Quark-Gluon Plasma of QCD.

It is also natural to ask what happens to the resulting many-body system when we put such a  theory at a finite $U(1)$ charge density (and zero temperature).
Immediate questions include:
What kind of quantum liquid is it?
Does the system have a Fermi surface?
If yes, is it a Landau Fermi liquid?
A precise understanding of the ground states of these finite density systems at strong coupling should help expand the horizon of our knowledge of quantum liquids, and may find applications to real condensed matter systems.

On the gravity side, such a finite density system is described by an extremal charged black hole in $d+1$-dimensional anti-de Sitter spacetime
(AdS$_{d+1}$)~\cite{Chamblin:1999tk}.  The metric of the extremal black hole
has two interesting features which give some clues regarding the nature of the system. The first
is that the black hole has a finite horizon area at zero temperature, suggesting
a large ground state degeneracy (or approximate degeneracy) in the large $N$ limit. The second is that the near horizon geometry is given by AdS$_2 \times \RR^{d-1}$, which
 appears to indicate that
 at low frequencies the boundary system should develop an enhanced symmetry group including scaling invariance. In particular, it is natural to expect that quantum gravity (or string theory) in this region may be described by a boundary CFT.
  It has been argued in~\cite{Lu:2009gj} that the asymptotic symmetry group of the near horizon AdS$_2$ region is generated by a single copy of Virasoro algebra with a nontrivial central charge, suggesting a possible description in terms of some chiral 2d CFT .

More clues to the system were found in~\cite{Liu:2009dm} from studying spectral functions of a family of spinor operators (following earlier work of~\cite{Lee:2008xf}):

\ben

\item The system possesses sharp quasi-particle-like fermionic excitations at low energies near some discrete shells in momentum space, which strongly suggests the presence of Fermi surfaces.\footnote{Note that since the underlying system is spherically symmetric, the Fermi surfaces are round.} In particular, the excitations exhibit scaling behavior as a Fermi surface is approached with scaling exponents {\it different}
from that of a Landau Fermi-liquid\footnote{In~\cite{Cubrovic:2009ye} a different family of operators were studied  at finite temperature.
The authors concluded there that the scaling behavior resembles that of a Fermi liquid.
We will discuss those operators in section~\ref{sec:zoo}.
}. The scaling behavior is consistent with the general behavior discussed by Senthil in~\cite{Senthil:0803} for a critical Fermi surface at the critical point of a continuous metal-insulator transition.

\item For a finite range of momenta, the spectral function becomes periodic in $\log \om$  in the low frequency limit. Such log-periodic behavior gives rise to a discrete
scaling symmetry which is typical of  a complex scaling exponent.

\een
Note that the above scaling behavior is emergent, a consequence of collective dynamics of
many particles, not related to the conformal invariance of the UV theory which is broken by finite density.

The results of~\cite{Liu:2009dm} were obtained by solving  numerically the Dirac equation for bulk spinor fields dual to boundary operators and it was not possible to identify the specific geometric feature of the black hole which is responsible for the emergence of the scaling behavior. Nevertheless, as speculated in~\cite{Liu:2009dm}, it is natural to suspect that the AdS$_2$ region of the black hole may be responsible.

In this paper, we show that at low frequencies\footnote{
By ``low frequency" we mean frequency close to the chemical potential.
How this arises from the AdS/CFT dictionary will be explained below.}, retarded Green functions\footnote{We focus on retarded Green function as its imaginary part directly gives the spectral function which reflects the density of states which couple to an operator. It is also the simplest observable to compute in the Lorentzian signature in AdS/CFT. The scaling behavior is of course  also present in other types of correlation functions.
We also expect similar scaling behavior to exist in higher-point functions which will be left for future study.} of generic operators in the boundary theory exhibit quantum critical behavior.
This critical behavior is determined by the AdS$_2$ region of the black hole;
assuming it exists, a CFT$_1$ dual to this region of the geometry (which we will call the `IR CFT') can be said to govern the critical behavior.

The spirit of the discussion of this paper will be similar to that of~\cite{Liu:2009dm};
we will not restrict to any specific theory. Since Einstein gravity coupled to matter fields captures universal features of a large class of field theories with a gravity dual, we will simply work with this universal sector, essentially scanning many possible CFTs.\footnote{The caveat is there may exist certain operator dimensions or charges which do not arise in a consistent gravity theory with UV completion.}

The role played by the IR CFT in determining the
low-frequency form of the
Green's functions of the $d$-dimensional theory requires some explanation.
Each operator $\sO$ in the UV theory gives rise to a tower of operators $\sO_\vk$ in the IR CFT labeled by spatial momentum $\vk$. The
small $\om$ expansion of the retarded Green function $G_R (\om, \vk)$ for $\sO$
contains an analytic part which is governed by the UV physics and a non-analytic part which is proportional to the retarded Green function of $\sO_\vk$ in the IR CFT.
 What kind of low-energy behavior occurs depends on the dimension $\delta_k$ of the operator $\sO_\vk$ in the IR CFT {\it and} the behavior of $G_R (\om =0, \vk)$.\footnote{The behavior at exactly zero frequency $G_R (\om =0, \vk)$ is controlled by UV physics, {\it not} by the IR CFT.}
For example, when $\delta_k$ is complex one finds the log-periodic behavior described earlier. When $G_R (\om =0, \vk)$ has a pole at some finite momentum $|\vk| = k_F$~(with $\delta_{k_F}$ real), one
then finds
gapless excitations around $|\vec k| = k_F$
indicative of a Fermi surface.

Our discussion is general and should be applicable to operators of any spin.
In particular both types of scaling behavior mentioned earlier for spinors also applies to scalars. But due to Bose statistics of the operator in the boundary theory,
this behavior is associated with instabilities of the ground state.
In contrast, there is no instability for spinors even when the dimension is complex.

Our results give a nice understanding of the low-energy scaling behavior around the Fermi surface. The scaling exponents are controlled by the dimension of the corresponding
operator in the IR CFT. When the operator is relevant (in the IR CFT), the quasi-particle is unstable. Its width is linearly proportional to its energy and the quasi-particle residue vanishes approaching the fermi surface. When the operator is irrelevant, the quasi-particle becomes stable, scaling toward the Fermi surface with a nonzero quasi-particle residue. When the operator is marginal the spectral function then has the form for a ``marginal Fermi liquid'' introduced in the phenomenological study of the normal state of high $T_c$ cuprates~\cite{varma}.

It is also worth emphasizing two important features of our system. The first is
that in the IR, the theory has not only an emergent scaling symmetry but an $SL(2,R)$ conformal symmetry (maybe even Virasoro algebra). The other is the critical behavior (including around the Fermi surfaces) only appears in the frequency, not in the spatial momentum directions.

The plan of the paper is as follows.
In \S2, we introduce the charged AdS black hole and
its AdS$_2$ near-horizon region.
In \S3, we determine the low energy behavior of Green's
functions in the dual field theory, using scalars as illustration.
The discussion for spinors is rather parallel and presented in Appendix~\ref{app:fer}.
In~\S\ref{sec:I}--\ref{sec:surf}, we apply this result to demonstrate three
forms of emergent quantum critical behavior in
the dual field theory:
scaling behavior of the spectral density~(\S\ref{sec:I}),
periodic behavior in $\log \omega$ at small momentum~(\S\ref{sec:IRp}),
and finally~(\S\ref{sec:surf}) the Fermi surfaces found in~\cite{Liu:2009dm}.
We conclude in~\S\ref{sec:dis} with a discussion of various results and possible future generalizations. We have included various technical appendices. In particular
in Appendix~\ref{app:ads2} we give retarded functions of charged scalars and spinors in the AdS$_2$/CFT$_1$ correspondence.

\section{Charged black holes in AdS and emergent infrared CFT }

\subsection{Black hole geometry}

Consider a $d$-dimensional conformal field theory (CFT) with a global $U(1)$
symmetry that has a gravity dual.
At finite charge density, the system can be described by a
charged black hole in $d+1$-dimensional anti-de Sitter spacetime
(AdS$_{d+1}$)~\cite{Chamblin:1999tk} with the current $J_\mu$ in the CFT mapped to a $U(1)$ gauge field $A_M$ in AdS.

The action for a vector field $A_M$ coupled to AdS$_{d+1}$ gravity can be written as
 \be \label{grac}
 S = {1 \ov 2 \kappa^2} \int d^{d+1} x \,
 \sqrt{-g} \le[\sR +  { d (d-1) \ov R^2} - {R^2 \ov g_F^2} F_{MN} F^{MN} \ri]
\ee
where $g_F^2$ is an effective dimensionless gauge coupling\footnote{It is defined so that for a
typical supergravity Lagrangian it is a constant of order $O(1)$. $g_F$ is related in the
boundary theory to the normalization of the two point function of $J_\mu$.} and $R$ is the
curvature radius of AdS. The equations of motion following from~\eqref{grac} are solved by the
geometry of a charged black hole~\cite{Romans:1991nq, Chamblin:1999tk}
 \be \label{bhmetric1}
 ds^2 \equiv g_{MN} dx^M dx^N =  {r^2 \ov R^2} (-f dt^2 + d\vec x^2)  + {R^2 \ov r^2} {dr^2 \ov f}
 \ee
 with
 \be \label{bhga2}
 f = 1 + { Q^2 \ov r^{2d-2}} - {M \ov r^d}, \qquad A_t = \mu \le(1- {r_0^{d-2} \ov  r^{d-2}}\ri) \ .
 \ee
$r_0$ is the horizon radius determined by the largest positive root of the redshift factor
\be \label{hord}
f(r_0) =0, \qquad \to \qquad M = r_0^d + {Q^2 \ov r^{d-2}_0}
\ee
and
 \be \label{chem}
\mu \equiv  {g_F Q \ov c_d R^2 r_0^{d-2}}, \qquad c_d \equiv \sqrt{2 (d-2) \ov d-1} \ .
 \ee

The geometry~\eqref{bhmetric1}--\eqref{bhga2} describes the boundary theory at a finite density
with the charge, energy and entropy densities respectively given by
 \bea \label{thqu}
 \rho = {2 (d-2) \ov c_d} { Q  \ov \kappa^2 R^{d-1} g_F}, \qquad \qquad\\ \ep = {d-1 \ov 2 \kappa^2} {M \ov R^{d+1}},
  \qquad s = {2 \pi \ov \kappa^2} \le({r_0 \ov R}\ri)^{d-1} \ .
 \eea
The temperature of system can be identified with the Hawking temperature of the black hole, which is
\be \label{rre1}
T = {d r_0 \ov 4 \pi R^2} \le(1 - {(d-2) Q^2 \ov d r_0^{2d-2}} \ri)
\ee
and $\mu$ in~\eqref{chem} corresponds to the chemical potential. It can be readily checked from the above equations that the first law of thermodynamics is satisfied
 \be
 d \ep = T ds + \mu d \rho \ .
 \ee

Note that $Q$ has dimension of $[L]^{d-1}$ and it is convenient to parameterize it as
 \be
 Q \equiv \sqrt{d \ov d-2} \, r_*^{d-1} \ .
 \ee
 by introducing a length scale $r_*$.  In order for the metric~\eqref{bhmetric1} not to have a naked singularity one needs
 \be \label{c1}
 M \geq {2 (d-1) \ov d-2} r_*^d  \quad \to \quad r_0 \geq r_* \ .
 \ee
In terms of $r_*$, the expressions for charge density $\rho$, chemical potential $\mu$, and
temperature can be simplified as
 \bea
 \rho ={1 \ov \kappa^2}   \le({r_* \ov R} \ri)^{d-1} {1 \ov e_d}, \qquad \\
\mu  =  {d (d-1) \ov d-2} {r_* \ov R^2} \le({r_* \ov  r_0} \ri)^{d-2} e_d,
\label{mudefJ}\\
 T = {d r_0 \ov 4
\pi R^2} \le(1 - {r_*^{2d-2} \ov  r_0^{2d-2}} \ri) \quad
 \eea
where we have introduced
 \be \label{defed}
 e_d \equiv {g_F \ov \sqrt{2d (d-1)}}  \ .
 \ee
Note that $r_*$ can be considered as fixed by the charge density of the boundary theory.

\subsection{AdS$_2$ and scaling limits}

In this paper we will be mostly interested in the behavior of the system at zero temperature, in
which limit the inequalities in~\eqref{c1} are saturated
 \be \label{valM}
 T=0 \quad \to \quad  r_0 = r_*  \quad {\rm and} \quad M = {2 (d-1) \ov d-2} r_*^d
 \ .
 \ee
Note that the horizon area remains nonzero at zero temperature and thus this finite charge density system has a nonzero ``ground state'' entropy density\footnote{Since
the semiclassical gravity
expression for the entropy is valid in the large $N$ limit, one only needs ``ground state degeneracy'' in the $N \to \infty$ limit.},
which can be expressed in terms of charge density as
 \be \label{zeroent}
 s = (2 \pi e_d) \, \rho \ . 
 \ee

In the zero temperature limit~\eqref{valM} the redshift factor $f$ in~\eqref{bhmetric1} develops
a double zero at the horizon
 \be
 f = d(d-1) {(r-r_*)^2 \ov r_*^2} + \cdots \ .
 \ee
As a result, very close to the horizon the metric becomes $AdS_2 \times \RR^{d-1}$ with the curvature radius of $AdS_2$ given by
 \be \label{ads21}
 R_2 = {1 \ov \sqrt{d (d-1)}}  R  \ . 
 \ee
More explicitly, considering the scaling limit
 \be \label{scalL}
 r-r_* = \lam {R_2^2 \ov \ze}, \quad t = \lam^{-1} \tau, \quad \lam \to 0 \  {\rm with} \
 \ze, \tau \  {\rm finite}
 \ee
we find that the metric~\eqref{bhmetric1} becomes $AdS_2 \times \RR^{d-1}$:
 \be \label{ads2}
 {ds^2 } = {R_2^2 \ov \ze^2 } \le(- d \tau^2 +  d \ze^2 \ri) + {r_*^2 \ov R^2} d\vec x^2
 \ee
 with
 \be \label{conEl}
 A_\tau 
 =  {e_d \ov  \ze} \ .
 \ee

The scaling limit~\eqref{scalL} can also be generalized to finite temperature by writing
in addition to~\eqref{scalL}
\be
r_0 - r_* = \lam {R_2^2 \ov \ze_0}  \quad {\rm with} \;\; \ze_0 \;\; {\rm finite} \
\ee
after which the metric becomes a black hole in $AdS_2$ times $\RR^{d-1}$:
 \be \label{ads2T}
 ds^2 =  {R^2_2 \ov \ze^2} \le( - \le(1- {\ze^2 \ov \ze_0^2} \ri)  d\tau^2 +
 {d \ze^2 \ov 1- {\ze^2 \ov \ze_0^2}} \ri)
 + {r_*^2 \ov R^2} d \vec x^2
 \ee
 with
\be
A_\tau =  {e_d  \ov \ze}  \le(1-{\ze \ov \ze_0} \ri)
\ee
and a temperature (with respect to $\tau$)
 \be
 T ={1 \ov 2 \pi \ze_0} \ .
 \ee

Note that in the scaling limit~\eqref{scalL}, finite $\tau$ corresponds to the long time limit of the original time coordinate. Thus in the language of the boundary theory~\eqref{ads2} and~\eqref{ads2T} should apply to the low frequency limit
 \be
 {\om \ov \mu} , {T \ov \mu} \to 0, \qquad \om \sim T \
 \ee
where $\om$ is the frequency conjugate to $t$.

\subsection{Emergent IR CFT}

One expects that gravity in the near-horizon AdS$_2$ region~\eqref{ads2} of an extremal charged
AdS black hole should be described by a CFT$_1$ dual. Little is known about this AdS$_2$/CFT$_1$
duality\footnote{For previous work on AdS$_2$/CFT$_1$ correspondence from other decoupling limits, see \eg\ \cite{adstworefs} and its citations.}. For example, it is not
clear whether the dual theory is a conformal quantum mechanics or a chiral sector of a
$(1+1)$-dimensional CFT.  It has been argued in~\cite{Lu:2009gj} that the asymptotic symmetry group of the near horizon AdS$_2$ region is generated by a single copy of Virasoro algebra with a nontrivial central charge, suggesting a possible description in terms of some chiral 2d CFT\footnote{The central charge is proportional to the
volume of the $d-1$-dimensional transverse space and is thus infinite for~\eqref{ads2}. To have
a finite central charge one could replace $\RR^{d-1}$ in~\eqref{ads2} by a large torus.}. Some of the problems associated with $AdS_2$, such as the fragmentation instability and the
impossibility of adding finite-energy excitations \cite{Maldacena:1998uz} are ameliorated by the infinite volume of the $\RR^{d-1}$ factor in the geometry \eqref{ads2}.

The scaling picture of the last subsection suggests that in the {\it low frequency limit}, the
$d$-dimensional boundary theory at finite charge density is described by this CFT$_1$, to which we will refer below as the {\it IR CFT} of the boundary theory. It is important to emphasize that the conformal symmetry of this IR CFT is {\it not} related to the microscopic conformal invariance of the higher dimensional theory (the UV theory) which is broken by finite charge density. It apparently emerges as a consequence of collective behavior of a large number of degrees of freedom.

In section~\ref{sec:corr} we will elucidate the role of this IR CFT
by examining the low frequency limit of two-point functions of the full theory.
Our discussion will not depend on the specific nature of the IR CFT,
but only on its existence.
In Appendix~\ref{app:ads2} we give
correlation functions for a charged scalar and spinor in the IR CFT as calculated from
the standard AdS/CFT procedure in AdS$_2$ \cite{WithNabil}. They will play an important role in our discussion of section~\ref{sec:corr}.

\section{Low frequency limit of retarded functions} \label{sec:corr}

In this section we elucidate the role of the IR CFT by examining the low frequency limit of correlation functions in the full theory. We will consider two-point retarded functions for simplicity leaving the generalization to multiple-point functions for future work.
We will mostly focus on zero temperature.

Our discussion below should apply to generic fields in AdS including scalars, spinors and  tensors. We will use a charged scalar for illustration. The results for spinors will be mentioned at the end with calculation details given in Appendix~\ref{app:fer}.
Vector fields and stress tensor will be considered elsewhere.

Consider a scalar field in AdS$_{d+1}$ of charge $q$ and mass $m$, which is dual to an operator $\sO$ in the boundary CFT$_d$ of charge $q$ and dimension
 \be
 \De = {d \ov 2} + \sqrt{m^2 R^2 + {d^2 \ov 4}} \ .
 \ee
In the black hole geometry~\eqref{bhmetric1}, the quadratic action for $\phi$ can be written as
\be \label{scaA}
S = -\int  d^{d+1} x \sqrt{-g} \, \le[(D_M \phi)^* D^M \phi + m^2 \phi^* \phi \ri]
\ee
with
\be
D_M \phi = (\p_M - i q A_M ) \phi \ .
\ee
Note that the action~\eqref{scaA} depends on $q$ only through
 \be \label{mme}
 \mu_q \equiv \mu q 
 \ee
which is the effective chemical potential for a field of charge $q$.
Writing\footnote{For simplicity of notation, we will distinguish $\phi (r, x^\mu)$ from its Fourier
transform $\phi (r, k_\mu)$ by its argument only.}
\be
\phi (r, x^\mu)= \int {d^d k \ov (2 \pi)^d} \, \phi (r, k_\mu) \, e^{i k_\mu x^\mu}, \qquad k_\mu = (-\om, \vec k) \
\ee
the equation of motion for $\phi (r, k_\mu)$ is given  by (below $k^2 \equiv |\vk|^2$)
\be  \label{pp2}
- {1 \ov \sqrt{-g}} \p_r (\sqrt{-g} g^{rr} \p_r \phi) + \le(g^{ii} (k^2 - u^2) + m^2 \ri) \phi
=0 \
\ee
where various metric components are given in~\eqref{bhmetric1} and
 \be \label{udef}
  u (r) \equiv \sqrt{g_{ii} \ov -g_{tt}} \le(\om + \mu_q \le(1-{r_0^{d-2} \ov r^{d-2}} \ri) \ri) \ .
\ee

In~\eqref{bhga2} we have chosen the gauge so that the scalar potential is zero at the horizon. As a result $A_t \to \mu$ for $r \to \infty$ and $u (r \to \infty) \to \om + \mu_q$.
This implies that $\om$ should correspond to the difference of the boundary theory frequency from $\mu_q$. Thus the low frequency limit really means very close to the effective
chemical potential $\mu_q$.

The retarded Green function for $\sO$ in the boundary theory can be obtained by
finding a solution $\phi$ which satisfies the in-falling boundary condition at the horizon, expanding it near the boundary as
\be \label{asum}
\phi (r,k_\mu) \buildrel{r \to \infty} \over {\approx} A (k_\mu) \, r^{\Delta - d} + B(k_\mu) \,
r^{- \Delta} ,
\ee
and then~\cite{Son:2002sd}
 \be \label{bore}
G_R (k_\mu) = K {B (k_\mu) \ov A(k_\mu)} \
\ee
where $K$ is a positive constant which depends on the overall normalization of the action, and is independent of $k_\mu$.

\subsection{Low frequency limit}

At $T=0$, expanding~\eqref{bore} in small $\om$ is not straightforward, as the $\om \to 0$ limit of  equation~\eqref{pp2} is singular. This is because $g^{tt}$ has a double pole at the horizon. As a result, the $\om$-dependent terms in equation~\eqref{pp2} always dominates sufficiently close to the horizon and thus cannot be treated as small perturbations no matter how small $\om$ is. To deal with this we divide the $r$ axis into two regions
\bea
\label{in1}
&& \textbf{Inner:} \quad r-r_* = \om {R_2^2 \ov \ze} \quad {\rm for} \quad
 \ep < \ze < \infty \\
&& \textbf{Outer:} \quad  {\om R_2^2 \ov \ep}  < r-r_*
\label{out1}
\eea
and consider the limit
 \be \label{limR2}
 \om \to 0, \quad \ze = {\rm finite} , \quad \ep \to 0, \quad {\om R_2^2 \ov \ep} \to 0 \ .
 \ee
Using $\ze$ as the variable for the inner region and $r$ as that for the outer region, small $\om$ perturbations in each region can now be treated straightforwardly, with
 \bea
&& {\rm inner:} \qquad  \phi_I (\ze)  = \phi_I^{(0)} (\ze) + \om \phi_I^{(1)} (\ze) + \cdots \\
&& {\rm outer:} \qquad  \phi_O (r)= \phi_O^{(0)} (r) + \om \phi_O^{(1)} (r) + \cdots \ .
 \eea
We obtain the full solution by matching $\phi_I$ and $\phi_O$
in the overlapping region, which is $\ze \to 0$ with $r-r_* = {\om R_2^2 \ov \ze} \to 0$. Note that since the definition of $\ze$ involves $\om$, the matching will reshuffle the perturbation series in two regions.

While the above scaling limit is defined for small $\om > 0$, all our later manipulations and final results can be analytically continued to generic complex values of $|\om| \ll 1$.

\subsubsection{Inner region: scalar fields in AdS$_2$}

The scaling limit~\eqref{in1},~\eqref{limR2} is in fact
identical to that introduced in~\eqref{scalL}~(with $\om$ replacing $\lam$) in which the metric reduces to that of AdS$_2 \times \RR^{d-1}$ with a constant electric field.
It can then be readily checked that in the inner region at leading order, equation~\eqref{pp2}~(\ie\ the equation for $\phi_I^{(0)}$)  reduces to equation~\eqref{eep2} in Appendix~\ref{app:ads2} for a charged scalar field in AdS$_2$ with
an effective AdS$_2$ mass
 \be \label{effM}
 m_k^2 = k^2 {R^2 \ov r_*^2} + m^2, \qquad k^2 = |\vk^2| \ .
 \ee
A single scalar field $\phi$ in AdS$_{d+1}$ gives rise a tower of fields $\phi_\vk$ in AdS$_2$ labeled by the spatial momentum $\vk$. From the discussion of Appendix~\ref{app:ads2}, the conformal dimension for the operator
$\sO_\vk$ in the IR CFT dual to $\phi_\vk$ is given by
 \be \label{nuk}
 \delta_k = \ha + \nu_k, \qquad  \nu_k \equiv \sqrt{m_k^2 R_2^2 - q^2 e_d^2 + {1 \ov 4}}
 \ .
 \ee
Note that momentum conservation in $\RR^{d-1}$ implies that operators corresponding to different momenta do not mix, \ie\
 \be
 \vev{\sO_{\vk}^\dagger (t) \sO_{\vk'} (0)} \propto \delta (\vk - \vk') \, t^{- 2 \delta_k} \ .
 \ee

To compute the retarded function~\eqref{bore} for the full theory, we impose the boundary
condition that $\phi_I^{(0)}$ should be in-falling at the horizon. Near the boundary of the
inner region (AdS$_2$ region), \ie\  $\ze = {\om R_2^2 \ov r-r_*} \to 0$, $\phi_I^{(0)}$ can then be expanded as (see~\eqref{exwp1})\footnote{For convenience for matching to outer region we have taken  a specific choice of normalization for
$\phi_I^{(0)}$ below. The calculation of retarded function~\eqref{bore} does not depend on the choice of normalization.}
 \bea
 \phi_I^{(0)} (\om, \vk; \ze) &= &   \le({R_2^2 \ov r-r_*} \ri)^{\ha - \nu_k} \le(1 + O(\ze)\ri) \;\; \cr
 & + & \;\; \sG_k (\om) \le({R_2^2 \ov r-r_*} \ri)^{\ha + \nu_k} \le(1 + O(\ze)\ri)  \ .  \label{exwpSca}
 \eea
The coefficient $\sG_k (\om)$ of the second term in~\eqref{exwp} is precisely the retarded
Green function for operator $\sO_\vk$ in the IR CFT. From~\eqref{exbG} it can be written as
 \be \label{exbG1}
 \sG_k (\om) = 2 \nu_k e^{ - i \pi \nu_k} \frac{ \Gamma (-2\nu_k ) \Gamma \left(\frac{1}{2}+ \nu_k-i q e_d\right)}{\Gamma
(2\nu_k )\Gamma \left(\frac{1}{2}- \nu_k-i q e_d\right)   } (2\om)^{2 \nu_k}\
 \ee
 with $\nu_k$ given by~\eqref{nuk}. Equation~\eqref{exwpSca} will be matched to the outer solution next.

\subsubsection{Outer region and matching}\label{sec:matching}

The leading order equation in the outer region is obtained by setting $\om =0$ in~\eqref{pp2}.
Examining the resulting equation near $r \to r_*$, one finds that it is identical to the inner
region equation for $\phi_I^{(0)}$ in the limit $\ze \to 0$. It is thus convenient to choose
the two linearly-independent solutions $\eta_\pm^{(0)}$ in the outer region using the two
linearly independent terms in~\eqref{exwpSca}, \ie\  $\eta_\pm^{(0)}$ are specified by the boundary
condition
  \be \label{bdc2}
 \eta_\pm^{(0)} (r) \approx \le({r-r_* \ov R^2_2}\ri)^{-\ha \pm \nu_k} + \cdots, \qquad
 r-r_* \to 0 \ .
 \ee
The matching to the inner region solution~\eqref{exwpSca} then becomes trivial and
the leading outer region solution $\phi_O^{(0)}$ can be written as
 \be \label{eoro}
 \phi_O^{(0)} = \eta_+^{(0)} (r)+ \sG_k (\om) \eta_-^{(0)} (r)
 \ee
  with $\sG_k (\om)$ given by~\eqref{exbG1}.

One can easily generalize~\eqref{eoro} to higher orders in $\om$.
The two linearly independent solutions to the full outer region equation can be expanded as
  \be \label{petS}
 \eta_\pm = \eta^{(0)}_\pm + \om \eta^{(1)}_\pm + \om^2 \eta^{(2)}_\pm + \cdots
 \ee
where higher order terms $\eta_\pm^{(n)}, \, n \geq 1$ can be obtained using the standard perturbation theory and are {\it uniquely} specified by requiring that when expanded near $r =r_*$, they do not contain any terms proportional to the zeroth order solutions. Each of the higher order terms satisfies an inhomogenous linear equation. The requirement amounts to choosing a specific special solution of the homogeneous equation. Note that it is important that the equations are  linear. Given that higher order terms in~\eqref{petS} are uniquely determined by  $\eta^{(0)}_\pm$, to match the full solution $\phi_O$ to the inner region it is enough to match the leading order term which we have already done. We thus conclude that perturbatively
 \be \label{solr}
 \phi_O = \eta_+ + \sG_k (\om)  \eta_- \ . 
 \ee

\subsubsection{Small $\om$ expansion of $G_R$}

We first look at the retarded function at $\om =0$. At $\om =0$ the inner region does not exist and the outer region equation reduces to that satisfied by $\phi_O^{(0)}$.
In~\eqref{eoro} we have chosen the normalization so that at $\om =0$, $\phi_O^{(0)} = \eta^{(0)}_+$.
For real $\nu_k$, this follows from the fact that $\eta^{(0)}_+$ gives the regular solution at $r \to r_*$. When $\nu_k$ is pure imaginary (\ie\  when $q$ is sufficiently large) we will define the branch of the square root by taking $m^2 \to m^2 - i \ep$ so that $\nu_k = - i \lam_k$
with $\lam_k$ positive.
Then $\eta^{(0)}_+$ is the in-falling solution at the horizon as is required by the
prescription for calculating retarded functions. Expanding $\eta^{(0)}_+ (r)$ near $r \to \infty$ as
 \be \label{expeta0}
 \eta_\pm^{(0)} (r,k) = a_\pm^{(0)} (k) r^{\De -d} \le(1 + \cdots\ri) + b_\pm^{(0)} (k) r^{-\De} \le(1 + \cdots \ri)
 \ee
then from~\eqref{bore} we find that
 \be \label{zerow}
 G_R (\om=0,k) = K {b_+^{(0)} \ov a_+^{(0)}} \ .
 \ee

Now consider a small nonzero $\om$. Expanding various functions in~\eqref{petS} ($n \geq 1$) near $r \to \infty$ as
 \be \label{verC}
 \eta_\pm^{(n)} (r,k) = a_\pm^{(n)} (k) r^{\De -d} \le(1 + \cdots\ri) + b_\pm^{(n)} (k) r^{-\De} \le(1 + \cdots \ri),
 \ee
from~\eqref{solr} and~\eqref{bore} we find that for small $\om$
\bwt
\be
\label{finG1}
 G_{R} (\om, k) = K {b_+^{(0)} + \om b_+^{(1)} + O(\om^2) + \sG_k( \om)
 \le(b_-^{(0)} + \om b_-^{(1)} + O(\om^2)\ri) \ov a_+^{(0)} + \om a_+^{(1)} + O(\om^2) + \sG_k (\om)
 \le(a_-^{(0)} + \om a_-^{(1)} + O(\om^2)\ri) } \ .
 \ee
\ewt
Equation~\eqref{finG1} is our central technical result. In next few sections we explore its implications for the low energy behavior of the finite density
boundary system. While its expression is somewhat formal, depending on various unknown functions $a_\pm^{(n)} (k), b_\pm^{(n)} (k)$ which can only be obtained by solving the full outer region equations order by order (numerically), we will see that a great deal about the low energy behavior of the system can be extracted from it
without knowing those functions explicitly.

\subsection{Generalization to fermions}

Our discussion above only hinges on the fact that in the low frequency limit the inner region wave equation becomes that in AdS$_2$. It applies also to spinors and other tensor fields even though the equations involved are more complicated. In Appendix~\ref{app:fer} we discuss equations and matching for a spinor in detail. After diagonalizing the spinor equations one finds that eigenvalues of the retarded spinor Green function (which is now a matrix) are also given exactly by equation~\eqref{finG1} with now $\sG_k (\om)$ given by equation~\eqref{effG}, which we copy here for convenience
 \bea \label{rprp}
 \sG_k (\om) &= & e^{ - i \pi \nu_k}\frac{\Gamma (-2 \nu_k ) \, \Gamma \left(1+\nu_k -i q e_d \right)}{\Gamma (2 \nu_k )\,   \Gamma \left(1-\nu_k -i q e_d\right)}
  \cr &\times & \; \frac{\le(m + {i k R \ov r_*} \ri)
 R_2 - i q e_d - \nu_k}{ \le(m + {i k R \ov r_*} \ri) R_2 - i q e_d +  \nu_k} \; (2\om)^{2 \nu_k}
 \eea
with
 \be \label{defNu}
 \nu_k = \sqrt{m_k^2 R_2^2 - q^2 e_d^2}, \qquad m_k^2 = {k^2 R^2 \ov r_*^2} + m^2 \ .
 \ee
Note the above scaling exponent can also be expressed as (using~\eqref{mudefJ})
 \be
 \nu_k = {g_F q \ov \sqrt{2 d (d-1)}} \sqrt{{2 m^2 R^2 \ov g_F^2 q^2} + {d (d-1) \ov (d-2)^2} {k^2 \ov \mu_q^2} - 1} \ .
 \ee
The conformal dimension of the operator $\sO_{\vk}$ in the IR CFT is again given by
 \be
 \delta_k = \ha + \nu_k \ .
 \ee

\subsection{Analytic properties of $\sG_k$} \label{sec:3C}

The analytic properties of $\sG_k (\om)$ will play an important role in our discussion of the next few sections. We collect some of them here for future reference. Readers should feel free to skip this subsection for now and refer back to it later.

We first introduce some notations, writing
 \be \label{defc}
 \sG_k (\om) \equiv c(k) \om^{2 \nu_k}, \qquad c(k) \equiv |c(k)| e^{i \ga_k}
 \ee
where $c(k)$ denotes the prefactor in~\eqref{rprp} for spinor and that in~\eqref{exbG1} for scalars.

For real $\nu_k$, the ratios in~\eqref{oo} and~\eqref{anid} of Appendix~\ref{app:ads2}
become a pure phase and we find that\footnote
{Since equations \eqref{anid} and \eqref{oo} determine $e^{2 i \gamma}$, 
they leave an additive ambiguity of $n\pi$ in the phase $\gamma$ of $\sG_R$.
In fact, this ambiguity is important for maintaining unitarity (in a small part of the parameter space);
we discuss this phenomenon further in Appendix \ref{app:piambiguity}.
The conclusion of that discussion is that 
the quantity appearing in denominator the Green's function multiplying $\omega^{2\nu}$ 
is $|h_2| e^{i \gamma_k}$ with $\gamma_k$ precisely as given in equation \eqref{phaseE}.
}
 \be \label{phaseE}
 \ga_k = \bca
 \arg\left( \Gamma(-2\nu_k) \le(e^{-2 \pi i \nu_k} - e^{- 2 \pi q e_d} \ri)\right) & {\rm spinor} \cr \cr
  \arg \left( \Gamma(-2\nu_k)\le(e^{-2 \pi i \nu_k} + e^{- 2 \pi q e_d} \ri) \right)
 & {\rm scalar}
 \eca
 \ee
It can be readily checked by drawing $e^{- 2 \pi i \nu_k}$ and $ e^{- 2 \pi q e_d}$ on the complex plane that the following are true:
 \bi

 \item For both scalars and spinors, $e^{i \ga_k}$ (and thus $c(k)$) always lies in the upper-half complex plane.

 \item For scalars $e^{i \ga_k + 2 \pi i \nu_k}$ always lies in the lower-half complex plane, while for spinors $e^{i \ga_k + 2 \pi i \nu_k}$ always lies in the upper-half complex plane.

 \item  For $\nu_k \in (0,\ha)$,
 \be
 \label{spinine}
 \begin{split}
 &{\rm spinor}: \qquad \pi-\ga_k > 2 \pi \nu_k   \\
  &{\rm scalar}: \qquad \pi-\ga_k < 2 \pi \nu_k  \ .
 \end{split}
 \ee

 \ei

For pure imaginary $\nu_k =- i \lam_k$ ($\lam_k > 0$), the ratios in~\eqref{oo} and~\eqref{anid} of Appendix~\ref{app:ads2} become real and give
 \be \label{modSp}
 |c_k|^2 = {e^{-2 \pi \lam_k} - e^{-2 \pi q e_d} \ov e^{2 \pi \lam_k} - e^{-2 \pi q e_d}}
 < e^{- 4 \pi \lam_k} \qquad {\rm spinor}
 \ee
 and
 \be \label{modSc}
 |c_k|^2 = {e^{-2 \pi \lam_k} + e^{-2 \pi q e_d} \ov e^{2 \pi \lam_k} + e^{-2 \pi q e_d}}
 > e^{- 4 \pi \lam_k} \qquad {\rm scalar} \ .
 \ee
It is also manifest from the above expressions that $|c(k)|^2 < 1$ for both scalars and spinors.

Also note that for generic $\nu_k$, $\sG_k (\om)$ and accordingly $G_R (\om, k)$ in~\eqref{finG1} have a logarithmic branch point at $\om=0$. We will define the physical sheet to be $\th \in (-{\pi \ov 2}, {3 \pi \ov 2})$, \ie\  we place the branch cut along the negative imaginary axis. This choice is not arbitrary. As discussed in Appendix~\ref{app:ads2}, when going to finite temperature, the branch cut resolves into a line of poles along the negative imaginary axis.

\subsection{Renormalization group interpretation of the matching}

The matching procedure described above has a natural interpretation
in terms of the renormalization group flow of the boundary theory.
The outer region can be interpreted as corresponding to UV physics while the inner
AdS$_2$ region describes the IR fixed point. The matching between in the inner and outer
regions can be interpreted as matching of the IR and UV physics at an intermediate scale.
More explicitly, coefficients $a_\pm^{(n)}, b_\pm^{(n)}$ from solving the
equations in the outer region thus encode the UV physics, but
$\sG_k (\om)$ is controlled by the IR CFT.

In this context $\om$ can be considered as a control parameter away from the IR fixed point. Equation \eqref{finG1} then shows a competition between analytic power corrections (in $\om$) away from the fixed point and contribution from
operator $\sO_\vk$. In particular when $\nu_k > \ha$ (\ie\  $\delta_k > 1$), $\sO_\vk$
becomes irrelevant in the IR CFT and its contribution becomes subleading compared to
analytic corrections. Nevertheless the leading {\it non-analytic} contribution is still given by $\sG_k (\om)$ and as we will see below in various circumstances $\sG_k$ does
control the leading behavior of the spectral function and other important physical quantities like the width of a quasi-particle.

It is interesting to note the similarity of our matching discussion to those used
in various black hole/brane emission and absorption calculations (see~\eg\cite{greybody}) which were important precursors to the discovery of AdS/CFT.
The important difference here is that in this asymptotically-AdS case
we can interpret the whole process (including the outer region) in terms of the dual field theory.

\section{Emergent quantum critical behavior I: Scaling of spectral functions} \label{sec:I}

In this and the next two sections we explore the implications of equation~\eqref{finG1} for the low energy behavior of the finite density boundary system.
In this section we look at the behavior of~\eqref{finG1} at a generic momentum for which $\nu_k$ is real and $a_+^{(0)} (k) $ is nonzero. Imaginary $\nu_k$ will be discussed in section~\ref{sec:IRp} and what happens when $a_+^{(0)} (k) =0$ will be discussed in section~\ref{sec:surf}.

When $\nu_k$ is real, the boundary condition~\eqref{bdc2} is real. Since the differential equation satisfied by $\phi_O^{(0)} $ is also real, one concludes that both $b_+^{(0)}$ and $a_+^{(0)}$ are real, which implies that
 \be \label{eoop}
 \Im G_R (\om=0, k) = 0, \qquad {\rm for \;\;real} \;\; \nu_k  \ .
 \ee
Similarly we can conclude that all coefficients in~\eqref{verC} are also real. Thus the only
complex quantity in~\eqref{finG1} is the Green function of the IR CFT, $\sG_k (\om)$. When
$a_+^{(0)} (k)$ is nonzero we can expand the denominator of~\eqref{finG1} and the spectral
function for $\sO$ can be written at small $\om$ as
  \be\label{irsc}
 {\rm Im} \, G_R (\om, k) = G_{R} (k, \om =0)  \, d_0 \, {\rm Im} \, \sG_k (\om) + \cdots
 \propto \om^{2 \nu_k}  \ .
 \ee
with
 \be
   d_0 =   {b_-^{(0)} \ov b_+^{(0)}} -{a_-^{(0)} \ov a_+^{(0)}} \ .
 \ee

 We thus see that the spectral function of the full theory has a
nontrivial scaling behavior at low frequency with the scaling exponent given by the conformal
dimension of operator $\sO_\vk$ in the IR CFT. Note that the $k$-dependent prefactor
in~\eqref{irsc} depends on $a_\pm^{(0)}, b_\pm^{(0)}$ and thus the metric of the outer region.
This is consistent with the RG picture we described at the end of last section; the scaling
exponent  of the spectral function is universal, while the amplitude does depend on UV physics
and is non-universal. By ``universal'' here, we mean the following. We can imaging
modifying the metric in the outer region without affecting the near horizon AdS$_2$ region. Then $a_\pm^{(0)}, b_\pm^{(0)}$ will change, but the exponent $\nu_k$ will remain the same. The real
part of $G_R$ is dominated by a term linear in $\om$ when $\nu_k > \ha$ and is non-universal,
but the leading nonanalytic term is again controlled by $\sG_k (\om)$.

\section{Emergent quantum critical behavior II: Log-periodicity} \label{sec:IRp}

In this section we examine the implication of~\eqref{finG1} when the $\nu_k$
becomes pure imaginary. We recover the log-oscillatory behavior for spinors first found numerically in~\cite{Liu:2009dm}.

\subsection{Log-periodic behavior: complex conformal dimensions}

When the charge $q$ of the field is sufficiently large (or $m^2$ too small)
\be
\label{ruop}
\begin{split}
&{\rm scalar}: \qquad m^2 R_2^2 + {1 \ov 4}  <  q^2 e_d^2 \\
&{\rm spinor}: \qquad m^2 R_2^2  < q^2 e_d^2
\end{split}
\ee
there exists a range of momenta
  \be \label{oenr}
 k^2 < k_o^2 \equiv \bca {r_*^2 \ov R^2} \le({q^2 e_d^2 - {1 \ov 4} \ov R_2^2} - m^2\ri) & {\rm scalar} \cr
   {r_*^2 \ov R^2} \le({q^2 e_d^2 \ov R_2^2} - m^2\ri) & {\rm spinor}
   \eca
 \ee
for which $\nu_k$ is pure imaginary
 \be \label{imnu}
 \nu_k = - i \lam_k , \qquad \lam_k = \bca \sqrt{q^2 e_d^2 - m_k^2 R_2^2 - {1 \ov 4}} & {\rm scalar} \cr
  \sqrt{q^2 e_d^2 - m_k^2 R_2^2 } & {\rm spinor}
 \eca
 \ .
 \ee
We have chosen the branch of the square root of $\nu_k$ by taking $m^2 \to m^2 - i \ep$.
The effective dimension of the operator $\sO_\vk$ in the IR CFT is thus
complex. Following~\cite{Liu:2009dm} we will call this region of momentum space~\eqref{oenr} the oscillatory region.\footnote{Note that the oscillatory region appears to be different from the Fermi ball discussed in~\cite{Lee:2008xf}.} For spinors we always have $m^2 \geq 0$ and the existence of the oscillatory region requires $q \neq 0$. For scalars,  equation~\eqref{ruop} can be satisfied for $q =0$ for $m^2$ in the range
 \be \label{tadkk}
 -{d^2 \ov 4} < m^2 R^2 < - {d (d-1) \ov 4}
 \ee
where the lower limit comes from the BF bound in AdS$_{d+1}$ and the upper limit
is the BF bound for the near horizon AdS$_2$ region.

For a charged field an imaginary $\nu_k$ reflects the fact that in the constant electric field~\eqref{conEl} of the AdS$_2$ region, particles with sufficiently large charge can be pair produced. It can be checked that equations~\eqref{ruop} indeed coincide with the threshold for pair production in AdS$_2$~\cite{Pioline:2005pf}.

With an imaginary $\nu_k$, the boundary condition~\eqref{bdc2} for $\eta_\pm^{(0)}$ is now complex. As a result
$b_+^{(0)}/a_+^{(0)}$ is complex and
 \be \label{meme}
 \Im G_R (\om=0, k) \neq 0 \ .
 \ee
Thus there are gapless excitations (since $\om=0$) for a range of momenta $k < k_o$. This should be contrasted with discussion around~\eqref{eoop}.

The leading small $\om$ behavior~\eqref{finG1} is now given by
 \be \label{osciP1}
 G_{R} (\om, k) \approx  { b_+^{(0)} + b_-^{(0)} c(k) \om^{-2 i \lam_k} \ov a_+^{(0)} +
 a_-^{(0)}  c(k) \om^{-2 i \lam_k}} + O(\om)
 \ee
where $c(k)$ was introduced in~\eqref{defc}. Note that here
 \be \label{realfoot}
 b_-^{(0)} = (b_+^{(0)})^*, \qquad a_-^{(0)} = (a_+^{(0)})^*
 \ee
since $\eta_-^{(0)} = (\eta_+^{(0)})^*$ at the horizon and that the differential
equation the $\eta$ satisfy is real.
Equation~\eqref{osciP1} is periodic in $\log \om$ with a period given by
\be \label{pero}
\tau_k ={\pi \ov \lam_k}  \ .
\ee
In other words~\eqref{osciP1} is invariant under a discrete scale transformation
\be
\om \to e^{n \tau_k} \om, \qquad n \in \ZZ, \quad  \om \to 0 \ .
\ee

We again stress that while the retarded function (and the spectral function) depends on UV physics (\ie\  solutions of the outer region), the leading nonanalytic behavior in $\om$ and in particular the period~\eqref{pero} only depends on the (complex) dimension of the operator in the IR CFT.

Here we find that the existence of log-periodic behavior at small frequency is
strongly correlated with~\eqref{meme}, \ie\  existence of gapless excitations. It would be desirable to have a better understanding of this phenomenon from the boundary theory side.

\subsection{(In)stabilities and statistics}

It is natural to wonder whether the complex exponent~\eqref{imnu} implies some instability. We will show now that it does for scalars but not for spinors. The scalar instability arises because the scalar becomes tachyonic in the AdS$_2$ region due to the electric field or reduced curvature radius. At zero momentum this is precisely the superconducting instability discussed before in~\cite{Gubser:2008px,Hartnoll:2008vx,Hartnoll:2008kx,Denef:2009tp}.\footnote{It was noted before in~\cite{Hartnoll:2008kx,Denef:2009tp} that there could be an instability even for neutral scalar in the mass range~\eqref{tadkk}. The condensation of such a neutral scalar
field does not break the $U(1)$ symmetry and thus is distinct from the superconducting instability of a charged field. It would be very interesting to understand better the boundary theory interpretation of such an instability. \label{ft:neur}}
That the log-oscillatory behavior does not
imply an instability for spinors was observed before in~\cite{Liu:2009dm}
by numerically showing there are no singularities in the upper half $\om$-plane.
Below we will give a unified treatment of both scalars and spinors, showing that
the difference between them can be solely attributed to statistics even though we have been
studying classical equations.

The spectral function following from~\eqref{osciP1} can be written as
  \be
  \begin{split}
 {{\rm Im} \, G_R (\om, k) \ov {\rm Im} \, G_R (\om =0,k)}  \buildrel{\om >0 }\over {=} &
   {1-|c (k)|^2  \ov |1 +
 |c(k)|  e^{i X}|^2}  \\
   \buildrel{\om < 0 }\over {=} &  {1-|c (k)|^2 e^{4 \pi \lam_k} \ov |1 +
 |c(k)| e^{2 \pi \lam_k } e^{i X}|^2}
 \end{split}
 \label{speM}
 \ee
where we have introduced
\be \label{varD1}
  X \equiv \ga_k - 2 \al - 2 \lam_k \log |\om|, \quad a_+^{(0)} = |a_+^{(0)}| e^{i \al}
 \ee
and $\ga_k$ was defined in~\eqref{defc}. In the boundary theory retarded Green functions for bosons are defined by commutators, while those for fermions by anti-commutators, which implies that  for $\om > 0$
 \bea
 \label{s1}
{\rm scalars:} \quad  \Im G_R (-\om, k) < 0, \quad \Im G_R (\om, k) > 0 \\\
 {\rm spinors:} \quad  \Im G_R (-\om, k) > 0, \quad \Im G_R (\om, k) > 0
 \label{f1}
 \eea
Applying~\eqref{s1} and~\eqref{f1} to~\eqref{speM} requires that
 \bea
  \label{cond1}
{\rm scalars:} \qquad |c(k)|^2 < 1, \qquad |c(k)| e^{2 \pi \lam_k} > 1 \\
{\rm spinors:} \qquad |c(k)|^2 < 1, \qquad |c(k)| e^{2 \pi \lam_k} < 1
 \label{cond2}
 \eea
which are indeed satisfied by~\eqref{modSc}--\eqref{modSp}.
It is important to stress that in the bulk we are dealing with classical
equations of scalars and spinors and have {\it not} imposed any statistics. However, the
self-consistency of AdS/CFT implies that classical equations for bulk scalars and spinors should
encode statistics of the boundary theory.

\begin{figure}[h]
 \begin{center}
\includegraphics[scale=0.70]{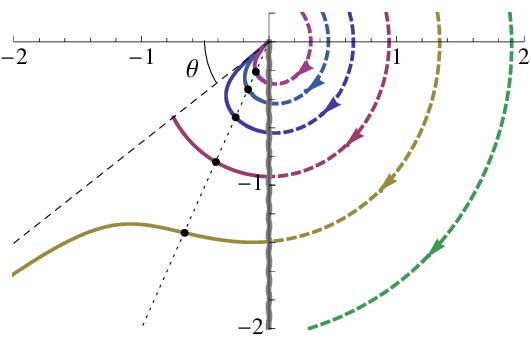}
\hskip0.5in
\includegraphics[scale=0.70]{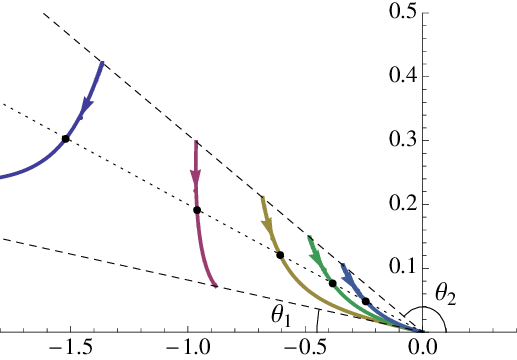}
\caption{\label{fig:poles1-osc}
The motion of poles of the Green functions~\eqref{osciP1} of spinors (left) and
scalars (right) in the complex frequency plane. For illustration purposes we have
chosen parameters and rescaled $|\omega|$( $\rightarrow |\omega|^\#$ with small $\#$) to give a better global picture.
The poles are exponentially spaced along a straight line (dotted line) with angle $\th_c$ given by~\eqref{pepO}. There are infinitely many poles, only a few of which are shown.
{\it Left plot:} The black dashed line crossing the origin corresponds to the value of $\th_c$ at $k=k_o$ (see~\eqref{oenr}): the boundary of the oscillatory region. As $k \to k_o$, most of the poles approach the branch point $\om=0$ except for a finite number of them which become quasi-particle poles for the Fermi surfaces at larger values of $k$. The color dashed lines in the right half indicate the motion of poles on another sheet of the complex frequency plane at smaller values of $k$ (see the end of sec.~\ref{sec:3C} for the choice of branch cut in the $\om$-plane). {\it Right plot:} the two dashed lines correspond to $k=0$ (upper one) and $k=k_o$ (lower one). Again most of the poles approach the branch point $\om=0$ as $k \to k_o$.
These plots are only to be trusted near $ \omega=0$.  }
\end{center}
\end{figure}

We now examine the poles of~\eqref{osciP1} in the complex $\om$-plane, which is given by
 \be \label{poLc}
 1 + |c(k)| e^{2 \lam_k \th} e^{i X} = 0 , \quad {\rm with} \; \om \equiv |\om| e^{i \th} \ .
 \ee
\eqref{poLc} implies a series of poles located along a straight line
with angle $\th_c$ (in the expression below the integer $n$ should be large enough for our small $\om$ approximation to be valid)
 \be \label{pepO}
 \om_n  = e^{\ga_k - 2 \al - (2 n +1) \pi \ov 2 \lam_k} e^{i \th_c}, \;\; n \in \ZZ, \;\;
 \th_c = - {1 \ov 2 \lam_k} \log |c(k)|  \ .
 \ee
Equations~\eqref{cond1} and~\eqref{cond2} then imply that
  \bea
  \label{sond1}
&& {\rm scalars:} \qquad \th_c \in (0, \pi) \\
&& {\rm spinors:} \qquad \th_c > \pi \ .
 \label{sond2}
 \eea
 Thus the poles for scalars lie in the upper half $\om$-plane
 \cite{Denef:2009tp}
while those for spinors are in the lower half $\om$-plane. Poles on the upper half $\om$-plane of a
retarded Green function on the one hand implies causality violation. On the other hand from
equation~\eqref{bore} it implies that there exist normalizable modes which have
a frequency with
positive imaginary part. This leads to a mode exponentially growing with time and thus an
instability in the charged black hole geometry. See figure~\ref{fig:poles1-osc} for
illustration of the locations of poles for scalars and spinors and their movement as $k$ is varied in the range~\eqref{oenr}.

The instability for a scalar can also be understood in terms of classical superradiance. To see this let us go back to equation~\eqref{exwpSca}
which for $\nu_k = -i \lam_k$ the first term can be interpreted as an incident wave into the AdS$_2$ region with the second term the reflected wave. Thus the reflection probability is given by
 \be
 |\sG_k (\om)|^2 = \bca |c(k)|^2 & \om > 0 \cr
  |c(k)|^2 e^{4 \pi \lam_k}  &  \om < 0
 \eca
 \ee
Equation~\eqref{cond1} then implies $|\sG_k (\om)| > 1$ for $\om < 0$. Recall that our $\om$ is defined to be the deviation from the effective chemical $\mu_q = q \mu$. Thus $\om < 0$ agrees with the standard frequency region for superradiance.
 As mentioned earlier, an imaginary $\nu$ corresponds to the parameter
 regime
 where charged particles can be pair-produced. While both scalars and spinors are
 pair-produced,
 the superradiance of scalars can enhance the pair production into a classical
 instability. The produced scalar particles are trapped by the $AdS$ gravitational potential well (of the full geometry), and return to
the black hole to induce further particle production. In contrast, since the reflection probability for a spinor falling into the black hole is smaller than $1$, after a few bounces back and forth the paired produced spinor particles should fall back into
the black hole.

In our context, the fact that a scalar superradiates while a spinor does not (\ie\  equations~\eqref{cond1}--\eqref{cond2}) can be seen as a consequence of statistics
of the operator in the {\it boundary} theory.

\bwt

\ewt

\section{Emergent quantum critical behavior III: Fermi surfaces} \label{sec:surf}

\subsection{Quasi-particle-like poles}

We now consider what happens to~\eqref{finG1} when $a_+^{(0)} (k)$ in the expansion of~\eqref{expeta0} is zero. This can only occur for {\it real} $\nu_k$ at discrete values of $k$, at which values the wave function $\eta_+^{(0)}$
becomes normalizable. The possible existence of such $k$'s can be visualized at a heuristic
level by rewriting~\eqref{pp2} with $\om=0$ in the form of a Schr\" odinger equation, and noticing that for a certain range of momentum the Schr\" odinger potential develops a well which may allow normalizable ``bound states''.
See Appendix~\ref{app:schro} for details and also the similar story for spinors.
For which values of $k$ such bound states indeed occur can then be determined by
solving~\eqref{pp2} (with $\om=0$) numerically.

Suppose that $a_+^{(0)}$ has a zero at $k= k_F$. Then from~\eqref{zerow}, for $k \sim k_F$
    \be
    G_{R} (k, \om =0) \approx {b_+^{(0)} (k_F) \ov \p_k a_+^{(0)} (k_F)} {1 \ov k_\perp}, \qquad k_\perp \equiv k - k_F \ .
    \ee
Since $\nu_k$  is real, $a_+^{(0)}, b_+^{(0)}$ are all real (see \eg\
discussion around~\eqref{eoop}). Thus $\Im G_R (\om=0, k)$ is identically zero around $k_F$, but the real part $\Re G_R (\om=0, k)$ develops a pole at $k=k_F$. Now turning on a small $\om$ near $k_F$, we then have to leading order
\bea
 G_{R} (k,\om) & \approx &  {b_+^{(0)} (k_F) \ov \p_k a_+^{(0)} (k_F) k_\perp  + \om a_+^{(1)} (k_F) + a_-^{(0)} (k_F) \sG_{k_F} (\om) } \cr
  &=&  {h_1 \ov  k_\perp -  {1 \ov v_F} \om - h_2 e^{i \ga_{k_F}} \om^{2 \nu_{k_F}}}
  \label{spinorG}
 \eea
where in the second line we have used~\eqref{defc} and introduced
 \bea \nonumber
 v_F \equiv - {\p_k a_+^{(0)} (k_F) \ov a_+^{(1)} (k_F)} ,\quad
 h_1 \equiv  {b_+^{(0)} (k_F) \ov \p_k a_+^{(0)} (k_F)} , \\
 h_2 \equiv  - |c(k_F)| {a_-^{(0)} (k_F) \ov \p_k a_+^{(0)} (k_F) }   \ . \qquad \qquad
 \label{thrIM}
 \eea
The term linear in $\om$ in the downstairs of~\eqref{spinorG} can be omitted if $\nu_{k_F} < \ha$. For $\nu_{k_F} > \ha$ we should still keep the term proportional to $\om^{2 \nu_{k_F}}$, since it makes the leading contribution to the imaginary part.
Note that the quantities in~\eqref{thrIM} are all real and as we will discuss below they are all {\it positive}.

The coefficients $v_F, h_{1,2}$ may be expressed in terms of integrals of bound state wave function $\eta_+^{(0)}$ at $k=k_F$ by
perturbing the Schr\"odinger problem at $\om=0, k=k_F$ in $\omega, k$,
similar to the demonstration of the Feynman-Hellmann theorem.
We present the details of this analysis in Appendix \ref{app:vF}.
In particular, from a combination of analytic and numerical analysis, we show that (for $q>0$):

\ben

\item For $\nu_{k_F} > \ha$, $v_F > 0$ for both scalars and spinors and in particular for a spinor $v_F < 1$.

\item For all $\nu_{k_F}$, for scalars
 \be \label{positive}
  h_1, h_2 > 0 \ .
 \ee
The above inequalities are established analytically in Appendix~\ref{app:c4}.
For spinors the story is more involved, and our conclusions rely on the numerics (see fig.~\ref{fig:h12}).
As for scalars, $h_1>0$ for all $\nu$.
As defined in \eqref{thrIM}, the sign of $h_2$ is indefinite.
However, we find that the sign of $h_2$ is precisely correlated with
the additive ambiguity in the phase $\gamma$ \eqref{phaseE} in such a way 
that  $h_2 e^{i \text{arg} \sG_R} = |h_2| e^{i \gamma}$.  

 \een

Equation~\eqref{spinorG} leads to a pole in the complex-$\om$ plane\footnote{Recall the
discussion at the end of sec.~\ref{sec:3C} for the choice of branch cut in the $\om$-plane.} located at
\bea
 \om_c (k) &\equiv & \om_* (k) - i \Ga (k) \cr
 &=& \bca \le({k_\perp \ov h_2}\ri)^{1 \ov 2 \nu_{k_F}} e^{- i {\ga_{k_F} \ov 2 \nu_{k_F}}} & \nu_{k_F} < \ha \cr
      v_F k_\perp - v_F h_2 e^{i \ga_{k_F}} (v_F k_\perp)^{2 \nu_{k_F}} & \nu_{k_F} > \ha
      \eca
 \label{poem}
 \eea
with residue at the pole given by
 \be \label{residue}
 Z = \bca - {\om _c h_1 \ov 2 \nu_{k_F} k_\perp} \propto k_\perp^{1- 2 \nu_{k_F} \ov 2 \nu_{k_F}} &  \nu_{k_F} < \ha \cr
    - h_1 v_F  &  \nu_{k_F} > \ha   \ .
 \eca
 \ee
Notice that both the real and imaginary part of the pole go to zero as $k_\perp \to 0$.
Thus~\eqref{poem} leads to a sharp quasi-particle peak in the spectral function $\Im G_R (\om,k)$ in the limit $k_\perp \to 0$ with a dispersion relation
 \be \label{disP}
 \om_* (k) \propto  k_\perp^{z} \quad {\rm with} \quad z = \bca {1 \ov 2 \nu_{k_F}} & \nu_{k_F} < \ha \cr
            1 & \nu_{k_F} > \ha
            \eca
 \ee
and
 \be \label{disP1}
 \Ga (k) \propto  k_\perp^{\delta} \quad {\rm with} \quad \delta = \bca {1 \ov 2 \nu_{k_F}} & \nu_{k_F} < \ha \cr
            2 \nu_{k_F} & \nu_{k_F} > \ha
            \eca \ .
 \ee

Note that when $\nu_{k_F} < \ha$, the pole
follows a straight line as $k_\perp$ is varied. More explicitly,
 \be \label{theCD}
 \th_c = {\rm arg}(\omega_c) = \bca
    - {\ga_{k_F} \ov 2 \nu_{k_F}} & k_\perp > 0 \cr
    {\pi - \ga_{k_F} \ov 2 \nu_{k_F}} & k_\perp < 0
    \eca
 \ee
and the width $\Ga$ is always comparable to
the frequency $\om_*$,\footnote{Note the concept of a width is only operationally meaningful when the pole lies in the physical sheet. See the end of sec.~\ref{sec:3C} for our choice of the physical sheet.}
\be \label{ane}
 {\Ga (k) \ov \om_* (k)} = -\tan \th_c = {\rm const}  \ .
 \ee
 In contrast, for $\nu_{k_F} > \ha$, ${\Ga(k) \ov
\om_* (k)} \to 0$ as $k_\perp \to 0$. See fig.~\ref{fig:poles2-nonfl} for examples of the motion of a spinor pole in the complex $\om$-plane.

At $\nu_{k_F} = \ha$, $a_+^{(1)}$ and $\sG_k (\om)$ in equation~\eqref{spinorG}
are divergent. Both of them have a simple pole at $\nu_{k_F} = \ha$. The pole in $\sG_k$ can be seen explicitly from the factor $\Ga (- 2 \nu_k)$ in~\eqref{exbG1} and~\eqref{rprp}. The pole in $a_+^{(1)}$ can be seen from the discussion~\eqref{nearha}--\eqref{paraM1} in Appendix~\ref{app:vF}. The two poles cancel each other and
leave behind a finite $\om \log \om$ term with a real coefficient. More explicitly, we have
 \be \label{eprM}
 G_R \approx {h_1 \ov k_\perp + \tilde c_1 \om \log \om + c_1 \om }
  \ee
where $\tilde c_1$ is real and $c_1$ is complex.

Similar logarithmic terms appear
for any $\nu_{k_F} = {n \ov 2}, \;\; n \in \ZZ_+$. For example, at $\nu_{k_F} =1$, one finds that
 \be \label{jejr}
 G_R (\om,k) \approx {h_1 \ov  k_\perp - {1 \ov v_F} \om + \tilde c_2 \om^2 \log \om + c_2 \om^2}
 \ee
with $\tilde c_2$ real and $c_2$ complex.

\subsection{A new instability for bosons}

The above discussion applies identically to both scalars and spinors with their respective parameters. We now show that for spinors the pole~\eqref{poem} never appears in the upper half plane of the physical sheet, while for scalars it always lies in the upper half plane for $k_\perp < 0$. The difference can again be attributed to the statistics of the corresponding boundary operators.

First note that $\Im G_R$ obtained from~\eqref{spinorG} should again satisfy~\eqref{s1} and~\eqref{f1} (which follow from the statistics of the full boundary theory), leading to~(using also~\eqref{positive})
 \bea
  \label{dond3}
{\rm scalars:} \quad \sin \ga_{k_F} > 0 , \quad  \sin \le(\ga_{k_F} + 2 \pi \nu_{k_F} \ri) <0 \quad \\
{\rm spinors:} \quad  \sin \ga_{k_F} > 0 , \quad  \sin \le(\ga_{k_F} + 2 \pi \nu_{k_F} \ri) >0
\quad
 \label{dond4}
 \eea
which indeed follow from discussion below~\eqref{phaseE}. Equations~\eqref{dond3} and~\eqref{dond4} are also consequences of (for $\om > 0$)
   \bea
 \label{s2}
&&{\rm scalars:} \quad  \Im \sG_k (-\om) < 0, \quad \Im \sG_k (\om) > 0  \quad \\
 &&{\rm spinors:} \quad  \Im \sG_k (-\om) > 0, \quad \Im \sG_k (\om) > 0 \  \quad
 \label{f2}
 \eea
which follow from the Bose and Fermi statistics of the IR CFT. This gives a self-consistency check of the statistics of the full theory and its IR CFT.

\begin{figure}[h!]
 \begin{center}
\includegraphics[scale=0.70]{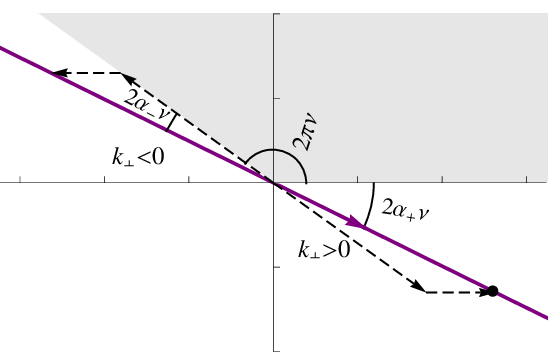}
\includegraphics[scale=0.70]{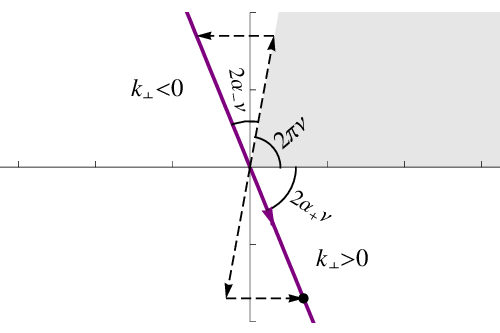}
\caption{\label{fig:poles2-cov-sheet}
A geometric illustration that poles of the spinor Green
function never appear in the upper-half $\omega$-plane of the physical sheet, for two choices of $\nu_{k_F} < \ha$.  Depicted here is the $\omega^{2\nu_{k_F}}$ covering space on which the Green function~\eqref{spinorG}, with the $\omega/v_F$ term neglected, is single-valued. The shaded region is the image of the upper-half $\omega$-plane of the physical sheet. The pole lies on the line $2 \nu_{k_F} \th_c = - \ga_{k_F} $ for $k_\perp > 0$ and on $2 \nu_{k_F} \th_c = \pi - \ga_{k_F} $ for $k_\perp <0$, which are indicated by the purple solid line in the figure. The triangle formed by dashed arrows and solid lines in the upper left quadrant gives the geometric illustration for the equation $\pi - \ga_k ={\rm arg} ( e^{2 \pi i \nu_k} - e^{-2 \pi q e_d})$ (following from the first equation of~\eqref{phaseE}), which makes it manifest that for $k_\perp < 0$ the pole lies outside the shaded region. In contrast for a scalar one needs to reverse the direction of the horizontal dashed !
 line and the pole lies inside the shaded region. Similarly, the triangle in the lower right quadrant gives the illustration for $- \ga_k ={\rm arg} (- e^{2 \pi i \nu_k} + e^{-2 \pi q e_d})$ which is relevant for $k_\perp > 0$. We also indicated the angles $\al_\pm$ which will be introduced and discussed in detail around~\eqref{aloenp1}.
}
\end{center}
\end{figure}

\begin{figure}[h!]
 \begin{center}
\includegraphics[scale=0.90]{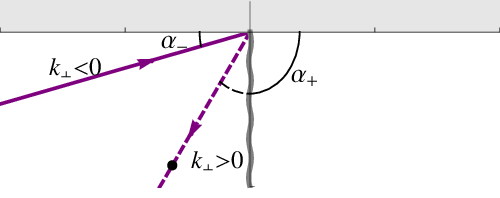}
\hskip 0.2cm
\includegraphics[scale=0.70]{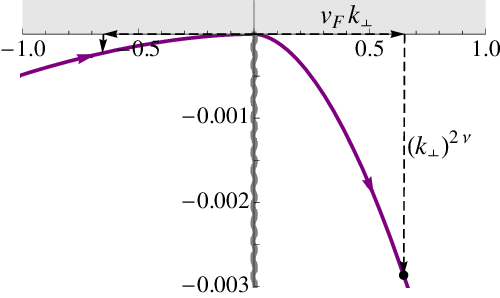}
\caption{\label{fig:poles2-nonfl}
Examples of the motion of the pole for a spinor as $k_\perp$ is varied (arrows indicating the directions of increasing $k_\perp$). {\it Left plot:} $\nu_{k_F}<\half$, for which
the pole moves in a straight line. The plot shows an example where the pole moves to another sheet of the Riemann plane for $k_\perp > 0$ (\ie\  $\al_+ > {\pi \ov 2}$);
$\al_\pm$ indicated there are introduced in~\eqref{aloenp1}.
{\it Right plot:} $\nu_{k_F} > \half$ for which the dispersion (real part of the pole) is linear.
}
\end{center}
\end{figure}

For $\nu_{k_F} > \ha$, applying equations~\eqref{dond3}--\eqref{dond4} to the last line of~\eqref{poem} we find that for spinors the pole always lies in the lower half plane while for scalars the pole is always on upper half complex $\om$-plane for $k_\perp < 0$.
By using in addition~\eqref{spinine} we again reach the conclusion that for spinors the pole never appears in the upper half plane of the physical sheet, while for scalar it always does for $k_\perp < 0$. In fig.~\ref{fig:poles2-cov-sheet} we give a geometric picture
to illustrate this. In fig.~\ref{fig:poles2-nonfl} we illustrate the motion of poles as $k_\perp$ is varied for spinors.

The pole in the upper half plane for scalars when $k_\perp < 0$ is intriguing.
As mentioned earlier poles in the upper-half $\om$-plane imply the existence of exponentially growing (in time) normalizable mode in AdS and lead to instability.
Again as in the case of the oscillatory region the instability for scalars can be attributed to the Bose statistics of the boundary theory. This instability is curious as
it occurs for real $\nu_k$ and is thus distinct from the instability discussed in last section, which is associated with an imaginary $\nu_k$. In particular, it appears that this instability can exist in a parameter range where previously considered superconducting instability does not occur. To understand the physical interpretation of such an instability we should examine the motion of the pole as $k$ is decreased, in particular, whether it persists to $k=0$. If it does, it seems likely the pole will have finite real part at $k=0$, \ie\  the growing mode also oscillates
in time. Such an instability appears to be novel and we will leave its interpretation and a detailed study for future work.

\bwt

\ewt

\subsection{Fermi surfaces}

\begin{figure}[h!]
 \begin{center}
\includegraphics[scale=0.70]{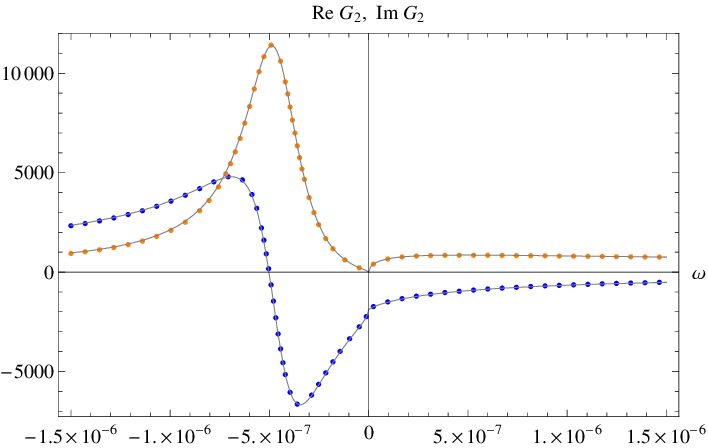}
\caption{\label{fig:fits} Spinor Green's function $G_2$ (as defined in Appendix~\ref{app:fer}) at $k= 0.918$ as a function of $\omega$, computed numerically. We have chosen parameters $r_*=R=g_F=q=1$, $m=0$ and $d=3$ for which the Fermi momentum is $k_F = 0.91853$.  The real and imaginary parts are shown in blue and orange dotted curves, respectively. Also shown is~\eqref{spinorG} (solid lines) with $h_1, h_2$ computed numerically using the method of Appendix~\ref{app:vF} and $\nu_k$ given by~\eqref{defNu}.
}
\end{center}
\end{figure}

We now focus on spinors, for which equations~\eqref{spinorG} and~\eqref{poem} give analytic expressions for, and generalize to any mass $m$ and charge $q$, numerical results of~\cite{Liu:2009dm}. Reference~\cite{Liu:2009dm} focused on $m=0$ and a few values of charge $q$. There $k_F$ was found by studying the scaling behavior of quasi-particle peaks as they become sharper and sharper as $k_F$ is approached.
In our current discussion $k_F$ are found from ``bound states'' of the Dirac equation
at $\om =0$~(which needs to be found numerically). We found perfect agreement between two approaches. Plugging explicit values of $k_F$ into~\eqref{spinorG} and~\eqref{poem} also leads to almost perfect agreement with numerical plots of the retarded function in~\cite{Liu:2009dm} including the scaling exponents~(see fig.~\ref{fig:fits}).

The sharp quasi-particle peaks in the
spectral function for spinors were interpreted in~\cite{Liu:2009dm} as strong indications of underlying Fermi surfaces with Fermi momentum $k_F$. The scaling behavior near a Fermi surface for the parameters range considered there, which all have $\nu_{k_F} < \ha$, is different from that of the Landau Fermi liquid, suggesting an underlying non-Fermi liquid.
Our current discussion allows us to obtain a ``landscape'' of non-Fermi liquids by
scanning all possible values of $m$ and $q$. Before doing that, let us first
make some general comments on the scaling behavior~\eqref{poem}--\eqref{ane}
and the possible underlying non-Fermi liquids:

\ben

\item The form of retarded Green function~\eqref{spinorG} and scaling behavior~\eqref{poem} again have a nice interpretation in terms of the RG picture described earlier; while the location of the Fermi momentum $k_F$ is governed by the UV physics (just like in real solids), the scaling exponents~\eqref{poem} near a Fermi surface are controlled by the dimension of the corresponding operators $\sO_{\vk}$ (with $|\vk| = k_F$) in the IR CFT.
    Note that while the specific value of $\nu_{k_F}$ depends on $k_F$ as an input parameter, its functional dependence on $k_F$ is fixed by the IR CFT.

\item For $\nu_{k_F} < \ha$, the corresponding operator in the IR CFT is relevant\footnote{Recall that the dimension in the IR CFT is $\delta_{k_F} = \ha + \nu_{k_F}$.}\footnote
{
Note added: The correct notion of `relevant' and `irrelevant' here 
has been explained in \cite{Faulkner:2010tq}.  
The pertinent issue is the dimension of the product of the operator in the IR CFT and a free fermion 
representing the boundstate at the Fermi surface.
}.
From~\eqref{ane}, the imaginary part of the pole is always comparable to the real part and thus the quasi-particle is never stable. Also note the ratio~\eqref{ane} depends only  on the IR data. Another important feature of the pole is that its residue~\eqref{residue} goes to zero as the Fermi surface is approached. In particular, the smaller $\nu_{k_F}$, the faster the residue approaches zero.

\item When $\nu_{k_F} > \ha$, the corresponding operator in the IR CFT is irrelevant.
The real part of the dispersion relation~\eqref{poem} is now controlled by the analytic UV contribution and becomes linear with a ``Fermi velocity'' given by $v_F$. $v_F$ is controlled by UV physics and will have to be found numerically by solving the outer region equations (see fig.~\ref{fig:vF} and discussion of the next subsection). The imaginary part of the pole is still controlled by the dimension of the operator in the IR CFT. In this case the width becomes negligible compared with the real part as the Fermi surface is approached $k_\perp \to 0$, so the corresponding quasi-particle becomes stable. Furthermore the quasiparticle residue is now non-vanishing~\eqref{residue}. Note the scaling exponent of the width $\Ga$ is generically different from the $\omega^2$-dependence of the Landau Fermi liquid.

\item When $\nu_k = \ha$, the corresponding operator in the IR CFT is marginal. Now the retarded function is given by~\eqref{eprM}, in which case the imaginary part of the pole is still suppressed compared to the real part as the Fermi surface is approached, but the suppression is only logarithmic. The quasi-particle residue now vanishes
    logarithmically as the Fermi surface is approached. Remarkably~\eqref{eprM} is precisely of the form postulated in~\cite{varma} for the ``Marginal Fermi Liquid'' to describe the optimally doped cuprates.
    In~\cite{varma} the term ``marginal'' referred to the logarithmic vanishing of the  quasi-particle weight~(residue) approaching the Fermi momentum. Here we see that this term ``marginal'' is indeed perfectly appropriate.

\item At $\nu_{k_F} = 1$, from~\eqref{jejr} the retarded Green function resembles that of a Landau Fermi liquid with the real part of the quasi-particle pole linear in $k_\perp$ and the imaginary part quadratic in $k_\perp$. Equation~\eqref{jejr}, however,
   has a logarithmic term $ \om^2 \log \om$ with a real coefficient which is not present for a Landau Fermi liquid\footnote{It is amusing to note that a Landau Fermi liquid in $(2+1)$-d has a logarithmic correction of the form  $ \om^2 \log \om$, but with a pure imaginary coefficient~(see~e.g.~\cite{nonaly}).}. In particular, this leads to
     a curious particle-hole asymmetry with a difference in width for a hole and a particle given by\footnote{The above asymmetry can also be expressed as $ {\Ga (\om_*<0) \ov \Ga (\om_* > 0)} = e^{-2 \pi q e_d}$.}
    \be
    \Ga (\om_*<0) - \Ga (\om_* > 0) = \pi \tilde c_2 \om_*^2 \ .
    \ee

\item Notice from~\eqref{disP} that for all values of $\nu_{k_F}$, $z \geq 1$.
This is consistent with an inequality proposed by Senthil in~\cite{Senthil:0803} for a critical Fermi surface.\footnote{Senthil derived the inequality $z \geq 1$ by approaching a non-Fermi liquid at the critical point of a continuous metal-insulator transition from the Landau Fermi liquid side and requiring the effective mass should not go to zero as the critical point is approached.} In the notation of~\cite{Senthil:0803} our Green function~\eqref{spinorG} also has scaling exponent $\al = 1$, with $\alpha$ defined as
$G_{R} (\lam^z \om, \lam k_\perp) = \lam^{- \alpha} G_R (\om, k_\perp)$. Thus we also have $z \geq \al$ for all cases, consistent with the other inequality proposed in~\cite{Senthil:0803}.

\item  For $\nu_{k_F} < \ha$, equation~\eqref{spinorG} (with the linear term in $\om$ omitted) exhibits a particle-hole asymmetry. To characterize this it is convenient to define
    \be
    \begin{split}
    \label{aloenp1}
    \al_+ & \equiv  -\th_c (k_\perp > 0) = {\ga_{k_F} \ov 2 \nu_{k_F}},
     \\
    \al_- & \equiv  \th_c (k_\perp < 0) - \pi =  {\pi - \ga_{k_F} \ov 2 \nu_{k_F}} - \pi
    \end{split}
    \ee
with $\th_c$ introduced in~\eqref{theCD}. $\al_+$ gives the angular distance away from the positive real $\om$-axis for the pole (\ie\  particle-type excitations) at $k_\perp > 0$, while $\al_-$ gives the angular distance away from the negative real $\om$-axis for the pole (\ie\  hole-type excitations) at $k_\perp < 0$. From fig.~\ref{fig:poles2-cov-sheet} and its caption, $\al_\pm$ are always positive. Their values determine how close the corresponding pole stays to the real axis and hence indicate the sharpness of the resulting peak in the spectral function (\ie\  imaginary part of $G_R$). In particular, when $\al_\pm$ exceeds ${\pi \ov 2}$ the corresponding pole moves to the other Riemann sheet\footnote{An example is given in the left plot of fig.~\ref{fig:poles2-nonfl}.}. The particle-hole asymmetry is then reflected in the relative magnitudes of $\al_\pm$. From equation~\eqref{phaseE} one
 finds that $2 \nu_{k_F} \al_- = {\rm arg} (1 - e^{-2 \pi q e_d- 2 \pi i \nu_{k_F}})$
which implies $\al_-$ quickly becomes small when $q$ increases as can also be visualized from fig.~\ref{fig:poles2-cov-sheet} since the horizontal line of the upper triangle becomes short. In contrast, with some thought one can conclude from fig.~\ref{fig:poles2-cov-sheet} that $\al_+$ can only be small when $\nu_{k_F}$ is close to $\ha$, where $\al_-$ is also small.
Away from $\nu_{k_F} = \ha$, one will generically (except for small $q$) have a particle-hole symmetry with a sharp peak on the hole side (\ie\  small $\al_-$), but a broad bump on the particle side (not so small $\al_+$). This was indeed what was observed in~\cite{Liu:2009dm} numerically. The above qualitative features will be confirmed by explicit numerical calculations represented in fig.~\ref{fig:mq_plot}.

\item Non-Fermi liquids have been described previously by coupling a Fermi surface to a propagating
bosonic mode, such as a transverse magnetic excitation of a gauge field,
\eg \cite{Holstein:1973zz,Polchinski:1993ii,Nayak:1993uh,Halperin:1992mh,altshuler-1994, Schafer:2004zf,Boyanovsky,sungsik-2009}. The forms of the fermion Green's functions thus obtained all fit into the set of functions we have found in~\eqref{spinorG}.
An important difference, however, is that each of these analyses required a small parameter\footnote{\eg\   $\alpha \sim {1\over 137}$ in~\cite{Holstein:1973zz}, $1/N$ in~\cite{Polchinski:1993ii}, the parameter~$x$ in the gauge boson dispersion in~\cite{Nayak:1993uh}... } to control the perturbation theory,
and the range of frequencies (or temperatures) over which the non-Fermi liquid behavior is
relevant is parametrically small in the control parameter. As a result the non-Fermi liquid behavior will only be visible at extremely low temperatures. In our non-perturbative calculation, this range is order unity, and so our non-Fermi liquids may be considered robust versions of these previously-identified phases.

\item The expression~\eqref{spinorG} for retarded function near a Fermi surface and the matching procedure from which~\eqref{spinorG} was derived suggest the following effective action of the UV $\sO_U$ and IR $\sO_I$ part of an operator $\sO$ (below $|\vk_F| = k_F$)
 \bea
 \nonumber
 S = \int d\om d\vk \, \bar \sO_U \, \Sig (\om, \vk_\perp )\,  \sO_U  \qquad\qquad \\
 + \int d\om d \vk_F  \,  D (\om, \vk_F) \bar \sO_U (\om, \vk_F)^\da \sO_I (\om, \vk_F ) + h.c. \
 \label{pheMp}
 \eea
where $\Sig (\om, \vk_\perp)$ represents the kinetic term for $\sO_U$ and $D(\om, \vk_F)$
denotes coupling between $\sO_I$ and $\sO_U$. Both $\Sig$ and $D$ are
controlled by UV physics. They are assumed to be real and depend analytically on
$\om$. We assume that the dynamics of the IR operator $\sO_I (\om, \vk_F)$ is controlled by the IR CFT with two point function given by
\be
\vev{\sO_I (\om, \vk_F)^\da \sO_I (\om', \vk_F')}_{IR}
= \sG_{k_F} (\om) \delta_{\vk_F, \vk_F'} \delta (\om-\om').
\ee
The action then implies that after summing a geometric series, the full correlation function of $\sO_U$ is then given by
 \be
 G_R (\om, \vk_\perp) = {1 \ov \Sig (\om, \vk_\perp) + D^2 (\om, \vk_F) \sG_{k_F} (\om)}
 \
 \ee
which has the form of~\eqref{spinorG}. Thus, action~\eqref{pheMp} gives a phenomenological model of the small-frequency matching procedure described in this
 paper as a coupling between UV and IR degrees of freedom. This argument is not dissimilar to that taken in~\cite{varma} to obtain the ``marginal Fermi liquid''. It might also be possible to reinterpret the discussion of~\cite{Holstein:1973zz,Polchinski:1993ii,Nayak:1993uh,Halperin:1992mh,altshuler-1994, Schafer:2004zf,Boyanovsky,sungsik-2009} in this language.

\een

\subsection{The zoo of non-Fermi liquids from gravity} \label{sec:zoo}

The discussion of previous subsection was based on qualitative features of~\eqref{spinorG}
which are controlled by the IR CFT. To obtain further information regarding properties of the Fermi surface and its low energy excitations we need to work out (numerically) the data which are controlled by UV physics. These include: $k_F$ (which then determines $\nu_{k_F}$ and $\ga_{k_F}$), $v_F$, $h_1$ and $h_2$. We will map out how they depend on the charge $q$ and dimension $\De$~(or equivalently mass $m$) of an operator, \ie\  the ``phase diagram''\footnote{This is an abuse of language because
the parameters we are varying are couplings in the bulk action,
and hence change the boundary system, not its couplings.}
of holographic non-Fermi liquids.

In this section we will often use the notations and equations developed in Appendix~\ref{app:fer}. Readers are strongly encouraged to read that part first.

We first solve the Dirac equation~\eqref{spinorequation} numerically with $\om=0$ to find $k_F$, for which $a_+^{(0)}$ defined in equation is~\eqref{spinAs} is zero. This is equivalent to finding the bound states of the Dirac equation. See Appendix~\ref{app:schro} for more details. For $m \in [0, \ha)$ we consider at the same time the alternative quantization of the bulk spinor field, whose boundary Green function $\tilde G_{1,2} (m,k)$ are given by $G_{1,2} (-m,-k)$ (from~\eqref{altenQ}). We will thus use negative mass to refer to the alternative quantization.
The results are presented in fig.~\ref{fig:qk}, where we plotted $k_F$ dependence on charge $q$ for three different masses $m = -0.4, \, 0, \, 0.4$. Note in this and all subsequent plots if not stated explicitly, we use without loss of generality $R=1$,  $r_* =1$, $g_F = 1$ and $d=3$, for which
the chemical potential is $\mu = \sqrt{3}$.

We see in fig.~\ref{fig:qk} that for a given mass when increasing $q$ new branches of Fermi surfaces appear as was observed before in~\cite{Liu:2009dm}. This can be understood from the point of view of the Dirac equation as that increasing $q$ allows more bound states. The lowest bound state has the largest $k_F$,\footnote{As discussed in Appendix~\ref{app:schro}, the bound state problem for the Dirac equation can be approximately thought of as a Schrodinger problem with eigenvalue $-k^2$.} which corresponds to the lowest curves in fig.~\ref{fig:qk}. We will refer it as the `primary Fermi surface'. For $q, m$ and $k$ large we can also use the WKB approximation to solve the Dirac equation to find the bound states, which is discussed in detail in Appendix~\ref{app:schro}. In particular, fig.~\ref{fig:allowedd} gives the parameter region where there exist Fermi surfaces for large $k_F ,m, q$ for various boundary theory space-time dimensions.

\begin{figure}[h!] \begin{center}
\includegraphics[scale=0.65]{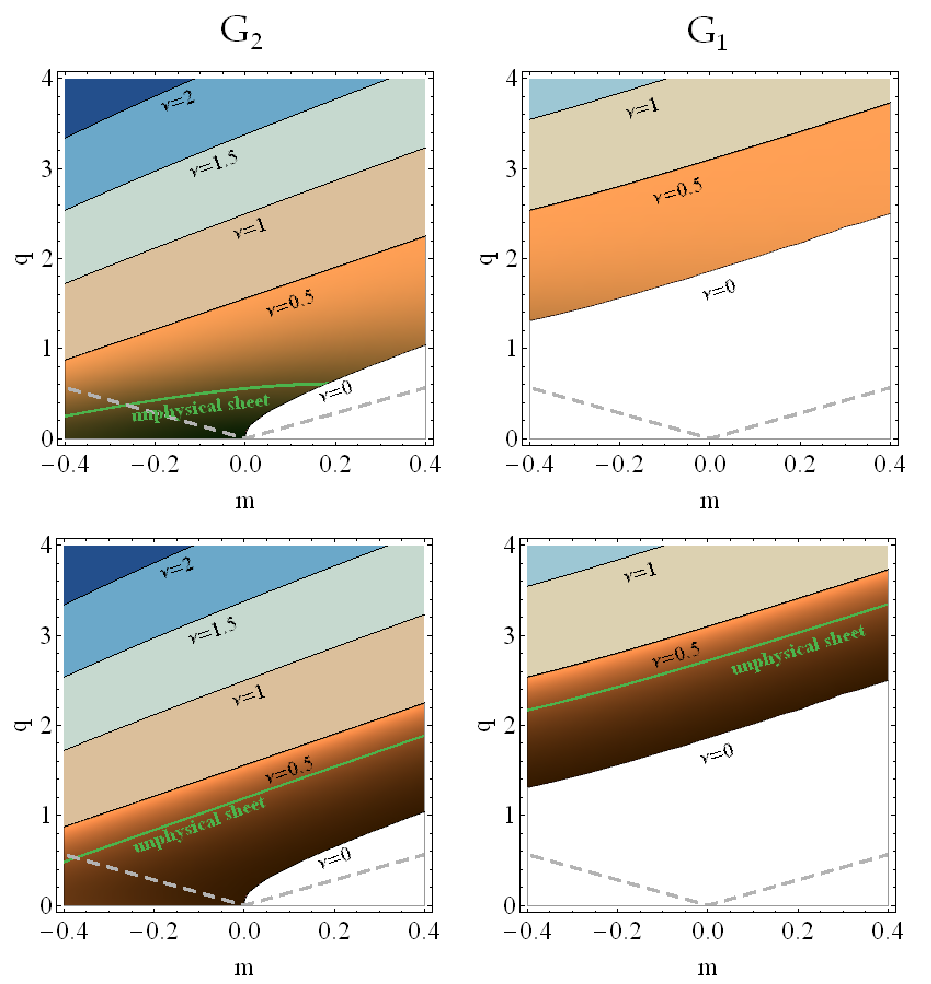}
\caption{\label{fig:mq_plot} The `phase diagram': Shown here are contour plots of the exponent $\nu_{k_F}$ evaluated on the ``primary Fermi surface" (the one with the largest $k_F$), as a function of $m$ and $q$, for each of the two components of the spinor operator. The top and bottom rows differ only in the function used to shade the region $\nu_{k_F} \in [0, \half]$. Top row is shaded according to $\al_-$  while the bottom row is
shaded according to $\al_+$.
$\al_\pm$ were introduced in~\eqref{aloenp1} and discussed in detail around there.
The darker region corresponds to larger values of the angles.
Both angles are zero at the line of $\nu_{k_F}= \ha$ and
increase with a decreasing $\nu_{k_F}$. When an angle exceeds ${\pi \ov 2}$ the corresponding pole moves into another Riemann sheet, the regions for which are indicated in the plots. As anticipated in the discussion below~\eqref{aloenp1}, $\al_-$ becomes $O(1)$ only for small $q$, which $\al_+$ becomes small only for $\nu_{k_F}$ close to $\ha$. We also indicated the region where there exists an oscillatory region for momentum satisfying~\eqref{oenr}. This region lies above the dashed lines.
It is clear from the plots that for $m > 0$, the region which allows a Fermi surface always lies inside the region which allows the oscillatory region. It is also interesting to note that in the left plot the dashed line in fact meets with the line for $\nu_{k_F} = \ha$
at $m = -\ha$ (not shown in figure). This happens for $d=3$ only. For general $d$ dimension, the dashed line intersects with $m =-\ha$ at $\nu_{k_F} = {1 \ov 2 (d-2)}$.
}
\end{center}
\end{figure}

\begin{figure}[h!]
 \begin{center}
\includegraphics[scale=0.45]{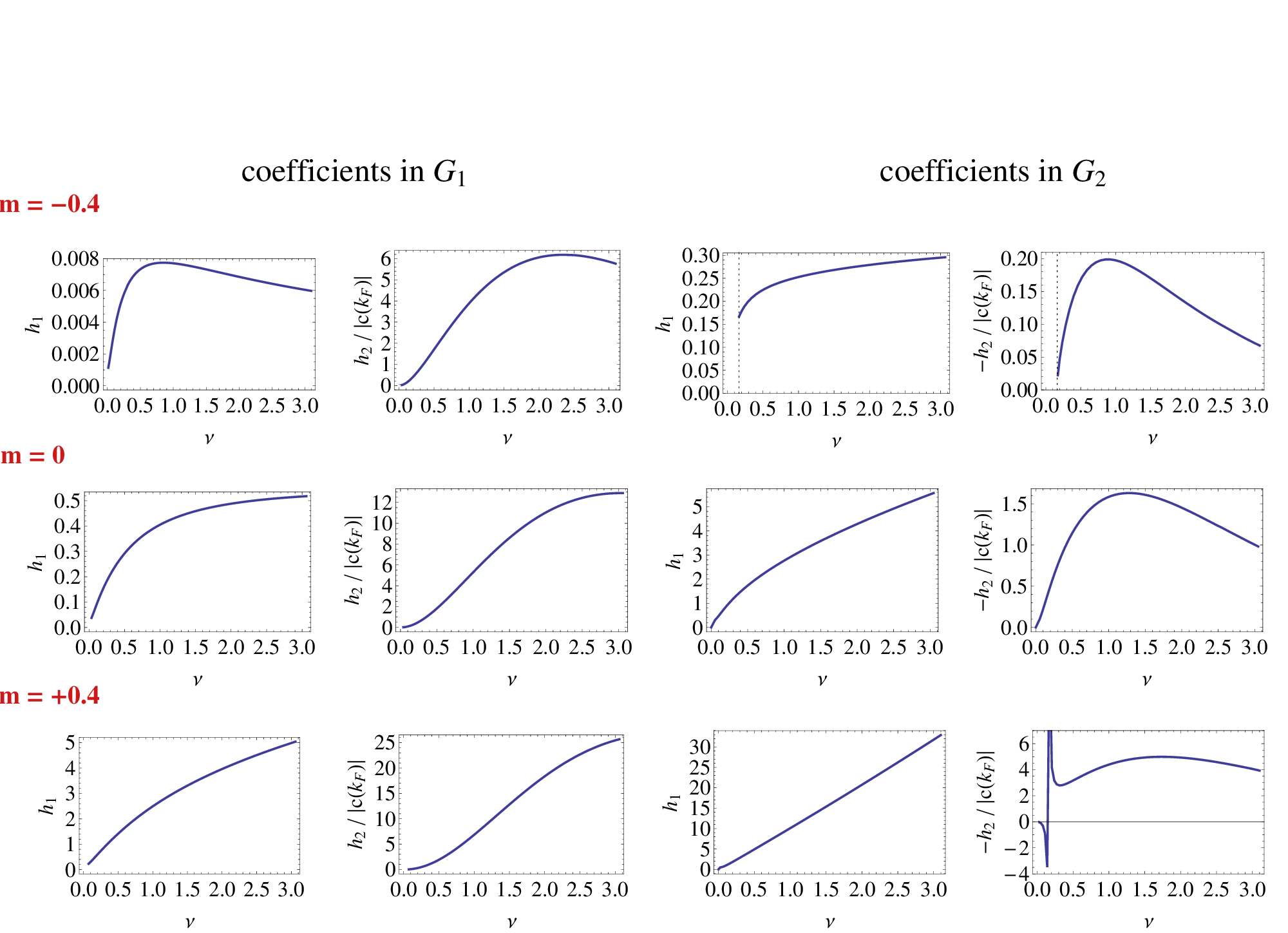} 
\caption{\label{fig:h12}
$h_1$ and $h_2/|c(k_F)|$ coefficients in~\eqref{spinorG} for the primary Fermi surface
of various components as a function of $\nu_{k_F}$. Various values of $m$ are indicated.
In the $h_2$ plot for $G_2$ at $m= -0.4$: 
there is a zero of $c(k_F)$ at $\nu = mR_2 \approx .16$, at which $h_2$ changes sign.
We explain the (lack of) significance of this phenomenon in Appendix \ref{app:piambiguity}.
For convenience we also plot $|c(k_F)|$ separately in fig.~\ref{fig:ck} below.}
\end{center}
\end{figure}

In fig.~\ref{fig:mq_plot} we map the values of $\nu_{k_F}$, which can be computed from~\eqref{defNu}, for the primary Fermi surface in the $q$-$m$ plane. As discussed further in Appendix~\ref{app:vF}, using the wave function for the bound state at $k_F$ we can evaluate $v_F$ and $h_1, h_2$. Their values for various masses as a function of $\nu_{k_F}$ are presented in fig.~\ref{fig:vF}--fig.~\ref{fig:h12}.


We now summarize some important properties which can be read from these plots and the WKB analysis in Appendix~\ref{app:schro}:

\ben

\item For any $m \geq 0$ and $q$, in the standard quantization, the existence of Fermi surfaces is always correlated with the existence of the oscillatory region which requires that for any dimension $d$ (c.f.~\eqref{ruop})
     \be \label{regOs}
     \De < {|q| g_F \ov \sqrt{2}} + {d \ov 2} \ .
     \ee
  This is clear from both fig.~\ref{fig:qk} and fig.~\ref{fig:mq_plot} for the masses plotted there. In fact, the allowed region for Fermi surfaces is more stringent than~\eqref{regOs}. This can be seen from fig.~\ref{fig:allowedd} (which follows from the WKB analysis in Appendix~\ref{app:schro}), which indicates that Fermi surfaces only exist for (for any $d$)
   \be \label{regFe}
     m^2 R^2< {q^2 g_F^2 \ov 3} , \quad {\rm \ie\ } \quad \De < {|q| g_F \ov \sqrt{3}} + {d \ov 2} \ .
     \ee
   At $m=0$, in units of the effective chemical potential $\mu_q$, the range for allowed $k_F$ is
   \be \label{kfrang}
  {d-2 \ov \sqrt{d (d-1)}} \leq {k_F \ov \mu_q} \leq 1
    \ee
   where the lower limit is the boundary of the oscillatory region.
   The upper limit in~\eqref{kfrang} also applies to other masses, achieved in the limit
   $m$ finite, $q \to \infty$. The lower limits for other masses are  smaller than that in~\eqref{kfrang} as can be seen from fig.~\ref{fig:allowed3} (but an analytic expression is not known). Note that for $m$ approaching the allowed limit~\eqref{regFe}, $k_F/\mu_q$ lies in a small region around
   \be
     {k_F \ov \mu_q} = {d-2 \ov \sqrt{3 d (d-1)}} \ .
     \ee

\item For $m R\in (-\ha,0)$, \ie\  for alternative quantization, there exists a single Fermi surface which does not enter the oscillatory region. This is the primary Fermi surface with the largest $k_F$. In fact for the small window
    \be
    {|q| g_F \ov \sqrt{2}} <  |m| R,
    \ee
\ie\  in terms of boundary theory dimension $\De = {d \ov 2} - |m| R$,
\be
 {d-1 \ov 2} < \De < {d \ov 2} - {|q| g_F \ov \sqrt{2}}
 \ee
 there exists a Fermi surface without oscillatory region.

\item For a given $m$, as one reduces $q$, $k_F$ (and $\nu_{k_F}$) decreases and eventually loses its Fermi surface identity by entering into the oscillatory region. Similarly, for a given $q$, as one increases $m$, $k_F$ (and $\nu_{k_F}$) decreases and eventually enters the oscillatory region.

\item For a Fermi surface, one can define a topological number using the Green's function~\cite{volovik}
\be
  n = \textrm{Tr} \oint_C {dl \over 2\pi i} G_R (k_\perp, i \omega) \partial_l G^{-1}_R (k_\perp, i \omega)
\ee
which measures the winding of its phase. Here $G_R$ should be considered as a matrix in the spinor space and $C$ is any closed loop in the $(k_\perp, i \omega)$ space around the origin. In our case, for $m=0$ since $\det G_R = 1$ (see discussion around~\eqref{eti}), \ie\  any pole is always canceled by a zero, any Fermi surface has a zero winding. For generic $m \neq 0$, then Fermi surfaces
here generically have $n = \pm 1$.

\item Except for the single primary Fermi surface for the alternative quantization, as indicated in the last plot of fig.~\ref{fig:qk}, the Fermi surfaces for the standard and alternative quantizations are paired with the standard quantization have a larger $k_F$.    In the limit $mR \to \ha$, the paired surfaces for two quantizations now have the same $k_F$.

\item From fig.~\ref{fig:vF}, for a given mass and $\nu_{k_F} > \ha$, $v_F$ decreases with $\nu_{k_F}$. In particular as $\nu_{k_F} \to \ha$, $v_F$ approaches zero. From the second line of~\eqref{residue} the residue also vanishes in this limit given that $h_1$ is regular there (see fig.~\ref{fig:h12}). Also note that as $\nu_{k_F} \to \infty$, $v_F \to 1$.
\een

\subsection{$mR \to -\ha$ limit and free fermions}

We will now consider the $mR \to -\ha$ limit in some detail (this corresponds to the alternative quantization of $mR = \ha$), as in this limit the dimension of the operator approaches that of a free fermion, \ie\  in $d$-dimension, $\De \to {d-1 \ov 2}$. Note at exactly $mR =-\ha$, the bulk wave function becomes non-normalizable and there does not exist a boundary operator corresponding to it.\footnote{In other words at $mR =\ha$ there is
only one quantization giving rise to a boundary operator of dimension $\De = {d+1 \ov 2}$.} It is natural to ask whether in the limit $m R \to -\ha$, the behavior near the Fermi surface approaches
that of a Landau Fermi liquid, as recently argued in~\cite{Cubrovic:2009ye}.

We first note the following features as $mR \to -\ha$:

\ben

\item  In this limit one finds numerically that for the primary Fermi surface, the Fermi momentum approaches the upper limit of~\eqref{kfrang}, \ie\
 \be \label{valkF}
 k_F = \mu_q \ .
 \ee

\item With $k_F$ constrained as in~\eqref{valkF}, by varying $q$,
$\nu_{k_F}$ can take any values greater than $\ha {1 \ov \sqrt{d (d-1)}}$.

\item From~\eqref{varPhP}, one finds that various coefficients in~\eqref{spinorG} behave as
\be
h_1, h_2 \propto \le(mR+\ha \ri) \to 0,  \quad v_F \to 1
\ee
This is due to the fact that at $mR = -\ha$, the bound state wave function
becomes non-normalizable with a logarithmic divergence. In the limit $mR \to -\ha$
both $J^1$ and $J^t$ defined in~\eqref{defJ1} and~\eqref{defJt} are proportional to ${1 \ov mR + \ha}$.

\een

With $h_2 \to 0$ and $v_F \to 1$, the Green function~\eqref{spinorG} approaches that of a free relativistic fermion, despite the fact that the non-analytic part can still have a nontrivial exponent $\nu_{k_F}$. Note that equation~\eqref{valkF} has a simple interpretation: it is simply the Fermi momentum for a free relativistic fermion with Fermi energy $\mu_q$!

With $h_1 \to 0$, the whole Green function vanishes, suggesting that at the same time
the fermion disappears in the limit. Note that nowhere along the limit does a Landau Fermi liquid emerge. This picture is consistent with general expectations: in the $mR \to -\ha$ limit the mode becomes a singleton mode (free fermion) living at the boundary and decoupling from everything else. There are no bulk degrees of freedom associated with it anymore. The fact that we do not see a Landau Fermi liquid emerging in the limit is also consistent with our current understanding of holography; we do not expect a weakly interacting boundary theory to have a bulk description in terms of low energy gravity.

Let us also mention that, at any given mass $mR$ close to $-\ha$, $h_1$ and $h_2$ are small but nonzero. Thus except at parametrically small frequencies (\ie\  very close to the Fermi surface), the linear analytic term in~\eqref{spinorG} will dominate over the non-analytic term. As a result the nontrivial exponent and the fact that the quasi-particle has a finite width
will not be easily visible. Turning on a temperature will generate a new width for the quasi-particle and could dominate over the zero-temperature width except at very low temperatures.

In~\cite{Cubrovic:2009ye} indications were found that there exists a Fermi surface which behaves like a Landau Fermi liquid for $mR$ close to $-\ha$ at a Fermi energy {\it smaller} than $\om =0$. This result is surprising, from the following points of view: (i) the Fermi energy is different from (in fact smaller than) the value of the effective chemical potential; (ii) it implies the existence of some kind of fermionic hair outside the black hole at $\om \neq 0$; (iii) from the general philosophy of holography mentioned earlier
it is surprising to see a gravity description of a weakly coupled theory. It would be nice to have a better understanding of it. (Note that the small $\om$-analysis performed in this paper does not apply to any possible Fermi surfaces at a Fermi energy different from $\om =0$.)

Finally, we mention in passing that there is another limit in which free fermions emerge
from the WKB analysis. In the limit $q \to \infty$, with $mR/q$ fixed, again one finds that $h_2 \to 0$. In this case, depending on the value $mR/q$, $v_F$ can take a range of values, see fig.~\ref{fig:vfwkb}.

\subsection{Double trace deformation and its effect on Fermi surfaces}

\begin{figure}[h!]
 \begin{center}
\includegraphics[scale=0.38]{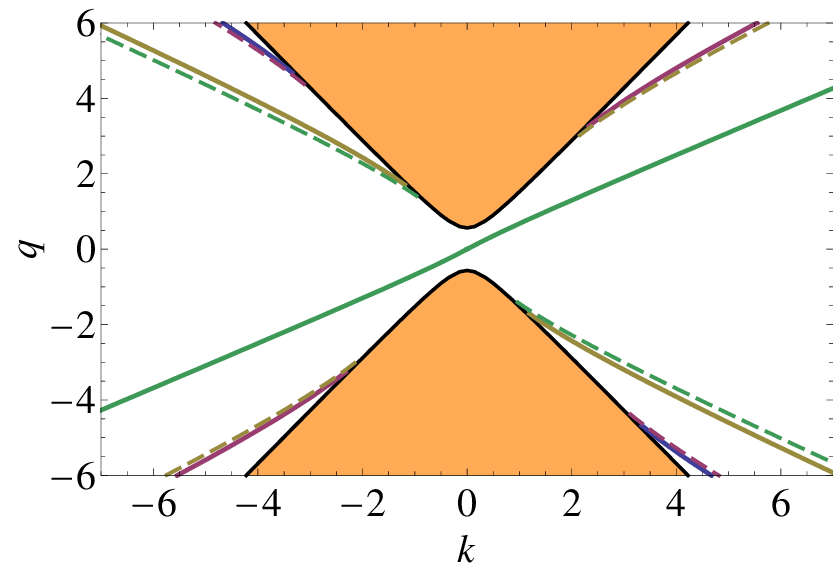}
\includegraphics[scale=0.38]{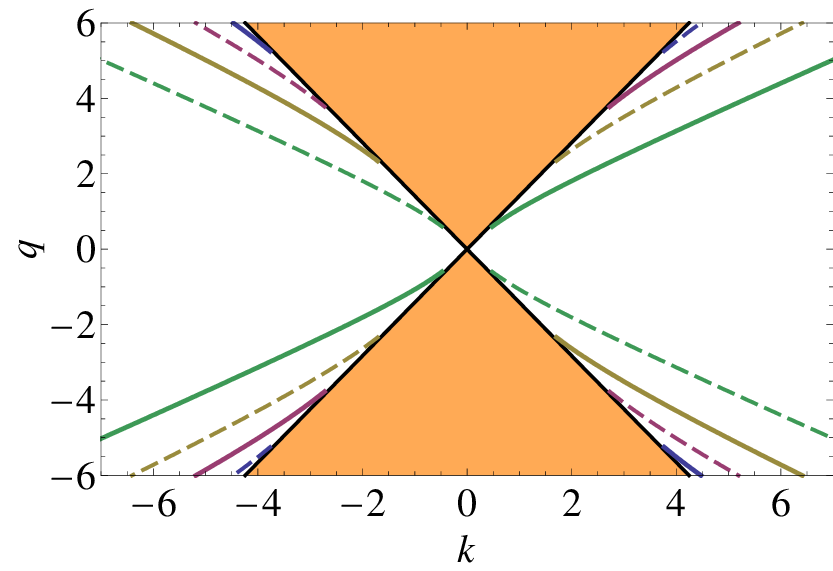}
\includegraphics[scale=0.38]{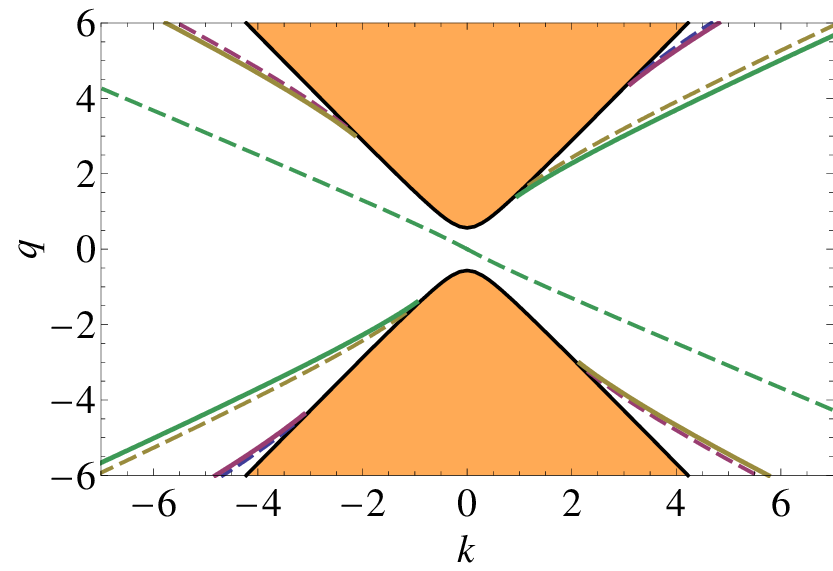}
\caption{\label{fig:qk} The values of $k_F$ as a function of $q$ for the Green function $G_{2}$ are shown by solid lines for $m = -0.4, \, 0, \, 0.4$. In this plot and ones below we use units where $R=1$, $r_* =1$, $g_F = 1$ and $d=3$.  The oscillatory region, where $ \nu_k =
{1 \ov \sqrt{6}} \sqrt{k^2 + m^2 -{q^2 \ov 2}}$ is imaginary, is shaded.
From~\eqref{ne3} in Appendix~\ref{app:a3}, $G_1 (k) = G_2 (-k)$, so $k_F$ for $G_1$ can be read from these plots by reflection through the vertical $k=0$ axis. The $m=-0.4$ plot corresponds to alternative quantization for $m =0.4$ following from equation~\eqref{altenQ}. For convenience we have
included in each plot the values of $k_F$ for the alternative quantization using
the dotted lines. Thus the first ($m=-0.4$) and the third plot ($m=-0.4$) in fact contain
identical information; they are related by taking $k \to -k$ and exchanging dotted and solid lines. Also as discussed after~\eqref{eep}, for $m=0$ the alternative quantization is equivalent to the original one. This is reflected in the middle plot
in the fact that the dotted lines and solid lines are completely symmetric. All plots are symmetric with respect to $q , k \to -q , -k$ as a result of equation~\eqref{jke}.}

\end{center}
\end{figure}

As discussed in Appendix~\ref{sec:IIa}, for $mR \in [0, \ha)$ we can turn on a double trace deformation $\sO^\da \sO$ in the unstable CFT (from the alternative quantization) to flow to the stable CFT. We note that the IR CFT of both CFTs appear to be the same. It is interesting to examine what happens to Fermi surfaces of $\sO$ under this flow. As an example, let us consider $m=0.4$. By examining the third plot of fig.~\ref{fig:qk} we observe that in flowing to the stable CFT, the primary Fermi surface (which has the biggest radius) of the unstable one disappears. As a result, for generic $q$, the number of Fermi surfaces in the stable CFT is one smaller than that in the unstable one.
For example, at $q=2$, one starts in the unstable theory with two Fermi surfaces with radii
given by $k_1 \gg k_2$. In the stable CFT one finds a single Fermi surface at a radius $k_3$ which is greater than but comparable to $k_2$, and much smaller than $k_1$.
It would be desirable to do a systematic study of more examples.


\bwt

\ewt

\begin{figure}[h!]
 \begin{center}
\hskip 1cm
\includegraphics[scale=0.6]{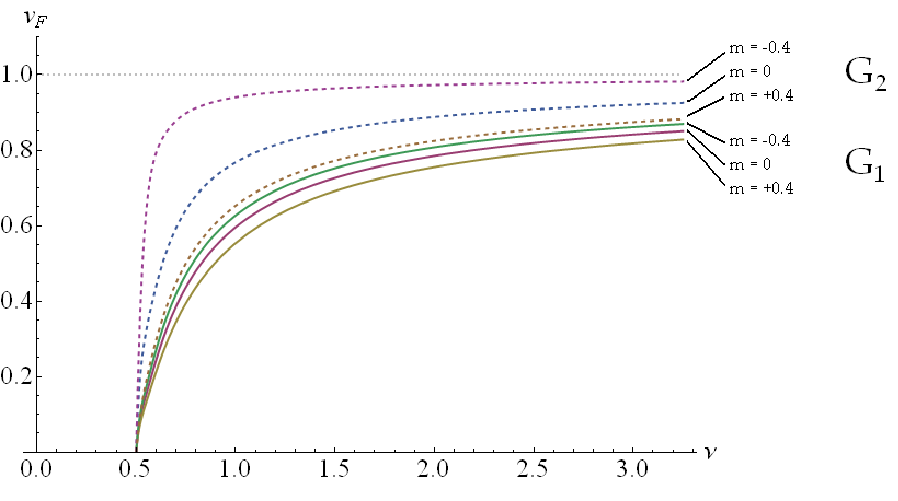} \caption{\label{fig:vF}
The Fermi velocity of the primary Fermi surface
of various components
as a function of $\nu_{k_F} > \ha$. Dotted lines are for $G_2$.
Various values of $m$ are indicated.
}
\end{center}
\end{figure}

\bwt

\ewt

\begin{figure}[h!]
 \begin{center}
\includegraphics[scale=0.75]{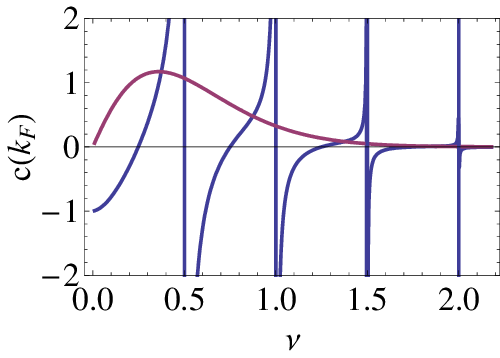} \caption{\label{fig:ck}
Real and imaginary part of $c(k_F)$ as a function of $\nu_{k_F}$ for $G_1$ at $m=0$. The plot for $G_2$ is very similar. Note that due to the Gamma function prefactor in~\eqref{rprp} the
real part diverges at half integers.
}
\end{center}
\end{figure}

\section{Discussion} \label{sec:dis}

In this paper we studied the low frequency expansion of retarded two-point functions of generic charged scalar and spinor operators in a CFT$_d$ at finite charge density using its gravity dual, following the earlier numerical study of~\cite{Liu:2009dm}. We showed that
the spectral functions exhibit various emergent critical behavior controlled by an infrared CFT described by the AdS$_2$ region of the black hole geometry.

Despite its classical nature, the bulk calculation we performed
turned out to have an intimate knowledge of the quantum statistics of the excitations
in the boundary theory.
In particular, the consistency with the boundary theory statistics dictates that a charged
scalar field in the bulk could have various instabilities including superradiance, which
are absent for a spinor field. We regard this as a nice indication of the robustness of the AdS/CFT correspondence. We also found a potentially new type of scalar instability which appears to be distinct from the standard tachyon instabilities (including those induced by an electric field).

Our results suggest a nice description for the low energy effective theory near a Fermi surface of a non-Fermi liquid. We find at each point on the Fermi surface, there lives a
CFT~\footnote{As mentioned earlier, the precise nature of the IR CFT is not yet understood. It might be a conformal quantum mechanics or a chiral sector of a $(1+1)$-dimensional CFT.}\footnote{In fact, the geometry we are studying displays such a CFT even for values of $k$
away from the Fermi surface.}, parameterized by the angle on the Fermi surface. This is reminiscent of the Fermi liquid picture, except that one replaces the free fermion CFT by a nontrivial one.
We found a direct relation between the dimension of an operator in the IR CFT and the scaling exponent of its spectral function; a relevant operator gives rise to unstable quasi-particle excitations at the Fermi surface with a zero quasi-particle weight, while an irrelevant operator gives rise to stable quasi-particles at the Fermi surface with a nonzero quasi-particle weight. We expect this description to be rather general, not restricted to
theories with a gravity dual, since it involves nothing more than general concepts of a fixed point. For example, the fact that for marginal operator our expression~\eqref{eprM} coincides with that of the ``marginal Fermi liquid'' description~\cite{varma} of the optimally doped cuprates may not be an accident and
may suggest that the ``electron'' operator is marginal at the possible quantum critical point describing a optimally doped cuprates.

Note that while in this paper we restricted to a charged black hole in AdS (which corresponds to CFT at a finite charged density), our results should apply to any
extremal solution with an AdS$_2$ region, e.g. that dual to a non-conformal theory at a finite density. One can also put the boundary theory at a finite chemical potential for some components of angular momenta, whose gravity dual is then given by an extremal Kerr-AdS black hole, which has an AdS$_2$ region.

It is worth comparing the form of our Green's functions with
the well-understood example of a non-Fermi liquid, namely Luttinger liquids in $1+1$ dimensions,  This is in some sense a realization of the picture described above, in that
to construct a Luttinger liquid
one replaces the free fermion CFT of each Fermi point
with a free boson of some radius other than
the free-fermion radius.
It differs from our case in two important ways:
First, we don't know how to obtain the behavior found here as a deformation
of the Landau theory.
Second, the Green function we find has a non-analyticity at $\omega=0$ for
arbitrary $k$ (unlike $G_{{\rm luttinger}} \sim { 1\over ( k_\perp - \omega)^\nu} $);
however, only at $k = k_F$ does this non-analyticity represent a peak in the spectral density.

We now discuss two caveats of our study. The first regards the stability of this extremal black hole geometry. While the black hole
is by itself thermodynamically and perturbatively stable,
as discussed earlier holographic superconductor instabilities\footnote{As mentioned in footnote~\ref{ft:neur}, a neutral scalar with a sufficient negative mass square can also condense, whose boundary theory interpretation is not yet known.} can occur if the gravity theory in which it is embedded contains charged scalars of sufficiently large charge or sufficiently small mass~\cite{Gubser:2008px,Hartnoll:2008vx,Hartnoll:2008kx,Denef:2009tp}.
This might not be of a concern since various physical systems to which we might try to
apply the mechanism described here including the normal state of high $T_C$ cuprates also exhibit a superconducting instability. Nevertheless, the criteria for a string vacuum which exhibits the fermi surfaces described here but {\it not} the superconducting
instability are reminiscent of those required of a string vacuum
which describes our universe: one doesn't want light scalar fields\footnote{We thank
Eva Silverstein for this analogy.}. In the latter context, a large machinery~\cite{Douglas:2006es} has been developed to meet the stated goal, and one can imagine that similar techniques would be useful
here. It would also be very interesting to understand how the condensate affects
the Fermi surfaces studied here.

As discussed around equation~\eqref{zeroent}, the black hole solution has a finite entropy at zero temperature. Since the semiclassical gravity
expression for the entropy is valid in the large $N$ limit, this ``ground state degeneracy'' may be a consequence of the $N \to \infty$ limit.  Given that the solution
is not supersymmetric, away from $N = \infty$ (i.e. beyond the gravity approximation)
likely these states are energetically closely-spaced,
rather than exactly degenerate. It may be useful to compare this situation to
that of systems with frustration. While in our geometry this non-vanishing ground-state
entropy comes together with the existence of the AdS$_2$ region which gives rise to
the IR CFT description, they in principle reflect different aspects of a
system and may not correlate with each other. For example, suppose the CFT$_1$ dual to AdS$_2$ can be considered as the right moving sector (with $T_R =0$) of a $1+1$-dimensional CFT. Then the nonzero ground state entropy should come from the left-moving sector of this $1+1$-dimensional CFT with a nonzero left temperature. One can certainly imagine a situation where the left-moving sector is absent, for which case one will then have an IR CFT without a zero-temperature entropy. We should caution that it might be hard for such a situation to arise as the near horizon limit of a classical gravity solution.

Our results can be generalized in a variety of ways. The most immediate is
finite temperature, which should shed further light on the structure of
CFT$_1$. It will be interesting to follow the instability for scalars
to the critical temperature~\cite{Denef:2009tp}.
It is also instructive to examine the two-point functions of charge current and density fluctuations. They correspond to fluctuations of the bulk gauge field $A_M$ in the transverse and longitudinal channel respectively. Here we expect even at zero temperature
there exist hydrodynamic modes like diffusion and sound modes as this is
a non-neutral plasma and dissipates even at zero temperature (\eg it has a finite entropy).  We expect our master formula~\eqref{finG1} still applies after
having diagonalized the perturbations. In particular the diffusion and sound modes will depend on the small $k$ expansion of $a_+^{(0)}$, \ie\  it should have a zero at $k=0$.

The matching behavior between the inner region and outer region is very reminiscent of
that for the D3-D7 system at small densities in~\cite{Faulkner:2008hm}. There the geometry
is more complicated and does not directly give a hint what would be the IR theory.
Nevertheless it appears likely the IR theory there is a non-relativistic CFT at finite density.

\vspace{0.2in}   \centerline{\bf{Acknowledgements}} \vspace{0.2in} We thank A.~Adams,
S.~Hartnoll, M.~Hermele, G.~Horowitz, N.~Iqbal, S.~Kachru, A.~Karch, Z.~Komargodski, M.~Lawler, P.~Lee, S-S. Lee, J.~Maldacena, J.~Polchinski, K.~Rajagopal, S.~Sachdev, N.~Seiberg, E.~Silverstein, G.~Shiu, S.~Sondhi, S.~Trivedi and in particular T.~Senthil for valuable discussions and encouragement. HL would also like to thank the physics department
at UCSB for hospitality during the last stage of this work. Work supported in part by funds provided by the U.S. Department of Energy
(D.O.E.) under cooperative research agreement DE-FG0205ER41360 and the OJI program.
The research was supported in part by the National Science Foundation under Grant No. NSF PHY05-51164.

\appendix

\section{Spinor calculation} \label{app:fer}

\subsection{Remarks on dictionary} \label{sec:IIa}

A bulk Dirac spinor field $\psi$ with charge $q$  is mapped to a fermionic operator $\sO$ in CFT of the same charge. $\sO$ is a Dirac spinor for $d$ odd, and  a chiral spinor for $d$ even~\cite{Iqbal:2009fd}. In both cases the dimension of the boundary spinor $\sO$ is half of that of $\psi$. Since $\psi$ has $2^{[{d+1 \ov 2}]}$ complex components, where $[x]$ denotes the integer part of $x$,
the boundary retarded Green function $G_R$ for $\sO$ is a $2^{[{d-1 \ov 2}]} \times 2^{[{d-1 \ov 2}]}$ matrix. The conformal dimension $\De$ of $\sO$  is given in terms of the mass $m$ of $\psi$ by\footnote{Without loss of generality in this paper we will take $m \geq 0$. For negative $m$ the discussion is exactly parallel, with $m$ replaced by $|m|$. For odd $d$
the Dirac equation for $-m$ is completely equivalent to $m$ as one can change the sign of mass by taking $\psi \to \Ga \psi$ where $\Ga$ is the $d+1$-dimensional chirality matrix.
For $d$ even, different signs of $m$ corresponds to different chirality of $\sO$.}
 \be  \label{dimS}
 \De = {d \ov 2} \pm m R \
 \ee
where $R$ is the AdS curvature radius. In~\eqref{dimS}, one should use the $+$ sign for $mR \geq  \ha$. For $mR \in [0 ,\ha)$, there are two ways to quantize $\psi$ by imposing different boundary conditions at the boundary, which corresponds to two different
CFTs. We will call the CFT in which $\sO$ has dimension $\De = {d \ov 2} + mR$ the ``stable'' CFT and the one with dimension $\tilde \De = {d \ov 2} - mR \in ({d-1 \ov 2}, {d \ov 2}]$ the ``unstable'' CFT.
In the ``unstable'' CFT, the double trace operator $\sO^\da \sO$
produces a relevant deformation under which the theory flows to the ``stable'' CFT~\cite{Witten:2001ua}.

The retarded Green function of $\sO$ at finite charge density
can be extracted by solving the Dirac equation for $\psi$ in the charged AdS black hole geometry.

\subsection{Dirac equation }

We consider a spinor field in the black hole geometry~\eqref{bhmetric1} with a quadratic action
 \be \label{actDi}
 S =  \int d^{d+1} x \sqrt{-g} \, i (\bar \psi \Ga^M \sD_M \psi  - m \bpsi \psi)
 \ee
where $\bpsi = \psi^\da \Ga^\ut$ and
 \be
 \sD_M  = \p_M + {1 \ov 4} \om_{ab M} \Ga^{ab} - i q A_M \
 \ee
with $\om_{ab M}$ the spin connection. Our notations are as follows. We will use $M$ and $a,b$ to denote abstract bulk spacetime and tangent space indices respectively, and $\mu, \nu \cdots$ to denote indices along
the boundary directions, \ie\  $M = (r, \mu)$. Underlined indices on Gamma matrices always refer to tangent space ones.

Writing
\be \label{tranS}
\psi =  (- g g^{rr})^{-{1 \ov 4}} e^{-i \om t + i k_i x^i}  \Psi
\ee
the corresponding Dirac equation for $\psi$ can be written as
  \be \label{Deq}
\sqrt{ g_{ii} \ov  g_{rr}} \le(\Ga^\ur \p_r  - m \sqrt{g_{rr}}\ri) \Psi
+ i K_\mu \Ga^{\underline \mu} \Psi = 0,
 \ee
with
\be
K_\mu (r) = \(-u(r),  k_i \)
\ee and $u (r)$ is given by
 \be \label{udef1}
  u= \sqrt{g_{ii} \ov -g_{tt}} \le(\om + \mu_q \le(1-{r_0^{d-2} \ov r^{d-2}} \ri) \ri) \ .
\ee
As in the case of a charged boson, equation~\eqref{Deq} depends on $q$ and $\mu$ only through the combination
 \be \label{mme1}
 \mu_q \equiv \mu q
 \ee
which is the effective chemical potential for a field of charge $q$. Similarly, $\om$ should be identified with frequency measured away from the
the effective chemical potential~\eqref{mme1}. Due to rotational symmetry in the spatial directions, we do not lose generality by setting
\be
k_1 = k, \qquad k_i = 0, \quad i \neq 1 \ .
\ee
Notice that equation~\eqref{Deq} then only depends on three Gamma matrices $\Ga^\ur, \Ga^\ut, \Ga^{\underline 1}$. As a result\footnote{For general $k_i$ the projector can be written as
 \be\label{proje1}
\Pi^{\hat k}_\pm \equiv \half \left(1 \pm \Gamma^{ \ur } \Gamma^{ \ut }  \hat k_i \Gamma^{ \ui } \right) ,
\ee
where $\hat k_i$ is the unit vector $ \hat k \equiv \vec k/ |\vec k|$.
}
projectors
 \be\label{projectors}
 \Pi_\al \equiv \half \left(1 - (-1)^\al \Gamma^{\ur } \Gamma^{\ut }  \Gamma^{\underline 1 } \right), \; \al =1,2 , \quad \Pi_1 + \Pi_2 = 1
\ee
commute with the Dirac operator of~\eqref{Deq} and
\be
\Phi_\al = \Pi_\al \Psi, \quad \al = 1,2
\ee
decouple from each other. It is then convenient to write $\Phi = \left(\begin{matrix} \Phi_1 \cr \Phi_2 \end{matrix}\right)$ and choose the following basis of Gamma matrices
 \bea
 && \Ga^\ur = \left( \begin{array}{cc}
-\sigma^3 \bone & 0  \\
0 & -\sigma^3 \bone
\end{array} \right), \;\;
 \Ga^\ut = \left( \begin{array}{cc}
 i \sigma^1 \bone & 0  \\
0 & i \sigma^1 \bone
\end{array} \right), \cr
&&
\Ga^{\underline 1} = \left( \begin{array}{cc}
-\sigma^2 \bone & 0  \\
0 & \sigma^2 \bone
\end{array} \right) ,
\qquad \cdots
\label{realbasis}
 \eea
under which the Dirac equation~\eqref{Deq} becomes
\be\label{spinorequation}
\left( \partial_r + m \sqrt{g_{rr}} \sigma^3 \right) \Phi_\al =\sqrt{g_{rr} \ov g_{ii}} \left(i \sigma^2 u  + (-1)^\al k  \sigma^1
\right) \Phi_\al \  .
\ee
In equation \eqref{realbasis}, $\bone$ is an identity matrix of size
$2^{d-3 \ov 2}$ for $d$ odd
(or size $2^{d-4 \ov 2}$ for $d$ even);
since the resulting Green's functions will
also be proportional to such an identity matrix, we will
suppress them below. Note that~\eqref{realbasis} is chosen so that equation~\eqref{spinorequation} is {\it real} for real $\om, k$.

Near the boundary,~\eqref{spinorequation} has two linearly independent solutions given by (with $1/r$ subleading terms for each solution suppressed)
\be\label{outerasymptoticspinor}
\yandz_\alpha \buildrel{r \to \infty}\over {\approx}
a_\al r^{mR} \left( \begin{matrix} 0 \cr  1 \end{matrix}\right) 
+ b_\al r^{-mR} \left( \begin{matrix}  1  \cr 0 \end{matrix}\right)
 \qquad
\al = 1,2  \ .
\ee
To compute
the retarded functions, one should impose the in-falling boundary condition
for $\Phi$ at the horizon. Then the boundary spinor Green functions have two sets of eigenvalues given by\footnote{That is, when diagonalized, the boundary retarded functions have the form
 \be
G_R (\om,k) =  \left( \begin{array}{cc}
G_1 (\om,k)  \bone & 0  \\
0  & G_2 (\om,k) \bone
\end{array} \right)  \ .
\ee
 }
\be \label{bdG}
G_{\alpha} (\om,k)=  {b_\alpha\over a_\alpha} \ , \quad \al =1,2 \ .
\ee
To see~\eqref{bdG}, consider spinors $\phi_\pm = \ha (1 \pm \Ga^\ur) \Phi$ with definite eigenvalues of $\Ga^\ur$. Then
 \be
  {\rm with} \quad \Phi_\al \equiv \left(\begin{matrix} y_\al \cr z_\al \end{matrix}\right),
  \qquad \phi_+ = \left(\begin{matrix} z_1 \cr z_2 \end{matrix}\right), \quad
   \phi_- = \left(\begin{matrix} y_1 \cr y_2 \end{matrix}\right)
 \ee
where we have suppressed zero entries in $\phi_\pm$. Equation~\eqref{bdG} then follows using the prescription of~\cite{Iqbal:2009fd} (see \eg\ sec.~IIIB).\footnote{A note on notation: compared
to the notation of \cite{Liu:2009dm},
$$
\yandz_1 \equiv  \left( \begin{matrix}  i y_- \cr z_+ \end{matrix}\right), ~~
\yandz_2 \equiv  \left( \begin{matrix} - i z_- \cr  y_+ \end{matrix}\right) ~.
$$
}

The dictionary \eqref{bdG} is for the conventional quantization which applies to any $m \geq 0$.
For $mR \in [0, \ha)$, there is also an alternative quantization, as discussed at the beginning of this section. Similar argument then leads to
 \be \label{bdG1}
\tilde G_{\alpha} =  -{a_\alpha\over b_\alpha} = - {1 \ov G_\al}.
\ee

\subsection{Some properties of the spinor correlators} \label{app:a3}

In~\eqref{spinorequation} the equation for $\Phi_2$ is related to that of $\Phi_1$ by $k \to -k$, so we immediately conclude that
  \be \label{ne3}
 G_{2} (\om, k) = G_{1} (\om,-k) \ .
  \ee
As a result the trace and determinant of $G_R$ are invariant under $k \to -k$ as should be
the case. Given~\eqref{ne3}, from now on we will focus solely on $G_1 (\om, k)$. For notational simplicity, we will also drop the subscript $\al =1$ below. Unless written explicitly all relevant quantities should be interpreted as having a subscript $\al =1$.

More properties of $G$ can be derived from~\eqref{spinorequation}. For this purpose it is convenient to write~\eqref{bdG} as\footnote{In~\eqref{mli} one should extract the finite terms in the limit.}
 \be \label{mli}
G = \lim_{\ep \to 0}  \ep^{-2mR}  \xi |_{r = {1 \ov \ep}},  \quad {\rm with} \quad
\xi \equiv { y \over z} \ .
\ee
From~\eqref{spinorequation}, as in \cite{Iqbal:2008by} one can then derive a flow equation for $\xi$,
\be \label{sor1}
 \sqrt{g_{ii} \ov g_{rr}} \p_r \xi = -2m \sqrt{g_{ii}} \xi + (u  - k )
+ (u + k) \xi^2
\ee
with in-falling boundary condition at the horizon given by (for $\omega\neq0$)
\be \label{tye}
\xi|_{r=r_0} =i \ .
\ee

Properties of $G$ can now be read from those of~\eqref{sor1}.
By taking $q \to -q, \om \to -\om, k \to -k$ and $\xi \to - \xi$ we find that the equation for $\xi$ goes back to itself, implying
\be \label{jke}
G (\om, k; q) = - G^* (-\om, -k; -q) \
 \ee
where the complex conjugation is due to that with $\om \to - \om$ the infalling horizon boundary condition turns into the outgoing one, which can then be changed back by a complex conjugation.

By dividing both sides of equation~\eqref{sor1} by $\xi^2$, we obtain an identical equation for $-{1 \ov \xi}$ if we also take $m \to -m, k \to -k$. This implies
that
 \be \label{ne4}
G (\om, k;-m) = - {1 \ov G (\om, -k; m)} \ .
\ee
Given equation~\eqref{bdG1} we conclude that for alternative quantization $\tilde G$
can be written as
 \be \label{altenQ}
 \tilde G(\om, k; m) = G (\om, -k; -m) \ .
 \ee
That is, alternative quantization can be included by extending the mass range for $G(\om,k;m)$ from $m \geq 0$ to $m R> -\ha$. Below and in the main text when we speak of negative mass it should be understood that it refers to the alternative quantization.

For $m=0$, from~\eqref{ne4} and~\eqref{ne3}, we find that
  \be \label{ne2}
G_{2} (\om, k) = - {1 \ov G_{1} (\om, k)}, \qquad m=0 \
\ee
which implies that
 \be \label{eti}
 \det G_R (\om, k)= 1, \qquad m =0
 \ee
Note that since a basis change and a Lorentz rotation do not change the determinant
of $G_R$, equation~\eqref{eti} applies to any basis of Gamma matrices and any momentum.
Combining~\eqref{ne3} and~\eqref{ne2} we also conclude that at $k=0$,
 \be \label{eep}
 G_{1} (\om,k=0) = G_{2} (\om,k=0) = i, \quad m=0 \ .
 \ee
Also note that equation~\eqref{ne2} implies that for $m=0$ the alternative quantization is equivalent to original one~\cite{Iqbal:2009fd}.

\subsection{Small-frequency expansion}

In this subsection, we present the low-frequency expansion
analysis of section \ref{sec:corr} adapted to the case of a spinor field. Now the equation is given by~\eqref{spinorequation} with $\al =1$ which we copy here for convenience
 \be\label{spinEq}
\left( \partial_r + m \sqrt{g_{rr}} \sigma^3 \right) \Phi =\sqrt{g_{rr} \ov g_{ii}} \left(i \sigma^2 u  - k  \sigma^1
\right) \Phi \ .
\ee
We will again divide the $r$-axis into two regions as~\eqref{in1}--\eqref{out1} and consider the low frequency limit~\eqref{limR2}.
The story is very much parallel, so we will be brief.

In the inner region, to leading order in $\om$-expansion equation~\eqref{spinEq}
reduces to equation~\eqref{spinads2} of a spinor field in AdS$_2$ with $\tilde m =
-(-1)^\alpha{k R \ov r_*}$. All the discussion in sec.~\ref{app:spinC} can now be carried over
with the replacement $\nu \to \nu_k$
 \be \label{spinornu}
\nu_k \equiv \sqrt{m_k^2 R_2^2 - e_d^2 q^2 - i \epsilon} , \quad m_k^2 \equiv m^2 +
{k^2 R^2 \ov r_*^2} \ .
 \ee
For example, near the boundary of the inner region, \ie\  ${\om R_2^2 \ov r-r_*} \to 0$, the leading order inner solution can be expanded as
  \be \label{exwp}
 \Phi_I^{(0)} (\om, \vk; \zeta) =   v_- \le({R_2^2 \ov r-r_*} \ri)^{ - \nu_k}  \;\; + \;\; \sG_k (\om) v_+ \le({R_2^2 \ov r-r_*} \ri)^{\nu_k}
 \ee
where $v_\pm$ and $\sG_k$ can be obtained respectively from~\eqref{eignV} and~\eqref{AeFG}. More explicitly,
\be \label{veorP}
v_\pm =   \left( \begin{matrix}
m R_2 \pm \nu_k
\cr
{k R \ov r_*} R_2 + q e_d
\end{matrix} \right) \ ,
\ee
and
  \bea \label{effG}
 \sG_k (\om) &= & e^{ - i \pi \nu_k}\frac{\Gamma (-2 \nu_k ) \, \Gamma \left(1+\nu_k -i q e_d \right)}{\Gamma (2 \nu_k )\,   \Gamma \left(1-\nu_k -i q e_d\right)}
  \cr &\times & \; \frac{\le(m - {i k R \ov r_*} \ri)
 R_2 - i q e_d - \nu_k}{ \le(m - {i k R \ov r_*} \ri) R_2 - i q e_d +  \nu_k} \; \om^{2 \nu_k} \ .
 \eea
As discussed around~\eqref{eignV}, $\sG_k$ depends on the normalizations of $v_\pm$ in~\eqref{veorP}. But it can be checked explicitly that the final correlation function~\eqref{finG1} is independent of the normalizations.

In the outer region we can then choose the two linearly independent solutions
for the zero-th order equation (\ie\ ~\eqref{spinorequation} with $\om=0$)
by the boundary conditions
 \be \label{spinZ}
 \eta_\pm^{(0)} = v_\mp \le(r-r_* \ov R_2^2 \ri)^{\pm \nu_k} + \cdots, \qquad r - r_* \to 0 \ .
 \ee
The matching, the generalization to higher orders in $\om$, and the low frequency expansion of $G_R$ now work completely in parallel as those for scalar fields. More explicitly, perturbatively in $\om$ the full outer solution $\Phi_O$ can be written as
 \be
\Phi_O =  \eta_+ + \sG_k (\om)  \eta_-
\ee
with
 \be \label{spiEx}
 \eta_\pm = \eta^{(0)}_\pm + \om \eta^{(1)}_\pm + \om^2 \eta^{(2)}_\pm + \cdots \ .
 \ee
$\eta^{(n)}_\pm, n \geq 1$ are obtained from solving~\eqref{spinEq} perturbatively in the outer region and are {\it uniquely} specified by requiring that when expanded near $r =r_*$, they do not contain any terms proportional to the zeroth order solutions~\eqref{spinZ}. Now expanding
various $\eta^{(n)}_\pm, n \geq 0$ near $r \to \infty$ as in~\eqref{outerasymptoticspinor}
 \be \label{spinAs}
\eta^{(n)}_\pm \buildrel{r \to \infty}\over {\approx}
a_\pm^{(n)} r^{mR} \left( \begin{matrix} 0 \cr  1 \end{matrix}\right)
+ b_\pm^{(n)} r^{-mR} \left( \begin{matrix}  1  \cr 0 \end{matrix}\right)
\ee
then the retarded function $G$ is again given by the master formula~\eqref{finG1}.

\section{Bound states at $\omega =0$ and WKB} \label{app:schro}

\subsection{Scalar bound states}

Setting $\omega=0$ in (\ref{pp2}) defines the \textbf{outer}
region differential equation for the scalar field.
We are then interested in examining normalizable solutions
to this equation, since this tells us about
the spectrum of excitations
of the boundary theory. This is equivalent
to studying solutions to this differential equation
with boundary conditions (\ref{bdc2}) for $\eta^{(0)}_+$
and $a_+^{(0)} = 0$ in (\ref{verC}). The
alternative quantization window can be achieved
by studying
$b_+^{(0)} = 0$.
As we will see, bound states
will exist for a discrete set of momenta, so (at least for the fermion
problem) these excitations will define a Fermi surface.

We will study the spinor problem in the next subsection. For now
the scalar problem will suffice since this problem is probably
be more intuitive.

By scaling the wave function and redefining the radial coordinate,
$\phi(r) = Z \psi(s)$,
the wave equation (\ref{pp2}) can be put in the form,
\begin{equation}
\label{schrodb}
- \partial_s^2 \psi  + V(s) \psi = (- k^2) \psi
\end{equation}
where the ``tortoise'' coordinate and the rescaling are,
\begin{equation}
\frac{d s}{d r} = \sqrt{\frac{g_{rr}}{g_{ii}}} \,,\qquad
Z = \left(\frac{ g_{rr} g_{ii}}{-g}\right)^{1/4}
\end{equation}
In doing this we have defined a
\emph{unique} Schr\"{o}dinger potential for this problem:
\begin{equation}
\label{potb}
V =  \left(- u^2 +  m^2 g_{ii}\right) 
+  \left( ( \partial_s \ln Z)^2 - \partial_s^2 \ln Z \right)
\end{equation}
\bwt

\begin{figure}[h!]
\begin{center}
\psfig{figure=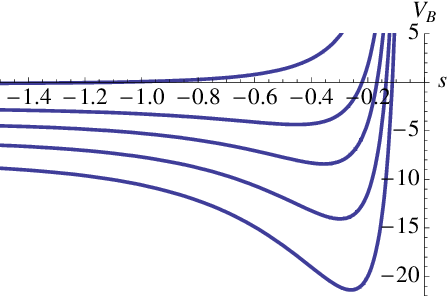,width=5.5cm}
\psfig{figure=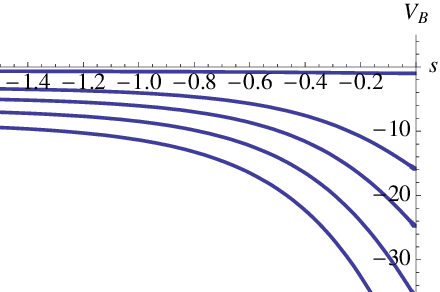,width=5.5cm}
\psfig{figure=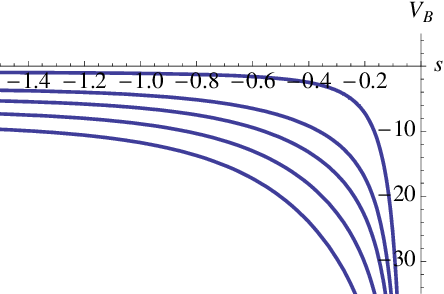,width=5.5cm}
\end{center}
\caption{ The  potential for the scalar field defined by (\ref{potb}) for
$m^2 = -3/2,-2,-9/4$. In terms of the tortoise coordinate $s$ the
horizon is located at $s=-\infty$ and the boundary
at $s=0$.
The oscillatory region is associated
with the continuum for $s \rightarrow - \infty$ and ``Fermi'' surfaces
are bound states in the potential well to the right of this
continuum.  Note the behavior of the potential close to the boundary is
$V(s) \sim ( 2 + m^2)/s^2$.
\label{fig:bosepots} }

\begin{center}
\psfig{figure=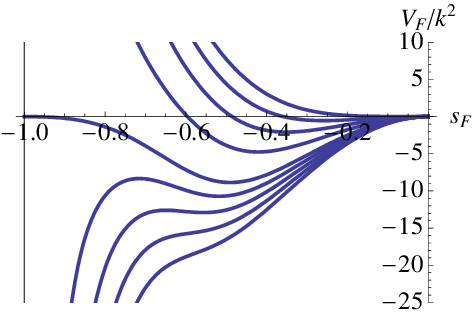,width=5.5cm}
\psfig{figure=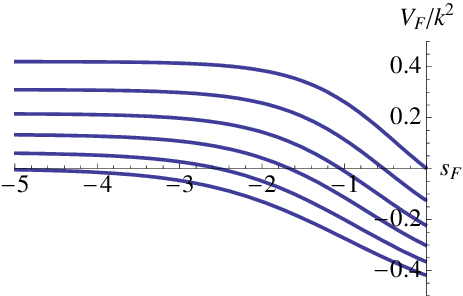,width=5.5cm}
\psfig{figure=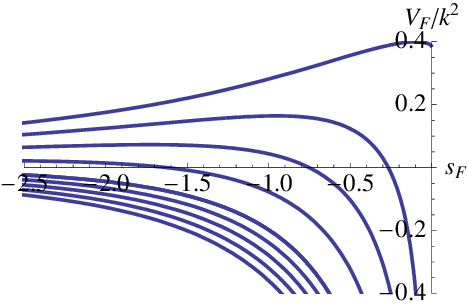,width=5.5cm}
\end{center}
\caption{A zero-energy Schr\"{o}dinger potential
which is equivalent to the spinor bound state
problem (\ref{spineq}). The three plots
are for $m=-1/4,0,+1/4$ respectively, for varying values
of the parameter $k/\mu_q$. For $m=-1/4$ we have
rescaled $s_F \rightarrow s_F/|s_*|, V_F \rightarrow V_F s_*^2$
so that for all values of $\mu_q/k$ the potential can be drawn
on the same interval $-1 < s_F <0$.
\label{fig:fermipots}}
\end{figure}

\ewt

The problem is to find bound states in this potential with negative
``energy'' $E = - k^2$. Pictures of this potential are shown in Fig.~\ref{fig:bosepots} for fixed
$m^2$ and various values of $q$. Examining
the $r\rightarrow r_*$ limit, which corresponds to $s\rightarrow -\infty$
we find that the potential goes to a constant
$$
V(s) \rightarrow \frac{r_*^2}{R^2 R_2^2} ( - (q e_d)^2 + m^2 R_2^2 + 1/4)
$$
Hence there exists a continuum near the horizon for
momentum satisfying
$ -k^2 >\frac{r_*^2}{R^2 R_2^2} ( - (q e_d)^2 + m^2 R_2^2 + 1/4) $.
This is the criterion for the oscillatory region. That is, we have
identified the oscillatory behavior as arising from
the existence of a continuum.

Bound states then exist if the potential well in Fig.~\ref{fig:bosepots}
close to $s=0$ is deep enough.
The ``sprouting'' of bound states out of the
oscillatory region in the $q-k$ plot of Fig.~\ref{fig:qk} has a nice interpretation
in terms of developing new bound states as the potential well varies.

\subsection{Spinor bound states}

The bound state problem is now the first order Dirac equation:
\begin{equation}
\label{spineq}
\sqrt{g_{ii} g^{rr}} \partial_r \Phi =
\begin{pmatrix} - (m/R) r  & -k + u \\ -k - u & (m/R) r \end{pmatrix} \Phi
\end{equation}
subject to normalizability conditions for $\eta^{(0)}_+$ in
(\ref{spinZ}) and $a_+^{(0)} = 0$ in (\ref{spinAs}).

It is harder to have an intuitive grasp over this equation,
as in the second order problem. We can of course square
this operator to obtain a Schr\"{o}dinger problem, but
there seems to be no way to define a \emph{unique}
potential, such as the one for the scalar. We proceed
with an arbitrary choice, in order to give a qualitative understanding
of the existence of bound states.

Taking $\Phi = (y,z)^{T}$ we can write (\ref{spineq}) as
\begin{eqnarray}
\sqrt{g_{ii} g^{rr}} Q^{-1}  \partial_r \left( Q y \right)  &=& ( u - k) z \\
\sqrt{g_{ii} g^{rr}} Q \partial_r \left( Q^{-1} z \right)  &=& -  (k+u) y
\end{eqnarray}
where $Q = \exp\left(m \int dr \sqrt{g^{ii} g_{rr}} mr/R \right)$. Then we can
write a second order differential equation for $\psi = z/Q$. Defining
a new tortoise coordinate:
\begin{equation}
\frac{d s_F}{dr} = \sqrt{ \frac{g_{rr}}{g_{ii}}} \frac{ 1 + u/k}{Q^2}
\end{equation}
we find a \emph{zero-energy} Schr\"{o}dinger equation,
\begin{equation}
- \partial_{s_F}^2 \psi + V_F \psi = 0, \qquad
(V_F)/k^2 = \frac{1 - u/k}{1+ u/k} Q^4
\end{equation}
Pictures of this potential are shown in Fig.~\ref{fig:fermipots} for 3 different fixed
values of $m$ and various values of $\mu_q/k$. Note in Fig.~\ref{fig:fermipots}
$k$ has been scaled out of the potential, so one should imagine
scaling the potential by dialing $k$ to find when a bound states energy
eigenvalue
crosses zero, this will define the fermi momentum $k_F$.
In particular for a given value of $\mu_q/k$ there are \emph{possibly}
an infinite set of bound states energies which cross zero as
$k$ is increased.

Examining the behavior close to the horizon
the tortoise coordinate behaves as $s_F \rightarrow -\infty$ for $m \geq 0$
and $s_F \rightarrow s_*$ for $m < 0$.
The potential is of the form
$c/s_F^2$ for $m > 0$ and $c/(s_F - s_*)^2$ for $m<0$ where
\begin{equation}
c = \frac{\left(k^2 R^2 R_2^2 / r_*^2 - (q e_d)^2 \right)}{ 4 m^2 R_2^2}
\end{equation}
so that the condition for being in the oscillatory region
is the usual condition for a singular $c/s^2$ potential
in quantum mechanics: $c < -1/4$.

\subsection{WKB analysis}

We can analyze both the scalar field and spinor
under various limits using WKB analysis.
We will focus here on the limit $q,k,m \rightarrow \infty$
 with ratios $k/q$ and $k/m$ fixed
and consider the scalar and spinor problems in parallel.
A useful reference for the application of WKB to the Dirac equation is 
\cite{lazur}.




\begin{figure}[h!]
\centerline{\hbox{\psfig{figure=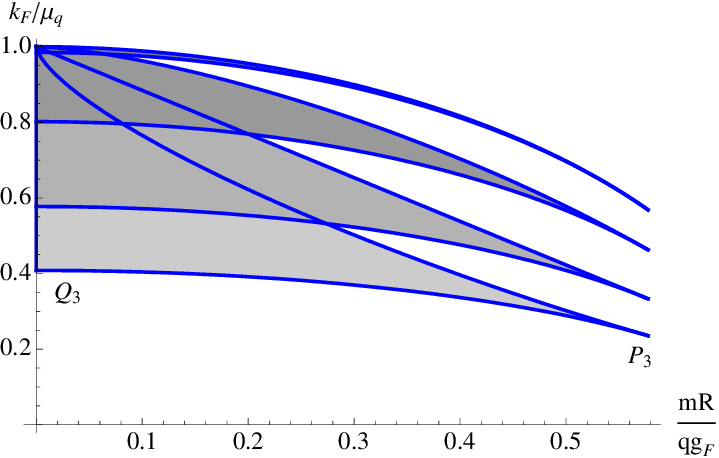,width=8cm}}}
\caption{ The parameter region where there exist
Fermi surfaces for large $k_F,m,q$ in the WKB approximation,
for different field theory space-time dimensions $d=3,4,6,50$. For
a given dimension the upper and lower boundary lines are defined
by the fermi surface moving into the oscillatory region (lower)
and the non-existence of classical orbits (upper.) See Fig.~\ref{fig:allowed3}
for an alternative way to state this in terms
of the radii of the two turning points. The extreme point
to the right of the allowed region is $P_d = \left(1/\sqrt{3},\, (d-2)/\sqrt{3 d(d-1)}\right)$
and the point at the lower left corner is $Q_d =\left(0,\, (d-2)/\sqrt{d(d-1)}\right)$.
\label{fig:allowedd} }
\end{figure}

As we will see, both problems are governed by the ``WKB momentum''
\begin{equation}
p^2 =  k^2 + (m^2/R^2) r^2 - u^2
\end{equation}
For the scalar field, this is simply the usual potential-minus-energy
term $V - (- k^2)$ from (\ref{potb}) where terms
which depend on $Z$ are small so should be dropped. For the
spinor it is the negative of the determinant of the matrix on the rhs
of (\ref{spineq}). The sign of $p^2$ will tell us if we
are in the classically allowed ($p^2<0$) or disallowed ($p^2>0$) region.
Note that $p^2 \sim \mathcal{O}(k^2)$ is large in the WKB approximation.
In the parameter space shown in Fig.~\ref{fig:allowedd},
there are two turning points $r_1,r_2$, with
a classically allowed region in between the two. The WKB approximation
to the wave function in the three regions is:
\begin{itemize}

\item $r<r_1$ and $r> r_2$
(we use a compact notation to write
both these regions together, with $1,2$ correlated
with $\pm$ in a self explanatory way)
\begin{eqnarray}
\label{wkbsoln}
\psi(r) &=&
\frac{C^B_{1,2}}{\sqrt{p}} \exp\left( \pm \int_{r_{1,2}}^r
dr' \sqrt{g^{ii} g_{rr}} p(r')
\right)  \\
\label{wkbsolnf}
\Phi(r) &=& \frac{C^F_{1,2}}{\sqrt{p (k + u)}} \begin{pmatrix} mr/R \mp p \\
k + u \end{pmatrix} \times  \\
\nonumber
& & \times \exp\left( \pm \int_{r_{1,2}}^r dr' \left( \sqrt{g^{ii} g_{rr}} p(r') + \chi(r') \right) \right)
\end{eqnarray}
where $\chi$ is an $\mathcal{O}(1)$ function
given by,
\begin{equation}
\chi(r')=
\frac{k+u(r')}{2 p(r') } \partial_{r'} \left( \frac{m r'/R}{k+u(r')} \right)
\end{equation}

\item $ r_1 < r < r_2$

\begin{eqnarray}
\label{allowb}
\hspace{-.7cm} \psi(r) \hspace{-.1cm} &=& \hspace{-.1cm} \frac{D_B}{\sqrt{\rho}} \mathrm{Re} \left\{
e^{i \theta_B(r_1,r) +
i \xi }  \right\}
\\
\hspace{-.7cm}\label{allowf}
\Phi(r) \hspace{-.1cm} &=& \hspace{-.2cm} \frac{D_F}{\sqrt{\rho ( k + u) }}
\mathrm{Re} \left\{
\begin{pmatrix} m r/R
- i \rho
\\ k + u
\end{pmatrix}
e^{i\theta_F(r_1,r) + i\xi}  \right\}
\end{eqnarray}
where $\rho^2 = - p^2$ and,
\begin{eqnarray}
\hspace{-.9cm} \theta_B(r_1,r) &=& \int_{r_1}^r dr'  \sqrt{g^{ii} g_{rr}} \rho(r') \\
\hspace{-.9cm} \theta_F(r_1,r) &=& \theta_B(r_1, r) \nonumber \\
&&- \int_{r_1}^r dr' \frac{ k + u(r')}{ 2 \rho(r')}
\partial_{r'} \left( \frac{ m r'}{ k + u(r')} \right)
\end{eqnarray}
\end{itemize}

We can formulate a quantization condition
by matching the integration constants (in particular $\xi = - \pi/4$)
across the turning points using Airy functions.
The quantization conditions for the scalar and
spinor problem turn out to be,
\begin{eqnarray}
\pi (n + 1/2) &=& \theta_B(r_1,r_2) \\
\label{eq:qf}
\pi (n +  1/2)&=& \theta_F(r_1,r_2)
\end{eqnarray}
respectively.
Note that the quantization condition
for the spinor only works for $m>0$, and no information on the alternative
quantization region for either spinor or scalar can be found with
this analysis.

\begin{figure}[h!]
\begin{center}
\psfig{figure=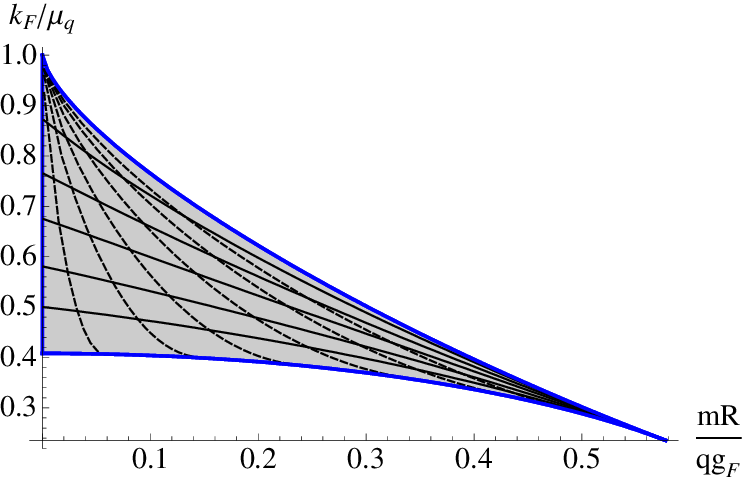,width=8cm}
\psfig{figure=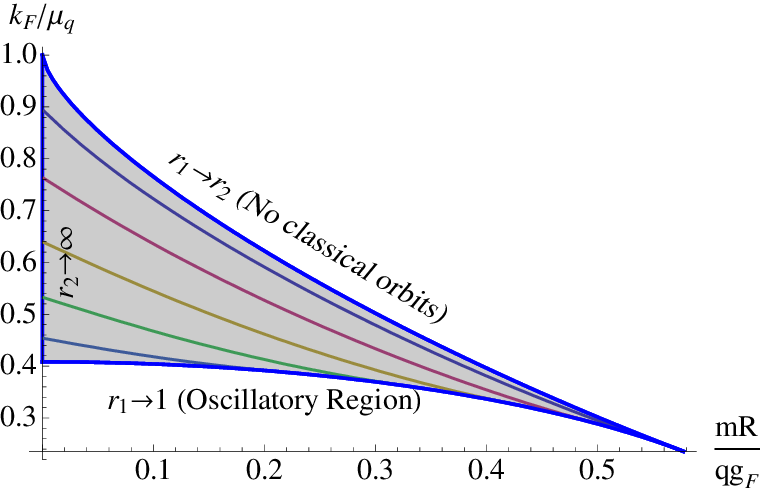,width=8cm}
\end{center}
\caption{The WKB allowed region for $d=3$ showing
contours of fixed $\nu$ and $m$ (above) and fixed $q$ (below)
based on the WKB quantization condition
(\ref{eq:qf}) for $n=0$. \emph{Upper plot:}
The lowest horizontal contour is $\nu=0$, increasing
towards $\nu \rightarrow \infty$ at the upper boundary.
The vertical contours
are for fixed $m$ with $m=0$ lying on the $k_F/\mu_q$  axis.
The contours in this plot demonstrate that the point $P_3$ controls
the asymptotic slope of the fixed $\nu$ contours in Fig.\ref{fig:mq_plot}.
\emph{Lower plot:} The contours of fixed $q$
move towards the upper boundary with increasing $q$. Note that
these contours end on the lower boundary (the oscillatory region).
In particular if we fix $q$ and increase $m$, following
along the contours in this plot, we see that $k_F$ decreases
until eventually the bound state enters the oscillatory region.
\label{fig:allowed3}}
\end{figure}

Using (\ref{eq:qf}) in Fig..~\ref{fig:allowed3}
we plot contours of fixed $\nu$ and $q$
in the WKB parameter
region for $n=0$. We note that the validity
of the WKB approximation is for $n$ large, however
it seems to work remarkably well for $n=0$ the ground state.
One might also imagine that it might be exact
for $n=0$ when the quantization condition forces
us into the limit $k_F,q,m \rightarrow \infty$. Such a situation
does occur at to the upper boundary of
Fig.~\ref{fig:allowed3} when $r_1\rightarrow r_2$.
Actually more care is required
in this limit: for the scalar field we can formulate a scaling
limit in which the potential becomes that of a simple
harmonic oscillator (SHO) located at the radius where $r_1 \rightarrow r_2$.
In this case, the WKB quantization condition should be exact
since it is exact for the SHO. Unfortunately it seems hard to find
the equivalent scaling limit for the spinor, see Sec.\ref{sec:qwkb}
for more discussion of this.

\section{Formulas for $v_F, h_1, h_2$}\label{app:vF}

\newcommand{\bdy}{{\rm bdy}}
\newcommand{\hor}{{\rm hor}}

\subsection{Spinors}

In this section we will
derive general formulas for various coefficients appearing
in equation~\eqref{spinorG}. These coefficients are of great importance in characterizing physical properties of the Fermi surface as they determine the Fermi velocity, the
locations of quasi-particle poles, and the residues at the poles.
The following discussion will be very similar
to the derivation of the Feynman-Hellmann
theorem, which (not coincidentally) is commonly used to determine dispersion relations
in {\it \eg\  } photonic crystals \cite{JoshWinn}.
We will focus on the spinor case, and
comment on the corresponding result
for the charged scalar at the end.

Consider the Dirac equation from~\eqref{actDi}
\be \label{zeDe0}
(\Ga^M \sD_M - m) \psi = 0
\ee
in Fourier space where the Dirac operator $\sD$ depends on $\om$ and $k$, which we will collectively denote as $\lam$. Now suppose that~\eqref{zeDe0} has a solution $\psi_0$ for $\lam = \lam_0$, \ie\
\be \label{zeDe}
(\Ga^M \sD_M |_{\lam_0} - m) \psi_0 = 0 \ .
\ee
Consider varying  $\lam_0 \to \lam_0 + \delta \lam$ with the corresponding solution to~\eqref{zeDe0} given by  $\psi_0 + \delta \lam \psi_1$. $\psi_1$ then satisfies
 \be \label{Diw}
 (\Ga^M \sD_M |_{\lam_0} - m) \psi_1 + \le. \Ga^M {\p \sD_M \ov \p \lam} \ri|_{\lam_0} \psi_0 =0 \ .
 \ee
Multiplying~\eqref{Diw} on the left by $\int_{r_*}^\infty dr \, \sqrt{-g} \bar \psi_0$, integrating by parts, and using~\eqref{zeDe}, we find that
 \be  \label{keyeq}
 W(\infty) - W (r_*) + \int_{r_*}^\infty dr \, \sqrt{-g}  \, \bar \psi_0  \Ga^M {\p \sD_M \ov \p \lam} \psi_0 =0
 \ee
where
 \be
 W = \sqrt{-g} \bar \psi_0 \Ga^r \psi_1  \ .
 \ee

We will now be more specific, taking $\lam= \lam_0$ corresponding to $\om =0, k = k_F$ and $\psi_0$ to be that corresponding to $\eta_+^{(0)}$ defined in~\eqref{spinZ}, \ie\
from~\eqref{tranS} $\psi_0 = (-g g^{rr})^{-{1 \ov 4}} \eta_+^{(0)}$.
Recall that at $k = k_F$, $\eta_+^{(0)}$ is normalizable with $a_+^{(0)} =0$. We will now consider small $\om$ and $k$ variations separately:

\ben

\item Take $\lam = \om$ in~\eqref{keyeq}. Then we have $\psi_1 =  (-g g^{rr})^{-{1 \ov 4}} \eta_+^{(1)}$, where $\eta_+^{(1)}$ was introduced in~\eqref{spiEx}, and
    $W = \bar \eta_+^{(0)}  \Ga^{\ur} \eta_+^{(1)}$.  Equation~\eqref{keyeq} then becomes
\be \label{FHex}
  i b_+^{(0)} a_+^{(1)}  - \le(\bar \eta_+^{(0)}  \Ga^{\ur} \eta_+^{(1)}\ri)\bigr|_{r_*}
  - i J^t|_{\psi_0} = 0
 \ee
with
\bea
 J^t|_{\psi_0}  & = &  \int_{r_*}^\infty dr \, \sqrt{-g} \, \bar \psi_0 \Ga^t \psi_0 \cr
 &=& -\int_{r_*}^\infty dr \, (-g_{rr} g^{tt})^\ha \, (\eta_+^{(0)})^\da \eta_+^{(0)}  \ .
 \label{defJt}
\eea
where in obtaining~\eqref{FHex} we have used~\eqref{spinAs} and~\eqref{realbasis}.
Note that near $r \to r_*$, $\eta_+^{(0)} \sim (r-r_*)^{\nu_k}$, thus
 \be  \label{nearha}
 J^t \propto {1 \ov 2 \nu_k -1} , \quad \nu_k \to \ha
 \ee
 and
 becomes divergent from integration near $r_*$ when $\nu_k \leq \ha$. Since the first term in~\eqref{FHex} is a finite constant, the divergence has to be canceled by the second term in~\eqref{FHex}.
As can be checked explicitly the second term in~\eqref{FHex} is indeed divergent\footnote{Near $r \to r_*$, $\eta^{(1)}_+ \sim (r-r_*)^{\nu_k -1}$.} for $\nu_k \geq \ha$ and precisely cancels that of $J^t$. For $\nu_k < \ha$ the boundary contribution from the horizon vanishes. Thus we find that
 \be \label{paraM1}
  a_+^{(1)} =  {J^t \ov b_+^{(0)}}
  \ee
where for $\nu_k \leq \ha$, $J^t$ should be regularized as discussed above.

\item Take $\lam = k_\perp$ in~\eqref{keyeq}. Then we have $\psi_1 =  (-g g^{rr})^{-{1 \ov 4}} \p_k \eta_+^{(0)} (k_F) $, and
    $W = \bar \eta_+^{(0)}  \Ga^{\ur} \p_k \eta_+^{(0)}$. In this case the horizon contribution in~\eqref{keyeq} vanishes and we find
\be
  i b_+^{(0)} \p_k a_+^{(0)}
  + i J^1|_{\psi_0} = 0
 \ee
with
\bea
 J^1|_{\psi_0} &=&  \int_{r_*}^\infty dr \, \sqrt{-g} \, \bar \psi_0 \Ga^1 \psi_0 \cr
&=& (-1)^\alpha \int_{r_*}^\infty  dr \, (g_{rr} g^{ii})^\ha \, (\eta_+^{(0)})^\da \sig^3 \eta_+^{(0)} \ .
\label{defJ1}
\eea
It can be checked explicitly that $J^1$ is always well-defined and finite. We then conclude that
  \be \label{paraM2}
  \p_k a_+^{(0)} = - {J^1 \ov b_+^{(0)}} \ .
 \ee

\item Denoting $\psi_0^- = (-g g^{rr})^{-{1 \ov 4}} \eta_-^{(0)}$ and
multiplying $\int_{r_*}^\infty dr \sqrt{-g} \bar \psi_{0}^-$ on equation~\eqref{zeDe} we find that
 \be
 \le(\sqrt{-g} \bar \psi_0^- \Ga^r \psi_0 \ri) \bigr|_{r_*} = \le(\sqrt{-g} \bar \psi_0^- \Ga^r \psi_0 \ri) \bigr|_{\infty}
 \ee
 which leads to
 \be \label{paraM3}
 a_-^{(0)} = {V \ov b_+^{(0)}}, \quad V = - i v_+^\da \sig^2 v_-
 \ee
where $v_\pm$ are given by~\eqref{veorP}.

\een

Using~\eqref{paraM1},~\eqref{paraM2} and~\eqref{paraM3} we thus find that the various quantities introduced in~\eqref{thrIM} can be written as (all expressions below  are evaluated at $k=k_F$, $\omega=0$)
 \be \label{varPhP}
 v_F =  {J^1 \ov J^t}, \quad h_1 = - {(b_+^{(0)})^2 \ov J^1}, \quad
 h_2 = {|c(k_F)| V \ov J^1} \ .
 \ee
It can be readily checked that the above expressions do not depend the normalizations of $v_\pm$ (even though $c(k_F)$ does, as discussed after~\eqref{effG}).

For $\nu_k > \ha$, writing $\eta^{(0)}_+  \equiv \left(\begin{matrix} y \cr z \end{matrix} \right)$, from~\eqref{defJt} and~\eqref{defJ1} we can express $v_F$ more explicitly as
\be
v_F =  { \int_{r_*}^\infty  dr \sqrt{g_{rr} g^{ii}} \left( |z|^2 - |y|^2 \right)   \over
\int_{r_*}^\infty  dr \sqrt{g_{rr} (-g^{tt})} \left( |y|^2 + |z|^2 \right)   } . \ee
Since $ { g^{ii} \over -g^{tt}} = f(r) \leq 1$,
the integrands of the numerator and denominator
pointwise have a ratio less than one,
from which it follows that $v_F \leq 1$.
This is borne out by the numerical results
displayed in figure \ref{fig:vF}.
Note that the diverging factor of $ {1\over 2\nu_k - 1}$ in equation~\eqref{nearha}
causes the Fermi velocity to vanish as $\nu_k \to {1\over 2}$.

\subsection{Scalars} \label{app:c4}

Completely parallel analysis can be applied to a scalar. One finds that
\begin{eqnarray}
J^t &=& q \int_{r_*}^\infty dr \sqrt{-g} (-g^{tt})  A_t (\eta^{(0)}_+)^2 \\
J^i &=& k_F \int_{r_*}^\infty dr \sqrt{-g} g^{ii}  (\eta^{(0)}_+)^2
\end{eqnarray}
and
\bea
h_1 &=& \frac{(\Delta - d/2)}{R^{d+1}} \frac{ (b^{(0)}_+)^2}{ J^i} \\
h_2 &=& |c(k_F)| \left( \frac{r_*}{R} \right)^{(d-1)} \frac{ \nu_{k_F} }{J^i}
\eea
and for $\nu_k > \ha$
\be
v_F = {J^i \ov J^t} = {k_F \over q }{ \int_{r_*}^\infty dr \sqrt g g^{ii} \, (\eta_+^{(0)})^2
\over  \int_{r_*}^\infty dr \sqrt g (-g^{tt}) A_t \, (\eta_+^{(0)})^2 } .
\ee
Note that the above equations make it manifest that (for $q > 0$)
\be
h_1, h_2, v_F > 0 \ .
\ee
But the value of $v_F$ is not obviously bounded.
The possibility that this `velocity' may exceed the speed of light is not problematic, since the scalar pole represents an instability
rather than a propagating mode.

\subsection{$v_F, h_1, h_2$ in the WKB approximation}
\label{sec:qwkb}

Here we work in the large $q,k,m$ limit.
In particular we
will take $n$ fixed (and large) such that in this limit
the two turning points come together at some radius
$(r_1 \rightarrow r_2) \equiv r_k$.
We see this by examining the
the quantization condition (\ref{eq:qf}) \emph{assuming} the two turning
points come close together. In this case
we may approximate
$\rho^2 \approx l^2 (r-r_1)(r_2 - r)$ with $l \sim k_F$
from which (\ref{eq:qf}) becomes,
\begin{equation}
(r_2 - r_1)^2 \sim n/k_F
\end{equation}
so indeed for fixed $n$ and large $k_F$, the two radii
come together.
Note in Fig.~\ref{fig:allowed3} this limit corresponds
to the upper boundary where
the radius $r_k$
is determined by the ratio $mR/qg_F$. The
radius $r_k$ moves from the boundary to the horizon
as one varies $mR/qg_F$ from $0$ to $1/\sqrt{3}$.

\begin{figure}[h!]
\centerline{\hbox{\psfig{figure=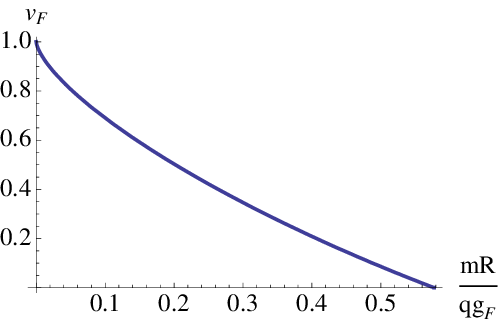,width=8cm}}}
\caption{The Fermi velocity in the limit $q,k_F,m$ large for $d=3$.
The limiting velocity is determined by the radius $r_k$
(which is in turn determined by $mR/q g_F$)
using formula (\ref{eq:vfwkb}).
For $mR/q g_F =0$ the wave function goes towards
the boundary ($r_k \rightarrow \infty$)
and $v_F \rightarrow 1$ . This is consistent with the asymptotic
behavior of $v_F$ in Fig. \ref{fig:vF}, since these
plots are for fixed $m$ and increasing $q$.
\label{fig:vfwkb} }
\end{figure}

The wave function
will then be localized at this radius and expressions
for $v_F, h_1,h_2$ can easily be derived.
We will use (\ref{varPhP}) to compute these quantities.
In the expectation values $J^1$ and $J^t$
we may drop the contribution from the disallowed regions
since it is exponentially small. In the allowed region we may
replace all oscillating functions with their averages,
since for large $n$ they are highly oscillatory:
$\sin^2(\theta_F - \pi/4), \cos^2(\theta_F - \pi/4) \rightarrow 1/2$
etc. In particular, using \eqref{allowf}, we make the replacement:
\begin{equation}
|y|^2 \pm |z|^2 \rightarrow \frac{ D_F^2 (u-k_F\pm ( u + k_F)) }{\rho}
\end{equation}
Finally the integrals we are left with are over
a small interval so we may evaluate all smooth
functions (such as $u, g_{rr},g_{ii} \ldots$) at $r=r_k$.
The result is:
\begin{eqnarray}
J^1 &=& -D_F^2 \left[2 \sqrt{g_{rr} g^{ii}} k_{F}\right]_{r_k}
\int_{r_1}^{r_2} \frac{dr}{\rho}  \\
J^t &=& - D_F^2 \left[2 \sqrt{-g_{rr} g^{tt}} u \right]_{r_k}
\int_{r_1}^{r_2} \frac{dr}{\rho}
\end{eqnarray}
The integral in the above equations evaluates to $ \pi/ l$.
Taking the ratio we find an expression for the Fermi
velocity:
\begin{equation}
\label{eq:vfwkb}
v_F =  \frac{k}{q A_t(r_k)} c_{{\rm light}}^2(r_k)
\end{equation}
where $c_{{\rm light}}(r_k) =  \sqrt{f(r_k)}$ is the local speed of light at radius
$r_k$.  If we interpret $A_t(r_k)$ as the local chemical potential
then this formula is consistent with that of a free
fermion with a relativistic dispersion relation (with the
speed of light replaced by the local speed of light).
Plots of $v_F$ are shown in Fig.~\ref{fig:vfwkb}.

To find $h_1$ and $h_2$ we need to match the various normalizations
in (\ref{wkbsolnf},\ref{allowf}) to the specified normalization
at the horizon (\ref{spinZ}) with $v_+$ given in (\ref{veorP}).
We will only be interested in their exponential behavior:
\begin{eqnarray}
D_F &\sim&  C^F_{2} \sim b_{(0)}^+ \exp( \Gamma_2) \\
D_F &\sim& C^F_{1} \sim  \exp( \Gamma_1)
\end{eqnarray}
so that
\begin{equation}
\label{eq:tun}
h_1 \sim  \exp( -2 \Gamma_2), \quad
h_2 \sim (k_F)^{-2 \nu} \exp( - 2 \Gamma_1 - 2 \Gamma_c)
\end{equation}
where the tunneling rates are given by,
\begin{eqnarray}
\label{eq:g2}
\Gamma_2 &=& \lim_{r_\Lambda
\rightarrow \infty}
\int_{r_2}^{r_\Lambda} dr \sqrt{ g^{ii} g_{rr} } \rho
- m R \log(r_\Lambda) \\
\label{eq:g1}
\Gamma_1 &=& \lim_{r_\epsilon
\rightarrow 0}
\int_{r_\epsilon}^{r_1} dr \sqrt{ g^{ii} g_{rr} } \rho
+ \nu \log(r_\epsilon)
\end{eqnarray}
and for $h_2$ we have included the exponential behavior
of $|c(k_F)| \sim k_F^{-2 \nu} e^{-2 \Gamma_c}$.
Together the factors in (\ref{eq:g1}) should be interpreted
as a tunneling amplitude from the bound state
to the inner region. In the limit considered here
$\nu$ is large so eventually the $k_F^{ - 2 \nu}$
term will dominate in $h_2$ which will always be asymptotically
small. On the other hand $\Gamma_2$
in (\ref{eq:g2}), once regulated, actually turns out to be
always negative. Hence $h_1$ will be asymptotically large
in this limit.

Unfortunately as it stands it is not clear these
results apply to the ground state $n=0$ with
asymptotically large $k,q,m$. However for
the scalar field a scaling limit exists (similar to the limits
considered in the next section) where the potential
as $r_1\rightarrow r_2$ can be approximated by that
of a simple harmonic oscillator. By matching the SHO wave function
onto the WKB wave function in the disallowed region, we can show that the
answers (\ref{eq:vfwkb},\ref{eq:tun}) also apply to any $n$ for the boson.
A similar result \emph{may} be derivable
for the spinor, but we have not been able to formulate it yet.

\section{Two-point functions for charged fields in AdS$_2$} \label{app:ads2}

Here we discuss retarded Green functions for operators in a CFT dual to a
charged scalar and a spinor field in AdS$_2$. We will give main results here, leaving derivations and more extended discussion elsewhere~\cite{WithNabil}.

\subsection{Scalars}

We consider the following quadratic scalar action
\be \label{2scaA}
S = -\int  d^{2} x \sqrt{-g} \, \le[(D^\al \phi)^* D_\al \phi  + m^2 \phi^* \phi \ri]
\ee
with $D_\al = \p_\al - i q A_\al $ and the background metric and gauge field given by
   \be \label{aads2}
 {ds^2 } = {R_2^2 \ov \ze^2 } \le(- d \tau^2 +  d \ze^2 \ri), \qquad  A_\tau  =  {e_d  \ov  \ze} \ .
 \ee
Writing $\phi (\tau, \ze) = e^{- i \om \tau} \phi(\om,\ze)$, the wave equation for $\phi$ can be written as
 \be \label{eep2}
 -\p_\ze^2 \phi + V (\ze) \phi =0
 \ee
 with
 \be \label{eoep}
 V(\ze) = {m^2 R_2^2 \ov \ze^2} - \le(\om + {q e_d \ov \ze} \ri)^2 \ .
\ee
To find the conformal dimension of the operator $\sO$ dual to $\phi$, we
solve~\eqref{eep2} near the boundary $\ze \to 0$ and find that
 \be \label{exwp1}
 \phi =A \ze^{\ha- \nu} \le(1 + O(\ze) \ri) + B
 \ze^{\ha + \nu} \le(1 + O(\ze) \ri),  \qquad \ze \to 0
 \ee
with
\be \label{dim1}
\nu = \sqrt{m^2 R_2^2 - q^2 e_d^2 + {1 \ov 4} - i \ep} \ .
\ee

Since $\om$ in~\eqref{eep2}--\eqref{eoep} can be scaled away from redefining $\ze$, we conclude that in~\eqref{exwp1}, $A \sim \om^{\ha - \nu}$ and $B \sim \om^{\ha + \nu}$ and thus (after imposing infalling boundary condition on $\phi$  at the horizon)
 \be \label{eemm}
 \sG_R (\om) \propto  {B \ov A} \sim \om^{2 \nu}
 \ee
 which implies a coordinate space correlation function
 \be \label{ero1}
 \sG_R (\tau) \sim {1 \ov \tau^{2 \delta}} \
 \ee
with the conformal dimension $\delta$ of $\sO$ given by\footnote{One can also reach the same conclusion by assuming a boundary coupling $\int d \tau \, \phi_0 \sO$ and considering a conformal scaling in the boundary theory.}
 \be \label{dim2}
 \delta = \ha + \nu \ .
 \ee
Notice that dimension $\delta$ also depends on charge $q$. In particular, it is possible for $\nu$ to become imaginary when $q$ is sufficiently large.
 Physically this reflects the fact that in the constant electric field~\eqref{aads2}
particles with a sufficiently large charge $q$ can be pair produced.\footnote{
As we discuss in the main text, when embedded in the full theory this causes an instability for scalars, but not for spinors.} When $\nu$ is imaginary, there is an ambiguity in specifying $\sG_R$ since one can in principle choose either term in~\eqref{exwp1} as the source term. We will follow the prescription
as determined by the $-i \ep$ term in~\eqref{dim1}, as this will be the choice one needs to use when patching the AdS$_2$
region to the outer region of the full black hole geometry.

Equation~\eqref{eep2} can in fact be solved exactly and one finds the full retarded Green function is given by~\cite{WithNabil}
 \be \label{exbG}
 \sG_R (\om) =  
 e^{-i \pi \nu} \frac{ \Gamma (-2\nu ) \Gamma \left(\frac{1}{2}+ \nu-i q e_d\right)}{\Gamma
(2\nu )\Gamma \left(\frac{1}{2}- \nu-i q e_d\right)   } (2\om)^{2 \nu}\ .
 \ee
Equation~\eqref{exbG} has the form of the retarded two-point function of
a scalar operator in a (1+1)-dimensional CFT with left/right-moving dimensions and momenta
 \be
 \delta_L = \delta_R = \ha +\nu, \qquad p_L = q, \qquad p_R = \om
 \ee
in a $(1+1)$-dimensional CFT with left/right temperatures given by
  \be \label{tl}
  T_L = {1 \ov 4 \pi e_d}, \qquad T_R =0  \ .
  \ee
Thus it is tempting to interpret the CFT$_1$ dual to AdS$_2$ as the right moving sector of a $(1+1)$-dimensional CFT.\footnote{We should caution, however, that many other aspects of this theory should be studied before one can really draw a conclusion.}

Also note that the advanced function is given by
 \be \label{BAG}
 \sG_A (\om) =
 e^{i \pi \nu} \frac{ \Gamma (-2\nu ) \Gamma \left(\frac{1}{2}+ \nu + i q e_d\right)}{\Gamma
(2\nu )\Gamma \left(\frac{1}{2}- \nu + i q e_d\right)   } (2\om)^{2 \nu}\
 \ee

\subsection{Spinors} \label{app:spinC}

We consider the following quadratic action for a spinor field $\psi$ in the geometry~\eqref{aads2}
  \be \label{ADact}
 S =  \int d^{2} x \sqrt{-g} \, i (\bar \psi \Ga^\al D_\al \psi  - m \bpsi \psi
 + i \tildem \, \bpsi \Gamma \psi)
 \ee
where we have included a time-reversal violating mass term proportional to $\tilde m$ which in our application will be related to momentum in $\RR^{d-1}$.

It is convenient to choose the following Gamma matrices\footnote{They are chosen
to be compatible with the choice made in appendix \ref{app:fer}, with
\eqref{aads2} arising as the near horizon limit.  Note the reversal of the gamma matrix for the radial coordinate which reflects the change in orientation between $r$ and $\zeta$.}
\be
\Ga^{\underline \ze}= \sigma^3 ,~~~\Ga^{\underline{\tau}} =  i  \sigma^1 ,~~~
\Ga =- \sigma^2
\ee
where underlined indices again denote those in the tanget frame and $\sig^i$ are standard sigma matrices. The equations of motion
for $\psi$ can be written in Fourier space as
\be\label{spinads2}
0 =  \partial_\ze \Phi + i \sigma^2 \le(\om + {q e_d \ov \ze} \ri) \Phi  -{R_2 \ov \ze}
  \le(m \sig^3 + \tilde m \sig^1
\right) \Phi \
\ee
with $\Phi = (- g g^{\ze \ze})^{-{1 \ov 4}} \psi$.  Near the boundary $\ze \to 0$, equation~\eqref{spinads2} becomes
 \be\label{defofU}
 \ze \p_\ze \Phi = U \Phi , \quad U =  \left( \begin{matrix}
m R_2 & \tilde m R_2 - q e_d
\cr
\tilde m R_2 + q e_d  & -m R_2
\end{matrix} \right) \ .
 \ee
Thus as $\ze \to 0$, $\Phi$ can be written as
 \be
 \Phi = A \, v_- \ze^{- \nu} \le(1 + O(\ze) \ri)+
 B \, v_+ \ze^{\nu} \le(1 + O(\ze) \ri)
 \ee
where $v_\pm$ are real eigenvectors of $U$ with eigenvalues $\pm \nu$ respectively and
\be \label{spinnu}
\nu = \sqrt{(m^2 + \tilde m^2) R_2^2 - q^2 e_d^2 - i \ep} \ .
\ee

Imposing the infalling boundary condition for $\Phi$ at the horizon, the retarded Green function for the boundary operator in the CFT$_1$ dual to $\psi$ can then be written as\footnote{As discussed earlier for a bulk spinor field the number of components of the boundary operator is always half of that of the bulk field.}
 \be  \label{eiie}
 \sG_R (\om) = {B \ov A} \sim \om^{2 \nu} \
 \ee
again suggesting the operator dimension to be given by $\delta = \ha + \nu$.
 There is, however, an ambiguity in~\eqref{eiie} as the ratio depends on the relative normalization of $v_\pm$; if we take $v_\pm \to \lam_\pm v_\pm$, then $\sG_R \to {\lam_- \ov \lam_+} \sG_R$. This ambiguity\footnote{It will be discussed in more detail in~\cite{WithNabil}.} will not be relevant for the present paper as we will see later the
ambiguity cancels in the matching procedure and the correlation function for the full geometry will not depend on the normalization.

As in the scalar case differential equation~\eqref{spinads2} can be solved exactly and
 with the choice of $v_\pm$ by\footnote
 {Note that the subscripts on $v_\pm$ are chosen to indicate the sign of the eigenvector of $U$;
 this leads to the unfortunate but innocuous notation clash in Eqn.~\eqref{spinZ} since $v_\mp$ appears in the 
 outgoing solution $\eta_\pm$.}
 \be \label{eignV}
v_\pm =   \left( \begin{matrix}
m R_2 \pm \nu
\cr
\tilde m R_2 + q e_d
\end{matrix} \right) \ ,
\ee
we find that $\sG_R$ can be written as~\cite{WithNabil}
 \bea
 \sG_R (\om) & = &  e^{- i \pi \nu} \frac{\Gamma (-2 \nu ) \, \Gamma \left(1+\nu -i q e_d \right)}{\Gamma (2 \nu )\,   \Gamma \left(1-\nu -i q e_d\right)} \cr
 &\times & \frac{ \le(m - i \tildem \ri) R_2- i q e_d - \nu}
{  \le(m - i \tildem \ri) R_2 - i q e_d +  \nu} \( 2\om\)^{2 \nu} \ .
 \label{AeFG}
 \eea
The advanced function is given by
 \bea
 \sG_A (\om) &=& e^{ i \pi \nu} \frac{\Gamma (-2 \nu ) \, \Gamma \left(1+\nu + i q e_d \right)}{\Gamma (2 \nu )\,   \Gamma \left(1-\nu +i q e_d\right)} \cr
 & \times & \frac{ \le(m + i \tildem \ri) R_2 +i q e_d - \nu}
 {  \le(m + i \tildem \ri) R_2 + i q e_d +  \nu} (2\om)^{2 \nu} \ .
  \label{AAG}
 \eea
Equation~\eqref{AeFG} is again suggestive of spin-$\ha$ operator with left/right-moving dimensions and momenta
 \be
 \label{dim4}
\delta_L = 1 + \nu, \qquad \delta_R = \ha + \nu, \qquad p_L = q, \qquad p_R = \om
 \ee
in a $(1+1)$-dimensional CFT with temperatures given by~\eqref{tl}.

Finally note as with scalar case,~\eqref{spinnu} can become imaginary when $q$ is sufficiently large, in which case the prescription for determining $\sG_R$ is again determined by the $- i \ep$ term.

\subsection{A useful formula} \label{app:ana}

Here we give a nice formula for $\sG_R$ which can be used to derive equations~\eqref{phaseE}--\eqref{modSp} in the main text.
For scalar, using~\eqref{exbG} and~\eqref{BAG} we find that
 \be \label{anid}
 {\sG_R (\om) \ov \sG_A (\om)} = e^{-2 \pi i \nu} {\cos \pi (\nu + iq e_d) \ov \cos \pi (\nu - i q e_d )} = {e^{-2 \pi i \nu} + e^{-2 \pi q e_d} \ov e^{2 \pi i \nu} + e^{-2 \pi q e_d}} \ .
 \ee
For spinor from~\eqref{AeFG} and~\eqref{AAG} we find
 \be \label{oo}
  {\sG_R (\om) \ov \sG_A (\om)} = - e^{-2 \pi i \nu} {\sin \pi (\nu + iq e_d) \ov \sin \pi (\nu -
i q e_d )} = {e^{-2 \pi i \nu} - e^{-2 \pi q e_d} \ov e^{2 \pi i \nu} - e^{-2 \pi q e_d}} \ .
 \ee
Writing $\sG_R = c \om^{2\nu}$, then for real $\nu$ equations~\eqref{anid} and~\eqref{oo} give the phase of $c$ and for imaginary $\nu$ they give the modulus of $c$.

\subsection{Finite temperature generalization}

One can in fact generalize the above discussion to finite temperature, {\it \ie\ } to the AdS$_2$ part of the metric~\eqref{ads2T}.  Details will be given in \cite{WithNabil};
we mention the result here because it gives useful information
about the analytic structure of the Green's functions in $\omega$ at zero temperature.
One finds respectively for a scalar and a spinor
 \bwt
 \be
 \sG_R (\om) = (4 \pi T)^{2 \nu} \frac{ \Gamma (-2\nu ) \Gamma \left(\frac{1}{2} +\nu-\frac{i \omega }{2 \pi T }+
i q e_d\right)\Gamma \left(\frac{1}{2}+ \nu-i q e_d\right)}{\Gamma
(2\nu )\Gamma \left(\frac{1}{2}-\nu -\frac{i \omega }{2 \pi T }+i q e_d
\right)\Gamma \left(\frac{1}{2}- \nu-i q e_d\right)   }
 \ee
\be
 \sG_R (\omega)= (4 \pi T)^{2 \nu} {\Gamma (-2 \nu )  \ov \Gamma (2 \nu )}
  {\Gamma (\ha+\nu -\frac{i \omega }{2 \pi T }+i q e_d )\,  \Gamma \left(1+\nu -i q e_d \right)\ov \Gamma \left(\frac{1}{2}-\nu -\frac{i \omega }{2 \pi T}+i
q e_d \right)\, \Gamma \left(1-\nu -i q e_d \right)}  \cdot \frac{(m - i \tilde m) R_2-
i q e_d - \nu}{ (m - i \tilde m) R_2- i q e_d +  \nu}
\ee
\ewt
Note that the branch point at $\om =0$ of zero temperature now disappears and
the branch cut is replaced at finite temperature by a line of poles parallel to the negative imaginary axis. In the zero temperature limit the pole line becomes a branch cut.
Similar phenomena have been observed previously \cite{Festuccia:2005pi}.

\subsection{The spinor pole is in the lower half plane} 
\label{app:piambiguity}

We remarked below equation \eqref{phaseE} that 
our expression for the phase of the IR CFT Green's function
suffers from a possible additive ambiguity by an integer multiple of $\pi$.  
Here we provide evidence that the additive $\pi$-ambiguity in the phase of $\sG_R$ and 
the sign variation of the UV coefficient ${a^{(0)}_- \over \partial_k a_+^{(0)} }$ 
precisely cancel each other to leave behind a smooth behavior of the coefficient of $\omega^{2\nu}$ 
in the spinor Green's function $G_R$.
Further, the phase of this coefficient is such that the quasiparticle
pole is always in the lower-half complex plane.
Recall that we argued that this conclusion followed from the expression \eqref{phaseE}
for the phase of $\sG_R$ combined with $h_1, h_2>0$.  
In fact, $h_2$ can be negative; when this happens, the phase of $\sG_R$ differs
from the expression in $\eqref{phaseE}$ by $\pi$, canceling the issue
in the full Green's function $G_R$.  

\begin{figure}[h!]
 \begin{center}
A ~\includegraphics[scale=0.3]{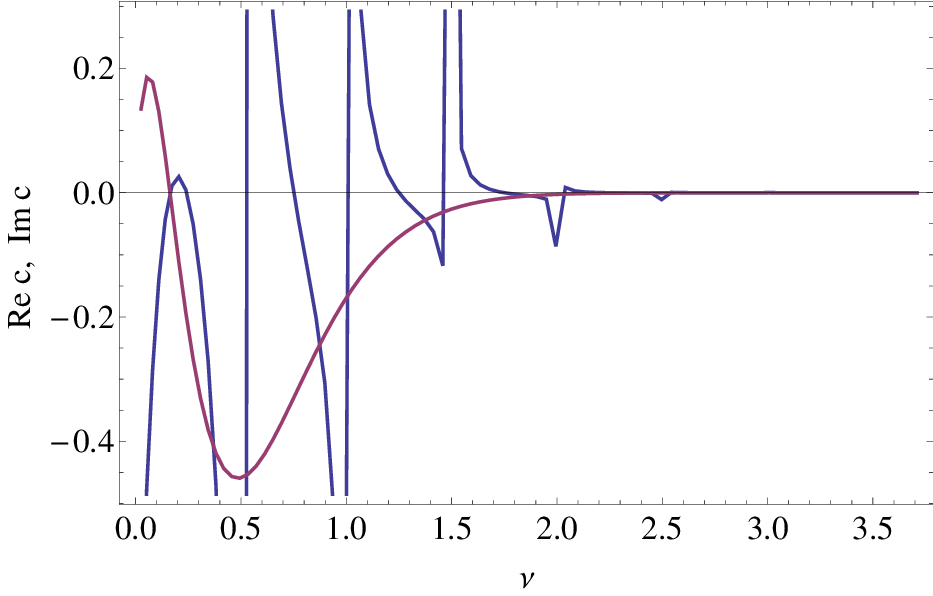}
~~~B~\includegraphics[scale=0.3]{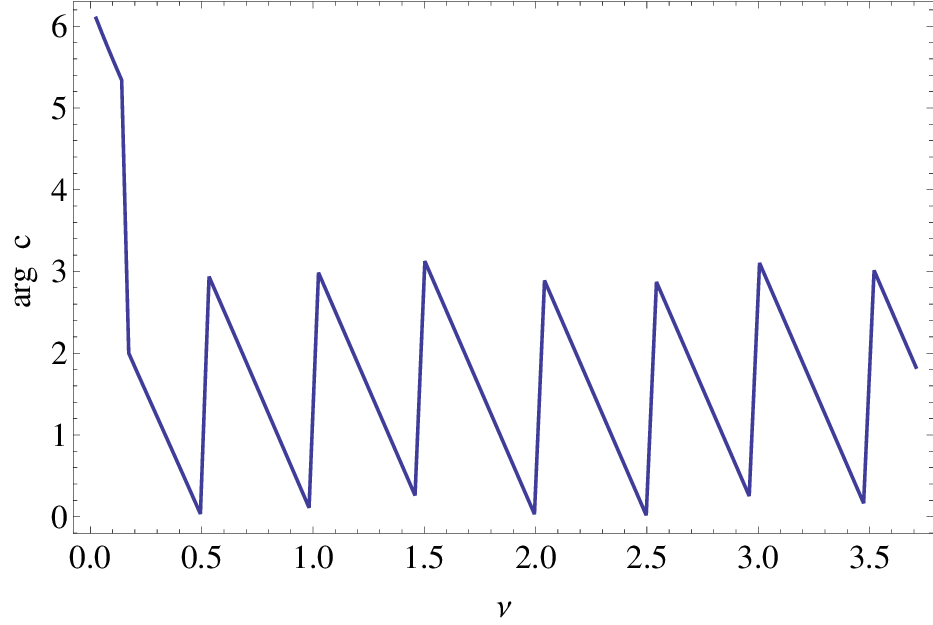}
~~~C~\includegraphics[scale=0.5]{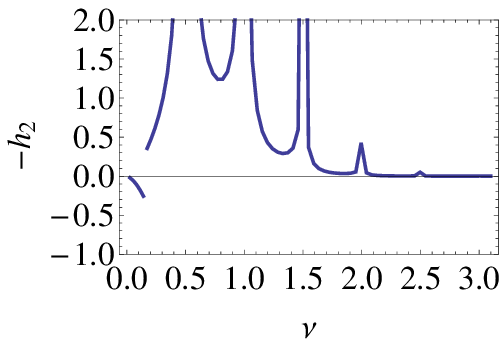}
\caption{\label{fig:anomaloush2ck} 
A: The phase and amplitude of $c(k_F)$ as a function of $\nu$ on the primary Fermi surface of $G_2$ for $m= .4$.
Note that the phase jumps by $\pi$ when $c(k_F)$ has a zero.
B: As defined in Eqn.~\eqref{spinorG}, the sign of $h_2$ (for $m=.4, \alpha=2$) jumps at $\nu = mR_2 \approx .16$.
In spite of the singularity in $h_2/|c(k_F)|$ visible in the lower right panel of Figure~\ref{fig:h12}, 
$|h_2|$ as a function of $\nu$ is completely smooth.  
}
\end{center}
\end{figure}

Note that for the case $m=.4, q=1, \alpha=2$, $h_2$ changes
sign as $\nu$ varies past $\nu = mR_2 \approx .16 $.  
As shown in Figure~\ref{fig:anomaloush2ck}A, 
the phase of $c(k_F)$ also jumps at this value of $\nu$.
Indeed, $c(k_F)$ has a zero, as a complex function, at this value of $\nu$
(recall that $\text{arg}(x)$ jumps by $ \pi$ as $x$ varies from $0^-$ to $0^+$). 
Note that the additive ambiguity in $\gamma_k$ is `topological', in the sense that 
only at zeros or singularities of $c(k)$ can the choice of branch change,
and the phase varies smoothly otherwise.
In Figure~\ref{fig:anomaloush2ck}B we show that 
the quantity $|h_2|$ (which is what enters the full Green's function)
is smooth near this value of $\nu$.
This phenomenon therefore is an artifact of the matching procedure; 
although the IR quantity $c(k_F)$ and the UV quantity 
${a^{(0)}_- \over \partial_k a_+^{(0)} }$ are each singular at this point, 
these singularities have no physical consequence,
since these quantities enter the Green's function only in the combination 
$|h_2|e^{i \gamma_k}$.

The origin of the zero of the IR CFT Green's function is simple to understand.
It is the prefactor 
\be
\frac{ \le(m - i \tildem \ri) R_2- i q e_d - \nu}
{  \le(m - i \tildem \ri) R_2 - i q e_d +  \nu} \ee
which is vanishing.
This happens when both real and imaginary parts of the numerator vanish.
This requires $ m R_2 = \nu $ and $ \tilde m R_2 = qe_d$;
the second equality follows from the first by the definition of $\nu$.
When the second equality is true, the matrix $U$ in 
Eqn.~\eqref{defofU} becomes diagonal.
In this limit, the choice of normalization of the eigenvectors $v_\pm$ given in Eqn.~\eqref{eignV}
ceases to be useful.
As we remarked below Eqn.~\eqref{eiie}, rescaling the choice of eigenvectors rescales the
answer for the IR CFT Green's function;
if we rescale $v_\pm$ to keep finite eigenvectors as $\nu \to mR_2$, 
the IR CFT Green's function $\sG_R$ will also stay finite.
This makes it clear that the singularity under discussion
here can have no physical effect.

In retrospect, it would have been preferable to {\it define} $h_2$ to be 
the magnitude of the coefficient of $\omega^{2\nu}$ in Eqn.~\eqref{spinorG}:
\be
 h_2^{\text{better}} e^{i \gamma_k} \equiv - \frac{a_-^{(0)}}{\partial_k a_+^{(0)}} c(k_F)~;
\ee
with $\gamma_k$ given as in \eqref{phaseE}, $h_2^{\text{better}}$ is positive.
Rather than changing our definition, we felt that it would be more useful to highlight this issue.

\begin{thebibliography}{9}
%



\bibitem{AdS/CFT}
J.~M.~Maldacena,
Adv.\ Theor.\ Math.\ Phys.\  {\bf 2}, 231 (1998);
S.~S.~Gubser, I.~R.~Klebanov and A.~M.~Polyakov,
Phys.\ Lett.\ B {\bf 428}, 105 (1998); 
E.~Witten,
Adv.\ Theor.\ Math.\ Phys.\ {\bf 2}, 505 (1998).


\bibitem{Lee:2008xf}
  S.~S.~Lee,
  arXiv:0809.3402 [hep-th].

\bibitem{Liu:2009dm}
  H.~Liu, J.~McGreevy and D.~Vegh,
  arXiv:0903.2477 [hep-th].






\bibitem{Cubrovic:2009ye}
  M.~Cubrovic, J.~Zaanen and K.~Schalm,
  arXiv:0904.1993 [hep-th].

\bibitem{Chamblin:1999tk}
  A.~Chamblin, R.~Emparan, C.~V.~Johnson and R.~C.~Myers,
  Phys.\ Rev.\  D {\bf 60}, 064018 (1999)
  arXiv:hep-th/9902170.


\bibitem{Lu:2009gj}
  H.~Lu, J.~w.~Mei, C.~N.~Pope and J.~F.~Vazquez-Poritz,
  Phys.\ Lett.\  B {\bf 673}, 77 (2009)
  arXiv:0901.1677 [hep-th].





\bibitem{Senthil:0803}
 T.~Senthil,
 arXiv:0803.4009 [cond-mat];
%
 arXiv:0804.1555 [cond-mat].


\bibitem{varma}
  C.~M.~Varma, P.~B.~Littlewood, S.~Schmitt-Rink, E.~Abrahams and A.~E.~Ruckenstein,
  Phys.\ Rev.\ Lett.\  {\bf 63}, 1996 (1989).

\bibitem{Romans:1991nq}
  L.~J.~Romans,
  Nucl.\ Phys.\  B {\bf 383}, 395 (1992)
  arXiv:hep-th/9203018.

\bibitem{Son:2002sd}
  D.~T.~Son and A.~O.~Starinets,
  JHEP {\bf 0209}, 042 (2002)
  arXiv:hep-th/0205051.

\bibitem{adstworefs}
  A.~Strominger,
  JHEP {\bf 9901} (1999) 007
  arXiv:hep-th/9809027;

\bibitem{Maldacena:1998uz}
  J.~M.~Maldacena, J.~Michelson and A.~Strominger,
  JHEP {\bf 9902}, 011 (1999)
  arXiv:hep-th/9812073.


\bibitem{greybody}
  J.~M.~Maldacena and A.~Strominger,
  Phys.\ Rev.\  D {\bf 55}, 861 (1997)
  [arXiv:hep-th/9609026];
  J.~M.~Maldacena and A.~Strominger,
  Phys.\ Rev.\  D {\bf 56}, 4975 (1997)
  [arXiv:hep-th/9702015];
  I.~R.~Klebanov,
  Nucl.\ Phys.\  B {\bf 496}, 231 (1997)
  [arXiv:hep-th/9702076];
  S.~S.~Gubser, I.~R.~Klebanov and A.~A.~Tseytlin,
  Nucl.\ Phys.\  B {\bf 499}, 217 (1997)
  [arXiv:hep-th/9703040].


\bibitem{Pioline:2005pf}
  B.~Pioline and J.~Troost,
  JHEP {\bf 0503}, 043 (2005)
  [arXiv:hep-th/0501169].

\bibitem{Gubser:2008px}
  S.~S.~Gubser,
  arXiv:0801.2977 [hep-th].


\bibitem{Hartnoll:2008vx}
  S.~A.~Hartnoll, C.~P.~Herzog and G.~T.~Horowitz,
  Phys.\ Rev.\ Lett.\  {\bf 101}, 031601 (2008)
  arXiv:0803.3295 [hep-th].



\bibitem{Hartnoll:2008kx}
  S.~A.~Hartnoll, C.~P.~Herzog and G.~T.~Horowitz,
  JHEP {\bf 0812}, 015 (2008)
  arXiv:0810.1563 [hep-th].


\bibitem{Denef:2009tp}
  F.~Denef and S.~A.~Hartnoll,
  arXiv:0901.1160 [hep-th].


\bibitem{WithNabil}
Work in progress with Nabil Iqbal.

\bibitem{nonaly}
A.~V.~Chubukov and D~L.~Maslov, Phys,\ Rev.\ B {\bf  68}, 155113 (2003).


\bibitem{volovik}
G.~E.~Volovik, 
 Pis'ma Zh. \ Eksp. \ Teor. \ Fiz., {\bf 53}, 208 (1991);
 G.~E.~Volovik, 
 [cond-mat/0601372]




\bibitem{Holstein:1973zz}
  T.~Holstein, R.~E.~Norton and P.~Pincus,
  Phys.\ Rev.\  B {\bf 8}, 2649 (1973).

\bibitem{Polchinski:1993ii}
  J.~Polchinski,
  Nucl.\ Phys.\  B {\bf 422}, 617 (1994)   arXiv:cond-mat/9303037.

\bibitem{Nayak:1993uh}
  C.~Nayak and F.~Wilczek,
  Nucl.\ Phys.\  B {\bf 417}, 359 (1994)
  arXiv:cond-mat/9312086,
  Nucl.\ Phys.\  B {\bf 430}, 534 (1994)
  arXiv:cond-mat/9408016.


\bibitem{Halperin:1992mh}
  B.~I.~Halperin, P.~A.~Lee and N.~Read,
  Phys.\ Rev.\  B {\bf 47}, 7312 (1993).

  \bibitem{altshuler-1994}
B.~L.~Altshuler, L.~B.~Ioffe and A.~J.~Millis (1994).

\bibitem{Schafer:2004zf}
  T.~Schafer and K.~Schwenzer,
  Phys.\ Rev.\  D {\bf 70}, 054007 (2004)
  arXiv:hep-ph/0405053.





\bibitem{Boyanovsky}
  D.~Boyanovsky and H.~J.~de Vega,
  Phys.\ Rev.\  D {\bf 63}, 034016 (2001)
  arXiv:hep-ph/0009172;


\bibitem{sungsik-2009}
  S.~S.~Lee, arXiv:0905.4532 [cond-mat].



\bibitem{Witten:2001ua}
  E.~Witten,
  arXiv:hep-th/0112258.

\bibitem{Klebanov:1999tb}
  I.~R.~Klebanov and E.~Witten,
  Nucl.\ Phys.\  B {\bf 556}, 89 (1999)
  arXiv:hep-th/9905104.



\bibitem{Iqbal:2008by}
  N.~Iqbal and H.~Liu,
  Phys.\ Rev.\  D {\bf 79}, 025023 (2009)
  arXiv:0809.3808 [hep-th].

\bibitem{Iqbal:2009fd}
  N.~Iqbal and H.~Liu,
  arXiv:0903.2596 [hep-th].

\bibitem{Mueck:1998iz}
    M.~Henningson and K.~Sfetsos,
  Phys.\ Lett.\  B {\bf 431}, 63 (1998)
  arXiv:hep-th/9803251;
  W.~Mueck and K.~S.~Viswanathan,
  Phys.\ Rev.\  D {\bf 58}, 106006 (1998)
  arXiv:hep-th/9805145.


\bibitem{Douglas:2006es}
  M.~R.~Douglas and S.~Kachru,
  Rev.\ Mod.\ Phys.\  {\bf 79}, 733 (2007)
  [arXiv:hep-th/0610102].


\bibitem{lazur}
O.~K.~Reity and V.~Y.~Lazur, 
Proceedings of Institute of Mathematics of NAS of Ukraine, 
{\bf 43}, 676 (2002).

\bibitem{Faulkner:2008hm}
  T.~Faulkner and H.~Liu,
  arXiv:0812.4278 [hep-th].











\bibitem{JoshWinn}
J.~D.~Joannopoulos, R.~D.~Meade, J.~N.~Winn,
{\it Photonic Crystals: Molding the Flow of Light}
(Princeton Univ. Press, Princeton, NJ, 1995)


\bibitem{Festuccia:2005pi}
  G.~Festuccia and H.~Liu,
  JHEP {\bf 0604}, 044 (2006)
  arXiv:hep-th/0506202.








\bibitem{Faulkner:2010tq}
  T.~Faulkner and J.~Polchinski,
  arXiv:1001.5049 [hep-th].



\end{thebibliography}
\end{document}